\documentclass[11pt,a4paper]{article}
\usepackage{jheppub}
\pdfoutput=1

\graphicspath{{Extra_plots/}}
\graphicspath{{EWBG_only/}}
\graphicspath{{Combined_leq/}}
\graphicspath{{Combined_eq/}}
\graphicspath{{GW_signals/}}

\usepackage{color,slashed,mathrsfs,float,bm,mathtools,xspace,makecell,blindtext}
\usepackage{booktabs,todonotes}

\title{Gravitational waves and electroweak baryogenesis in a global study of the extended scalar singlet model}

\author[a,b,1]{Ankit Beniwal,\note{ORCID ID: \href{https://orcid.org/0000-0003-4849-0611}{0000-0003-4849-0611}}}
\author[b,c,d,2]{Marek Lewicki,\note{ORCID ID: \href{https://orcid.org/0000-0002-8378-0107}{0000-0002-8378-0107}}}
\author[b]{Martin White}
\author[b,3]{and Anthony G. Williams\note{ORCID ID: \href{https://orcid.org/0000-0002-1472-1592}{0000-0002-1472-1592}}}

\affiliation[a]{The Oskar Klein Centre for Cosmoparticle Physics, Department of Physics, \\
Stockholm University, AlbaNova, SE-106 91 Stockholm, Sweden}
\affiliation[b]{ARC Centre of Excellence for Particle Physics at the Terascale (CoEPP) and CSSM, \\
Department of Physics, University of Adelaide,
South Australia 5005, Adelaide, Australia}
\affiliation[c]{Kings College London, Strand, London, WC2R 2LS, United Kingdom}
\affiliation[d]{Faculty of Physics, University of Warsaw ul.\ Pasteura 5, 02-093 Warsaw, Poland}

\emailAdd{ankit.beniwal@fysik.su.se}
\emailAdd{marek.lewicki@kcl.ac.uk}
\emailAdd{martin.white@adelaide.edu.au}
\emailAdd{anthony.williams@adelaide.edu.au}

\abstract{We perform a global fit of the extended scalar singlet model with a fermionic dark matter (DM) candidate.~Using the most up-to-date results from the \emph{Planck} measured DM relic density, direct detection limits from the XENON1T (2018) experiment, electroweak precision observables and Higgs searches at colliders, we constrain the 7-dimensional model parameter space.~We also find regions in the model parameter space where a successful electroweak baryogenesis (EWBG) can be viable.~This allows us to compute the gravitational wave (GW) signals arising from the phase transition, and discuss the potential discovery prospects of the model at current and future GW experiments.~Our global fit places a strong upper \emph{and} lower limit on the second scalar mass, the fermion DM mass and the scalar-fermion DM coupling.~In agreement with previous studies, we find that our model can simultaneously yield a strong first-order phase transition and saturate the observed DM abundance. More importantly, the GW spectra of viable points can often be within reach of future GW experiments such as LISA, DECIGO and BBO.}

\preprint{ADP-18-27-T1075, KCL-PH-TH/2018-54}


\arxivnumber{1810.02380}

\definecolor{aogreen}{rgb}{0.0, 0.5, 0.0}

\everymath{\displaystyle}

\newcommand{\PP}{\Phi^\dagger}
\newcommand{\pp}{\Phi}

\newcommand{\calM}{\mathcal{M}}
\newcommand{\calN}{\mathcal{N}}
\newcommand{\calX}{\mathcal{X}}
\newcommand{\calY}{\mathcal{Y}}

\newcommand{\lagr}{\mathscr{L}}
\newcommand{\like}{\mathcal{L}}
\newcommand{\ovr}{\overline}

\newcommand{\sa}{\sin\alpha\,}
\newcommand{\ca}{\cos\alpha\,}

\newcommand{\micro}{\textsf{micrOMEGAs}\xspace}
\newcommand{\microver}{\textsf{micrOMEGAs\_v4.3.5}\xspace}

\newcommand{\lhepver}{\textsf{LanHEP\_v3.2.0}\xspace}
\newcommand{\chep}{\textsf{CalcHEP}\xspace}

\newcommand{\DE}{\textsf{Diver\_v1.0.4}\xspace}
\newcommand{\HBver}{\textsf{HiggsBounds\_v4.3.1}\xspace}
\newcommand{\HB}{\textsf{HiggsBounds}\xspace}
\newcommand{\HSver}{\textsf{HiggsSignals\_v1.4.0}\xspace}

\newcolumntype{?}[1]{!{\vrule width #1}}

\begin{document}

\maketitle

\flushbottom

\section{Introduction}
The discovery of the Higgs boson at the LHC \cite{Aad:2012tfa, Chatrchyan:2012xdj} has finally completed the Standard Model (SM) of particle physics.~Not only does it provide a new way to study the properties of the Higgs boson, it also offers a way to investigate the details of electroweak symmetry breaking (EWSB). Meanwhile, a more recent observation of the first gravitational wave (GW) signal \cite{Abbott:2016blz} and subsequent discoveries \cite{Abbott:2016nmj,Abbott:2017ylp,Abbott:2017vtc,Abbott:2017oio,TheLIGOScientific:2017qsa,Monitor:2017mdv,Abbott:2017gyy} have opened up a new window to probe the early history of our universe.~In particular, rather violent events such as the first-order electroweak phase transition (EWPT) would necessarily leave GW imprints. With the current and future generations of ground/space-based GW experiments, we can hope to observe such signals~\cite{Caprini:2015zlo,Weir:2017wfa,Caprini:2018mtu}.~The existence of dark matter (DM) also offers a way to probe the early history of our universe. With the current generation of direct DM searches, experiments are probing the DM-nucleon interaction with increasing sensitivity and placing strong limits on the allowed particle DM models. 

Motivated by the above experimental probes that are constantly developing, we revisit an extended scalar singlet extension of the SM in this paper.~In particular, we focus on two main features of this model. Firstly, it helps to facilitate electroweak baryogenesis (EWBG) \cite{Kuzmin:1985mm, Cohen:1993nk, Riotto:1999yt, Morrissey:2012db}, a mechanism that aims to explain the observed matter-antimatter asymmetry via a strong first-order EWPT. In the SM, this phase transition is not first-order \cite{Arnold:1992rz,Kajantie:1996qd} and thus requires a modification.~With an extra scalar, a potential barrier can be generated between the symmetric high-temperature minimum and the EWSB one as the universe cools down \cite{Curtin:2014jma,Kotwal:2016tex}.~This leads to a strong first-order EWPT which can be probed using GWs and standard collider searches \cite{Choi:1993cv,Ashoorioon:2009nf,Enqvist:2014zqa,Kakizaki:2015wua,Huang:2016odd,Hashino:2016rvx,Chala:2016ykx,Tenkanen:2016idg,Kobakhidze:2016mch,Huang:2016cjm,Artymowski:2016tme,Hashino:2016xoj,Vaskonen:2016yiu,Baldes:2017rcu,Beniwal:2017eik,Kobakhidze:2017mru,Cai:2017tmh,Croon:2018erz,Baldes:2018emh,Hashino:2018wee,Ahriche:2018rao}.~Secondly, the new scalar mixes with the SM Higgs boson and provides a portal for a fermion DM to saturate the observed DM abundance \cite{Silveira1985136,PhysRevD.50.3637,Burgess:2000yq}.

Simple DM models with a Higgs portal type interaction are still viable and enjoy a rich interest in the particle physics community \cite{Silveira1985136,PhysRevD.50.3637,Burgess:2000yq,Espinosa:2008kw,Alanne:2014bra,Martin-Lozano:2015dja,Falkowski:2015iwa,Buttazzo:2015bka,Heikinheimo:2016yds,Balazs:2016tbi,Lewis:2017dme,Ghorbani:2017jls,Chen:2017qcz,Kamon:2017yfx,Ettefaghi:2017vbh,Baker:2017zwx,Baum:2017enm,Bernal:2018ins}.~In our study, we focus on a singlet fermion DM model which was first introduced in Ref.~\cite{Kim:2008pp} and subsequently improved in Ref.~\cite{Baek:2011aa}. After the discovery of a SM-like Higgs boson at the LHC, the model was revisited in Ref.~\cite{Baek:2012uj} in the context of vacuum stability (see also Ref.~\cite{Espinosa:2011ax}). Here it was pointed out that the model is stable and perturbative up to the Planck scale for a $125$\,GeV Higgs boson. In light of EWBG, the model was first studied in Ref.~\cite{Fairbairn:2013uta} and more recently in Ref.~\cite{Li:2014wia}. Using a Monte Carlo scan of the model parameter space, the model was shown to realise a strong first-order phase transition without conflicting with any bounds from direct DM searches, electroweak precision observables (EWPO) and latest Higgs data from the LHC.

We aim to perform the most comprehensive and up-to-date study of the extended scalar singlet model with a fermionic DM candidate. In our global fit, we include the latest results from the \emph{Planck} measured DM relic density \cite{Ade:2015xua}, direct detection limits from the XENON1T (2018) experiment \cite{Aprile:2018dbl}, EWPO \cite{Haller:2018nnx} and Higgs searches at colliders \cite{Bechtle:2013wla,Bechtle:2013xfa}.~We also find regions in the model parameter space where a successful EWBG can be viable, compute the resulting GW spectra, and check the discovery prospects of the model at current and future GW experiments.~In agreement with previous studies, we confirm that our model with additional couplings to the SM Higgs boson can \emph{simultaneously} explain the observed DM abundance and matter-antimatter asymmetry; this was not possible in the $\mathbb{Z}_2$ symmetric case studied in our previous work \cite{Beniwal:2017eik}.~We also find that our global fit places a strong upper \emph{and} lower limit on the second scalar mass $m_H$, fermion DM mass $m_\psi$ and the scalar-fermion DM coupling $g_S$. In addition, the GW spectra of viable points can often be within reach of future GW experiments such LISA, DECIGO and BBO.

The rest of the paper is organised as follows. In section~\ref{sec:model}, we introduce the extended scalar singlet model with a fermionic DM candidate. After taking note of the free parameters of our model, we describe a set of constraints and likelihoods used in our global fit in section~\ref{sec:const_like}.~Our model results and conclusions are presented in sections~\ref{sec:results} and \ref{sec:conclusions} respectively. Appendices \ref{app:tree-potential}, \ref{app:mass-eg}, \ref{app:DM-nucleon} and \ref{app:effpot} provide supplementary details for understanding various expressions in the paper.

\section{Singlet fermion dark matter model}\label{sec:model}
We extend the SM by adding a new real scalar singlet $S$ and a Dirac fermion DM field $\psi$. The fermion DM is assumed to be living in the hidden sector and communicates with the SM particles only via the new scalar $S$. The model Lagrangian is given by \cite{Baek:2011aa} 
\begin{equation}\label{eqn:model_lagr}
    \lagr = \lagr_{\textnormal{SM}} + \lagr_S + \lagr_\psi + \lagr_{\textnormal{portal}},
\end{equation}    
where $\lagr_{\textnormal{SM}}$ is the SM Lagrangian, 
\begin{align}
    \lagr_S &= \frac{1}{2} (\partial_\mu S) (\partial^\mu S) + \frac{1}{2} \mu_S^2 S^2 + \frac{1}{3} \mu_3 S^3 - \frac{1}{4} \lambda_S S^4, \label{eqn:S-lagr} \\
    \lagr_\psi &= \ovr{\psi} (i \slashed{\partial} - \mu_{\psi})\psi - g_S \ovr{\psi} \psi S, \label{eqn:psi-lagr} \\
    \lagr_{\textnormal{portal}} &= -\mu_{\pp S} \PP \pp S - \frac{1}{2} \lambda_{\pp S} \PP \pp S^2. \label{eqn:portal_lagr}
\end{align}
In general, a linear term in the $S$ field is allowed by symmetry.~However, such a term can be removed by performing a constant shift in $S$ which also redefines $\mu_S^2$, $\mu_{\pp}^2$, $\mu_3$, $g_S$ and $\mu_{\pp S}$.\footnote{The parameter $\mu_\pp^2$ appears in the SM Higgs potential, see Eq.~\eqref{eqn:H-pot}.}~In writing the above Lagrangians, we have assumed that these parameters are defined after a constant shift in $S$. If we set $\mu_3=g_S=\mu_{\Phi S}=0$, we can see that the above Lagrangian becomes $\mathbb{Z}_2$ symmetric under $S \rightarrow -S$, i.e., it is even in $S$.~In this case, the fermion DM $\psi$ is decoupled and becomes a hidden DM candidate, whereas the scalar $S$ serves as a new DM candidate and reproduces the scalar Higgs portal model \cite{Cline:2013gha,Beniwal:2015sdl,He:2016mls,Escudero:2016gzx,Wu:2016mbe,Banerjee:2016vrp,Casas:2017jjg,Beniwal:2017eik,Athron:2017kgt,Hoferichter:2017olk,Athron:2018ipf}.

With an extra scalar field, the tree-level scalar potential is given by
\begin{equation}\label{eqn:pot-part}
    V_{\textnormal{tree}} = V_{\textnormal{SM}} + V_{S} + V_{\textnormal{portal}},
\end{equation}
where $V_{S}$ and $V_{\textnormal{portal}}$ can be read directly from Eqs.~\eqref{eqn:S-lagr} and \eqref{eqn:portal_lagr} respectively. The SM part of the potential reads
\begin{equation}\label{eqn:H-pot}
    V_{\textnormal{SM}} = -\mu_\pp^2 \PP \pp + \lambda_\pp (\PP \pp)^2,
\end{equation}
where
\begin{equation}\label{eqn:H-doublet}
    \pp = 
    \begin{pmatrix}
        G^+ \\
        \frac{1}{\sqrt{2}} \left(\phi + iG^0 \right)
    \end{pmatrix}
\end{equation}
is the SM Higgs doublet and $(G^\pm, G^0)$ are the Goldstone bosons. 

In general, both $\phi$ and $S$ can develop non-trivial vacuum expectation values (VEVs). At $T = 0$, these are denoted by $v_0$ and $s_0$ respectively, i.e.,
\begin{equation}\label{eqn:vevs}
    \left.\langle 0| \phi |0 \rangle\right|_{T = 0} \equiv \left.\langle \phi \rangle\right|_{T=0} = v_0, \quad \left.\langle 0| S |0 \rangle\right|_{T = 0} \equiv \left.\langle S \rangle\right|_{T=0} = s_0.
\end{equation}
After EWSB, we can expand $\pp$ and $S$ in the unitary gauge as
\begin{equation}
    \pp = \frac{1}{\sqrt{2}} 
    \begin{pmatrix}
        0 \\
        v_0 + \varphi
    \end{pmatrix}, \quad
    S  = s_0 + s,
\end{equation}
where $(\varphi, s)$ fields represent quantum fluctuations around the $T = 0$ VEVs.~Using the results presented in Appendix~\ref{app:tree-potential}, we arrive at the following EWSB conditions
\begin{align}
    \mu_\pp^2 &= \lambda_\pp v_0^2 + \mu_{\pp S} s_0 + \frac{1}{2} \lambda_{\pp S} s_0^2, \\ 
    \mu_{S}^2 &= - \mu_3 s_0 + \lambda_S s_0^2 + \frac{\mu_{\pp S} v_0^2}{2 s_0} + \frac{1}{2} \lambda_{\pp S} v_0^2. 
\end{align}

The portal interaction Lagrangian in Eq.~\eqref{eqn:portal_lagr} induces a mixing between the $\varphi$ and $s$ fields. Thus, the squared mass matrix 
\begin{equation}\label{eqn:sq-mass}
    \calM^2 = 
    \begin{pmatrix}
        \calM_{\varphi \varphi}^2 & \calM_{\varphi s}^2 \\[2mm]
        \calM_{s \varphi}^2 & \calM_{ss}^2
    \end{pmatrix}
\end{equation}
is non-diagonal. As shown in Appendix~\ref{app:tree-potential}, its elements are given by
\begin{equation}
    \calM_{\varphi\varphi}^2 = 2 \lambda_\pp v_0^2, \quad \calM_{ss}^2 = - \mu_3 s_0 + 2 \lambda_S s_0^2 - \frac{\mu_{\pp S} v_0^2}{2 s_0}, \quad \calM_{\varphi s}^2 = \calM_{s \varphi}^2 = \mu_{\pp S} v_0 + \lambda_{\pp S} v_0 s_0.
\end{equation}
The squared mass matrix in Eq.~\eqref{eqn:sq-mass} can be diagonalised by rotating the interaction eigenstates $(\varphi, s)$ into the physical mass eigenstates $(h, H)$ as
\begin{equation}
    \begin{pmatrix}
        h \\
        H
    \end{pmatrix} = 
    \begin{pmatrix}
        \cos\alpha & -\sin\alpha \\
        \sin\alpha & \cos\alpha
    \end{pmatrix}
    \begin{pmatrix}
        \varphi \\
        s
    \end{pmatrix},
\end{equation}
where $\alpha$ is the mixing angle. Thus, for small mixing, $h$ is a SM-like Higgs boson, whereas $H$ is dominated by the scalar singlet. 

For the tree-level scalar potential in Eq.~\eqref{eqn:pot-part} to be bounded from below, the following conditions must be satisfied (see Appendix~\ref{app:tree-potential})
\begin{equation}\label{eqn:conditions}
    \lambda_\pp > 0, \quad \lambda_S > 0, \quad \lambda_{\pp S} > - 2\sqrt{\lambda_{\pp} \lambda_S}.
\end{equation}
After EWSB, the fermion DM Lagrangian in Eq.~\eqref{eqn:psi-lagr} becomes
\begin{equation}\label{eqn:final-fermion}
    \lagr_\psi = \ovr{\psi} (i\slashed{\partial} - m_\psi) \psi - g_S \ovr{\psi} \psi s,
\end{equation}
where 
\begin{equation}
    m_\psi = \mu_\psi + g_S s_0
\end{equation}
is the physical fermion DM mass.

\section{Constraints and likelihoods}\label{sec:const_like}
In light of the recent discovery of a SM-like Higgs boson at the LHC \cite{Aad:2012tfa,Chatrchyan:2012xdj}, we set
\begin{equation}
    m_h = 125.13\,\textnormal{GeV}, \quad v_0 = 246.22\,\textnormal{GeV}.
\end{equation}
Thus, the model is completely described by the following 7 free parameters
\begin{equation}\label{eqn:free-scalar}
    m_H, \quad s_0, \quad \mu_3, \quad \lambda_S, \quad \alpha, \quad m_\psi, \quad g_S.
\end{equation}
The remaining parameters in Eqs.~\eqref{eqn:S-lagr}, \eqref{eqn:portal_lagr} and \eqref{eqn:H-pot} can be expressed as (see Appendix~\ref{app:mass-eg})
\begin{align}
    \lambda_\pp &= \frac{1}{2 v_0^2} \left( m_h^2 \cos^2\alpha + m_H^2 \sin^2 \alpha \right), \label{eqn:lP} \\
     \mu_{\pp S} &= -\frac{2 s_0}{v_0^2} \left(m_h^2 \sin^2 \alpha + m_H^2 \cos^2\alpha  + \mu_3 s_0 - 2 \lambda_S s_0^2 \right), \label{eqn:muP} \\ 
     \lambda_{\pp S} &= \frac{1}{v_0 s_0} \Big[(m_H^2 - m_h^2) \sin\alpha \cos\alpha - \mu_{\pp S} v_0 \Big], \label{eqn:lPS} \\
    \mu_\pp^2 &= \lambda_\pp v_0^2 + \mu_{\pp S} s_0 + \frac{1}{2} \lambda_{\pp S} s_0^2, \\
    \mu_{S}^2 &= - \mu_3 s_0 + \lambda_S s_0^2 + \frac{\mu_{\pp S} v_0^2}{2 s_0} + \frac{1}{2} \lambda_{\pp S} v_0^2. 
\end{align}

To study the phenomenology of our model, we implement the extended scalar singlet and fermion DM model in the \lhepver~\cite{Semenov:2014rea} package.~For the calculation of the fermion DM relic density and Higgs decay rates, we use \microver \cite{Belanger:2014vza} which relies on the \chep \cite{Belyaev:2012qa} package.  

We make parameter inferences by adopting a frequentist approach and performing 7-dimensional scans of the model parameter space using the \DE~\cite{Workgroup:2017htr} package.\footnote{\url{http://diver.hepforge.org}}~The combined log-likelihood used in our global fit is
\begin{align}\label{eqn:tot_like}
    \ln \like_{\textnormal{total}} (\bm{\theta}) &= \ln \like_{\Omega h^2} (\bm{\theta}) + \ln \like_{\textnormal{XENON1T}} (\bm{\theta}) + \ln \like_{v_c/T_c} (\bm{\theta}) \nonumber \\
    &\hspace{4mm} + \ln \like_{\textnormal{EWPO}} (\bm{\theta}) + \ln \like_{\textnormal{HB}} (\bm{\theta}) + \ln \like_{\textnormal{HS}} (\bm{\theta}),
\end{align}
where
\begin{itemize}
    \item $\ln \like_{\Omega h^2}(\bm{\theta})$:~log-likelihood for the \emph{Planck} measured DM relic density, see subsection~\ref{subsec:relic};
    \item $\ln \like_{\textnormal{XENON1T}} (\bm{\theta})$: log-likelihood for the direct detection limits from the XENON1T (2018) experiment, see subsection~\ref{subsec:DD};
    \item $\ln \like_{v_c/T_c} (\bm{\theta})$: log-likelihood for the EWBG constraint, see subsection~\ref{subsec:vcTc};
    \item $\ln \like_{\textnormal{EWPO}} (\bm{\theta})$: log-likelihood for the electroweak precision observables (EWPO) constraint, see subsection~\ref{subsec:EWPO}; 
    \item $\ln \like_{\textnormal{HB}} (\bm{\theta})$: log-likelihood for the direct Higgs searches performed at the LEP, Tevatron and the LHC, see subsection~\ref{subsec:HB_HS};
    \item $\ln \like_{\textnormal{HS}} (\bm{\theta})$: log-likelihood for the Higgs signal strength and mass measurements performed at the LHC, see subsection~\ref{subsec:HB_HS}.
\end{itemize}
Here $\bm{\theta} \equiv (m_H,\,s_0,\,\mu_3,\,\lambda_S,\,\alpha,\,m_\psi,\,g_S)$ denotes the free parameters of our model.~These are uniformly sampled over their ranges shown in Table~\ref{tab:par_ranges} in either flat or logarithmic space.

In the following subsections, we outline the details of all constraints and likelihoods used in our global fit.

\begin{table}[t]
    \centering    
    \begin{tabular}{cccc}
        \toprule
        Parameter & Minimum & Maximum & Prior type \\ \midrule
        $m_H$ & $10\,\textnormal{GeV}$ & $10\,\textnormal{TeV}$ & log \\[1.2mm]
        $s_0$ & $-1\,\textnormal{TeV}$ & $1\,\textnormal{TeV}$ & flat \\[1.2mm]
        $\mu_3$ & $-1\,\textnormal{TeV}$ & $1\,\textnormal{TeV}$ & flat \\[1.2mm]
        $\lambda_S$ & $10^{-3}$ & $10$ & log \\[1.2mm]
        $\alpha$ & $0$ & $\pi$ & flat \\[1.2mm]
        $m_\psi$ & $10\,\textnormal{GeV}$ & $10\,\textnormal{TeV}$ & log \\[1.2mm]
        $g_S$ & $10^{-3}$ & $10$ & log \\ 
        \bottomrule
    \end{tabular}    
    \caption{Ranges and priors for the free parameters of our model.~All parameters are uniformly sampled over their ranges in either flat or logarithmic space. For the mixing angle $\alpha$, all constraints are symmetric under $\alpha \rightarrow -\alpha$, thus we only scan over $\alpha \in [0, \pi]$.}
    \label{tab:par_ranges}
\end{table}

\subsection{Thermal relic density}\label{subsec:relic}
From the \emph{Planck} satellite's observation of the temperature and polarization anisotropies in the cosmic microwave background (CMB), a strong bound on the present-day abundance of the DM particles can be extracted. The latest results indicate \cite{Ade:2015xua} 
\begin{equation}\label{eqn:planck-omega}
    \Omega_{\textnormal{DM}} h^2 = 0.1188 \pm 0.0010,
\end{equation}
where $\Omega_{\textnormal{DM}} = \rho_{\textnormal{DM}}/\rho_c$ is the density parameter, $\rho_c = 3H_0^2 M_p^2$ is the critical mass density and $h = H_0/(100\,\textnormal{km}\,\textnormal{s}^{-1}\,\textnormal{Mpc}^{-1})$ is the reduced Hubble constant.

\begin{figure}[t]
    \centering
    \includegraphics[width=0.28\textwidth]{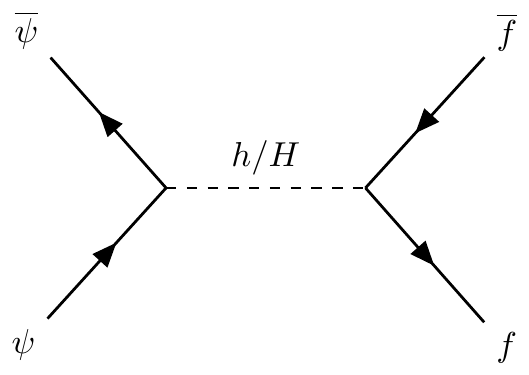} \hfil
    \includegraphics[width=0.3\textwidth]{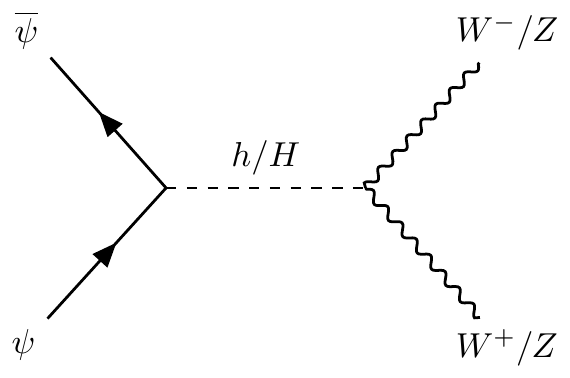} \hfil
    \includegraphics[width=0.3\textwidth]{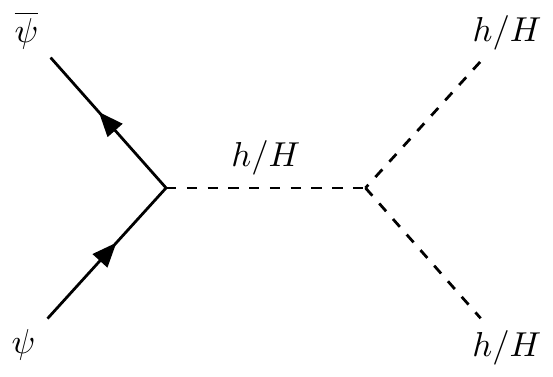} 
	\\[2mm]
    \includegraphics[width=0.22\textwidth]{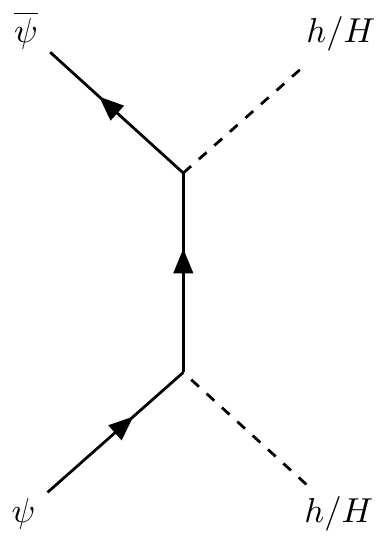} \quad
    \includegraphics[width=0.24\textwidth]{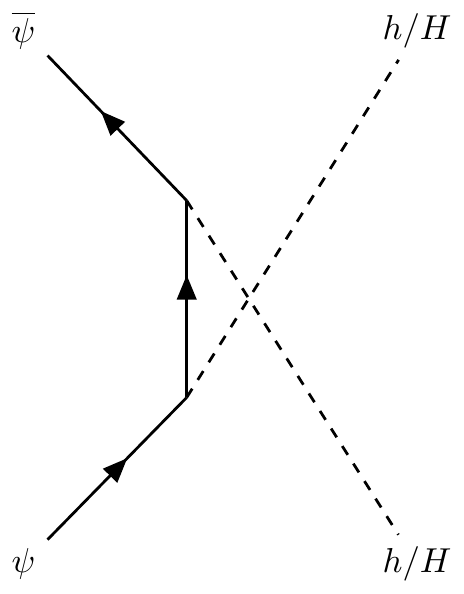}        
    \caption{Feynman diagrams for the fermion DM annihilation into SM and $h/H$ particles when $\alpha \neq 0,\,\pi/2$.~Here $f$ refers to a SM fermion.}
    \label{fig:fermion-ann-diag}
\end{figure}

In our model, the Dirac fermion $\psi$ is the DM candidate.~Its relic density is mainly determined by an $s$-channel annihilation into SM particles via an $h/H$ exchange. Annihilation into $hh$, $HH$ and $hH$ final states are also possible via the $t$- and $u$-channels. Due to a mixing between the interaction eigenstates $(\varphi, s)$, the decay rates go as
\begin{align}
    \Gamma(h \rightarrow \ovr{\psi} \psi) &\propto g_S^2 \sin^2\alpha, \quad \Gamma(h \rightarrow \ovr{\calX}\calX) \propto \cos^2\alpha, \label{eqn:psi-h-SM} \\
    \Gamma(H \rightarrow \ovr{\psi} \psi) &\propto g_S^2 \cos^2\alpha, \quad \Gamma(H \rightarrow \ovr{\calX}\calX) \propto \sin^2\alpha, \label{eqn:psi-H-SM}
\end{align}
where $\calX$ is a general SM final state, e.g., quarks, leptons or gauge bosons.~Depending on the mixing angle $\alpha$, various SM and non-SM final states are allowed in the $s$, $t$ and $u$ channels. 
\begin{enumerate}
    \item $\alpha = 0$: In this case, $h$ is a SM-like Higgs boson, whereas $H$ is a scalar singlet. Thus, the only allowed final states from the fermion DM annihilation are $hh$, $HH$ and $hH$ via an $s$-channel $H$ exchange.
    \item $\alpha = \pi/2$: In this case, $h$ is a scalar singlet, whereas $H$ is a SM-like Higgs boson. Similar to the $\alpha = 0$ case, the only allowed final states from the fermion DM annihilation are $hh$, $HH$ and $hH$ via an $s$-channel $h$ exchange.
    \item $\alpha \neq 0,\,\pi/2$: In these cases, all final states shown in Fig.~\ref{fig:fermion-ann-diag} are allowed via either an $h$ or $H$ exchange.
\end{enumerate}

With two scalar mediators $h$ and $H$, the annihilation rate of the fermion DM into SM particles is enhanced when $m_\psi \sim m_{h,H}/2$. At these two resonances, the fermion DM relic density $\Omega_\psi h^2$ drops rapidly with increasing scalar-fermion DM coupling $g_S$. For the fermion DM to account for the observed DM abundance, i.e., $\Omega_\psi h^2 = \Omega_{\textnormal{DM}} h^2$, smaller values of $g_S$ are required to compensate for the enhanced DM annihilation rate into SM particles.

In order to address the strong possibility of a multicomponent dark sector, we define a relic abundance parameter \cite{Cline:2012hg,Cline:2013gha,Beniwal:2015sdl} as
\begin{equation}
    f_{\textnormal{rel}} = \frac{\Omega_\psi}{\Omega_{\textnormal{DM}}},
\end{equation}
where $\Omega_{\textnormal{DM}} h^2 = 0.1188$ is the \emph{Planck} measured central value in Eq.~\eqref{eqn:planck-omega}. Consequently, the indirect and direct detection rates must be scaled by $f_{\textnormal{rel}}^2$ and $f_{\textnormal{rel}}$ respectively.\footnote{In our study, we do not include any indirect detection limits as the fermion DM annihilation rate into SM particles is $p$-wave suppressed \cite{Agrawal:2010fh}.~However, when a pure pseudoscalar, parity-violating interaction term $(\propto \ovr{\psi} i\gamma_5 \psi)$ is introduced, the resulting indirect detection limits can be sizeable \cite{Esch:2013rta,Bagherian:2014iia,Franarin:2014yua,Kim:2016csm,Ghorbani:2014qpa,Balazs:2015boa,Athron:2018hpc,Kim:2018uov}.}~In regions of the model parameter space where $f_{\textnormal{rel}} > 1$,  parameter points are robustly excluded by the relic density constraint. 

We investigate both possibilities of our model to either account for all or part of the observed DM abundance. In the former case, we use a Gaussian likelihood function with a central value equal to the \emph{Planck} measured one and a combined uncertainty equal to the \emph{Planck} measured uncertainty with a 5\% theoretical error.\footnote{A possible source of theoretical uncertainty is in our relic density calculations as performed in \micro.}~In the latter case, we instead use a Gaussian likelihood function as an \emph{upper} limit and require the parameter points to satisfy $f_{\textnormal{rel}} \leq 1$. The results for both of these scenarios will be discussed in more detail in section~\ref{sec:results}.

\subsection{Direct detection}\label{subsec:DD}
Direct detection experiments aim to measure the recoil of a nucleus from an elastic scattering off a DM particle. Such an event generates a typical recoil energy $E_R$ on the order of a few keV. As most radioactive elements and high-energy cosmic rays induce nuclear recoils with energies well above this value, direct DM searches must be conducted in deep underground laboratories to shield them from potential background sources.

In our model, the DM-quark interaction proceeds via a $t$-channel exchange of $h/H$ particles.~With two neutral scalar mediators $(h, H)$, the resulting DM-nucleus interaction is nuclear spin-independent (SI). The SI DM-nucleus cross-section is given by
\begin{equation}\label{eqn:DM-nucleus}
     \sigma_{\textnormal{SI}}^{\psi N} = \frac{\mu_{\psi N}^2}{\pi} \Big[Z G_p + (A - Z) G_n \Big]^2,
\end{equation}
where $\mu_{\psi N} = m_\psi m_N/(m_\psi + m_N)$ is the DM-nucleus reduced mass and $Z\,(A - Z)$ are the number of protons (neutrons) in the target nucleus $N$.~The dimensionful parameters $(G_p,\,G_n)$ are the effective DM-nucleon couplings. These are given by (see Appendix~\ref{app:DM-nucleon})
\begin{equation}\label{eqn:G_coupling}
    G_\calN = \frac{g_S \sin\alpha \cos\alpha}{v_0} \left(\frac{1}{m_h^2} - \frac{1}{m_H^2} \right) m_\calN f_\calN,
\end{equation}
where $\calN \in (p, n)$,
\begin{equation}
    f_\calN = \frac{2}{9} + \frac{7}{9} \sum_{q = u,\,d,\,s} f_{Tq}^{(\calN)}
\end{equation}
is the Higgs-nucleon coupling and 
\begin{equation}
f_{Tq}^{(\calN)} \equiv \frac{m_q}{m_\calN} \langle \calN | \ovr{q}q | \calN \rangle
\end{equation}
are the hadronic matrix elements.

For isospin conserving couplings $(G_p \simeq G_n)$, the DM-nucleus cross-section in Eq.~\eqref{eqn:DM-nucleus} is enhanced by a factor of $A^2$. This is expected as the matrix element for a SI interaction involves a coherent sum over the individual protons and neutrons in the target nucleus $N$. For this reason, direct detection experiments rely on heavy target materials with large $Z$ to better constrain the DM-nucleon cross-section $\sigma_{\textnormal{SI}}^{\psi \calN}$. In our model, it is given by
\begin{equation}
    \sigma_{\textnormal{SI}}^{\psi \calN} = \frac{\mu_{\psi\calN}^2}{\pi} \left(\frac{g_S \sa \ca}{v_0}\right)^2 \left(\frac{1}{m_h^2} - \frac{1}{m_H^2} \right)^2 m_\calN^2 f_\calN^2,
\end{equation}
where $\mu_{\psi\calN} = m_\psi m_\calN/(m_\psi + m_\calN)$ is the DM-nucleon reduced mass, $m_\calN = 939$\,MeV and $f_\calN = 0.3$ \cite{Cline:2013gha}. 

Currently, the best upper limits on the SI DM-nucleon cross-section comes from the XENON1T (2018) experiment \cite{Aprile:2018dbl}.~To constrain the model parameter space from the XENON1T experiment, we use a one-sided Gaussian likelihood function, i.e., we require the parameter points to satisfy\footnote{The official XENON1T (2018) limits are only available for DM masses up to $1$\,TeV. Beyond $1$\,TeV, we perform a linear extrapolation of the limit due to the reduced DM number density.}
\begin{equation}\label{eqn:compare_Panda}
    \sigma_{\textnormal{SI}}^{\textnormal{eff}} \leq \sigma_{\textnormal{XENON1T}},
\end{equation}
where $\sigma_{\textnormal{XENON1T}}$ is the 90$\%$ C.L.~upper limit from the XENON1T experiment and
\begin{equation}\label{eqn:eff_xsection}
    \sigma_{\textnormal{SI}}^{\textnormal{eff}} = 
    \begin{cases}
        \sigma_{\textnormal{SI}}^{\psi \calN} f_{\textnormal{rel}}, & f_{\textnormal{rel}} < 1, \\[1.5mm]
        \sigma_{\textnormal{SI}}^{\psi \calN}, & f_{\textnormal{rel}} \geq 1,
    \end{cases}
\end{equation}
is the effective SI DM-nucleon cross-section. The scaling of $\sigma_{\textnormal{SI}}^{\psi \calN}$ by $f_{\textnormal{rel}}$ is done to suppress signals when $f_{\textnormal{rel}} < 1$. In regions of the model parameter space where $f_{\textnormal{rel}} > 1$, parameter points are already ruled out by the relic density constraint. 

We also include a theoretical uncertainty of 5$\%$ in our analysis. This can easily arise from the uncertainties associated with the nuclear physics, DM halo and velocity distribution parameters. For a recent review, see Ref.~\cite{Workgroup:2017lvb}.

\subsection{Electroweak baryogenesis (EWBG)}\label{subsec:vcTc}
In our model, the VEV of the new scalar $S$ does not initially have to be zero.~Thus, the transition pattern can be $(\langle \phi \rangle,\langle s \rangle)=(0, s_i) \rightarrow (v,s)$.~At low temperatures, the latter minimum evolves slowly to become the electroweak minimum at $T = 0$, i.e., $(\langle \phi \rangle,\langle s \rangle)=(v_0, s_0)$. The initial transition can break the electroweak symmetry by tunnelling through a potential barrier to the broken phase minimum. This transition can proceed via nucleation of bubbles of the broken phase which results in a departure from thermal equilibrium \cite{Kuzmin:1985mm,Cohen:1993nk,Riotto:1999yt,Morrissey:2012db}. In addition, it can generate a significant gravitational wave (GW) signal \cite{Grojean:2006bp}.
 
Using the standard notation, we define a strong first-order phase transition by 
\begin{equation}\label{eqn:sphbound}
    \frac{v}{T} \gtrsim 1,
\end{equation}
where $v$ is the Higgs VEV at temperature $T$. However, one has to keep in mind that the calculation of the baryon asymmetry remaining after the transition is quite complicated. This leads to a slightly different exact lower bound on $v/T$ \cite{Quiros:1999jp,Funakubo:2009eg,Curtin:2014jma,Katz:2014bha,Fuyuto:2014yia}.

To find regions in the model parameter space where a successful EWBG is potentially viable, we first find the minima of the effective potential $V_{\textnormal{eff}}(\phi, S, T)$ (see Appendix~\ref{app:effpot}) numerically, and compute the critical temperature $T_c$ at which the initial and symmetry breaking minima are degenerate.~This allows us to compute the dimensionless parameter $v_c/T_c$ (the Higgs VEV $v_c$ at the critical temperature $T_c$) and constrain parts of the 7-dimensional model parameter space, i.e., parameter points are excluded if they lead to a too weak phase transition.~Specifically, we use a one-sided Gaussian likelihood function and require the parameter points to satisfy
\begin{equation}\label{eqn:large_vcTc}
    \frac{v_c}{T_c} \geq 0.6
\end{equation}
as a conservative limit.~A theoretical uncertainity of 5$\%$ on the resulting $v_c/T_c$ values is assumed to obtain a smooth likelihood function. The actual uncertainty can be much larger as the value of $v_c/T_c$ required to facilitate EWBG is not yet settled. 

In addition, parameter points are also excluded if they exhibit any of the following three features. 
\begin{enumerate}
    \item \emph{Incorrect minimum at $T = 0$}: This situation arises when the electroweak vacuum $(\langle \phi \rangle, \langle S \rangle) = (v_0, s_0)$ is not the true minimum of the potential at $T = 0$.
    \item \emph{Runaway directions in the potential}: This occur when the $\phi$ and $S$ field values in the symmetric or broken phase are too large, or if the potential in Eq.~\eqref{eqn:pot-part} is unbounded from below in the general $\phi$ and $S$ directions, i.e., when $\lambda_{\pp S} \leq -2\sqrt{\lambda_\pp \lambda_S}$. 
    \item \emph{Non-perturbative couplings}: This situation arises when $|\lambda_\pp|,\,|\lambda_{\pp S}| \geq 4\pi$. In this case, our 1-loop treatment of the effective potential is not reliable. 
\end{enumerate}

\begin{figure}[t]
    \centering    
    \includegraphics[width=0.65\textwidth]{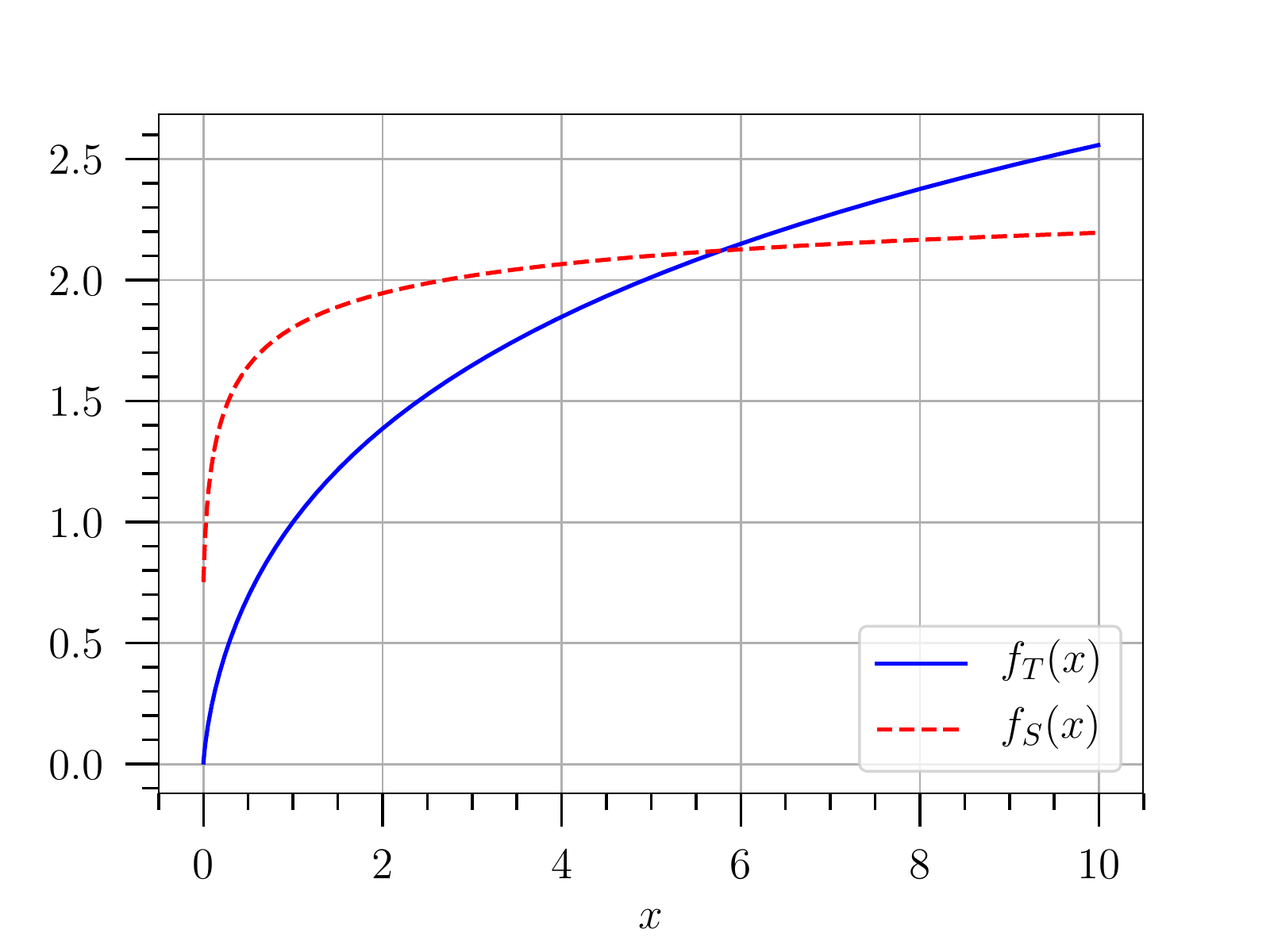}    
    \caption{Loop functions $f_T(x)$ (solid blue) and $f_S(x)$ (dashed red).}
    \label{fig:loop_fn}
\end{figure}

We also perform a complete analysis of the phase transition in this model by following our previous work on the $\mathbb{Z}_2$ symmetric case, i.e., scalar Higgs portal \cite{Beniwal:2017eik} and a very recent update on the calculation of the phase transition dynamics~\cite{Ellis:2018mja}. In particular, we find the percolation temperature $T_p$ at which the phase transition truly completes.~This is used to compute the GW signals arising from the phase transition, and discuss the potential discovery prospects of the model at current and future GW experiments. For more details, see section~\ref{sec:results}.

Let us also point out that we only check one of the necessary conditions for a successful EWBG, while other difficulties might still arise.~For instance, the standard mechanism of generating a baryon yield requires a sufficiently slow speed of the expanding bubble walls~\cite{Bodeker:2009qy,Kozaczuk:2015owa,Kurup:2017dzf}.~We do not compute the bubble wall velocity to check this requirement (in fact, we assume it to be very high) while calculating the GW spectra.~While there are mechanisms which could generate the asymmetry even for very fast walls~\cite{No:2011fi,Caprini:2011uz,Katz:2016adq}, we also do not explicitly make sure that other conditions they carry are fulfilled.

\subsection{Electroweak precision observables (EWPO)}\label{subsec:EWPO}
With an extra scalar, our model can induce corrections to the gauge boson self-energy diagrams. Its effect on the electroweak precision observables (EWPO) can be parametrised by the oblique parameters $S$, $T$ and $U$ \cite{Peskin:1991sw}. The $\gamma\gamma$ and $\gamma Z$ self-energies ($\Pi_{\gamma\gamma}$ and $\Pi_{\gamma Z}$ respectively) are not modified as the new scalar is electrically neutral. Thus, only the $W$ and $Z$ boson self-energies are subject to corrections.

In our model, the oblique parameters are shifted from their SM values by \cite{Baek:2011aa}
\begin{align}
    \Delta T &= \frac{3}{16\pi s_W^2} \left[\cos^2\alpha \left \{f_T \left(\frac{m_h^2}{m_W^2}\right) - \frac{1}{c_W^2}f_T\left(\frac{m_h^2}{m_Z^2}\right) \right\} + \sin^2\alpha \left\{f_T\left(\frac{m_H^2}{m_W^2} \right) \right. \right. \nonumber \\
    &\hspace{4mm} \left. \left. - \frac{1}{c_W^2} f_T\left(\frac{m_H^2}{m_Z^2} \right) \right\} - \left\{f_T\left(\frac{m_h^2}{m_W^2}\right) - \frac{1}{c_W^2}f_T\left(\frac{m_h^2}{m_Z^2}\right) \right\} \right], \label{eqn:del_T} \\[1.5mm]
    \Delta S &= \frac{1}{2\pi} \left[\cos^2\alpha f_S \left(\frac{m_h^2}{m_Z^2}\right) + \sin^2\alpha f_S \left(\frac{m_H^2}{m_Z^2}\right) - f_S \left(\frac{m_h^2}{m_Z^2}\right) \right], \label{eqn:del_S} \\[1.5mm]    
    \Delta U &= \frac{1}{2\pi} \left[\cos^2\alpha f_S \left(\frac{m_h^2}{m_W^2}\right) + \sin^2\alpha f_S \left(\frac{m_H^2}{m_W^2}\right) - f_S \left(\frac{m_h^2}{m_W^2}\right) \right] - \Delta S, \label{eqn:del_U}
\end{align}
where $\Delta \mathcal{O} \equiv \mathcal{O} - \mathcal{O}_{\textnormal{SM}}$ for $\mathcal{O} \in (S, T, U)$, $m_W\,(m_Z)$ is the $W\,(Z)$ boson mass, $c_W^2 = m_W^2/m_Z^2$ and $s_W^2 = 1 - c_W^2$. The loop functions $f_T(x)$ and $f_S(x)$ are given by \cite{Grimus:2008nb}
\begin{align}
    f_T(x) &= \frac{x\log x}{x-1}, \\[1.5mm]
    f_S(x) &= 
    \begin{dcases}
        \frac{1}{12} \left[ -2 x^2 + 9 x + \left((x-3) \left(x^2-4 x+12\right)+\frac{1-x}{x}\right) f_T(x) \right. \\ 
        \left. + 2 \sqrt{(4-x) x} \left(x^2-4 x+12\right)    
        \tan ^{-1}\left(\sqrt{\frac{4 - x}{x}}\right) \right], \quad 0 < x < 4, \\
        \frac{1}{12} \left[-2 x^2 + 9 x + \left((x-3) \left(x^2-4 x+12\right)+\frac{1-x}{x}\right)f_T(x) \right. \\
        \left. + \sqrt{(x-4) x} \left(x^2 - 4x + 12\right)  \log \left( \frac{x - \sqrt{(x-4)x}}{x + \sqrt{(x-4)x}} \right)\right], \quad x \geq 4. 
    \end{dcases} 
\end{align} 
These are also plotted in Fig.~\ref{fig:loop_fn}. From Eqs.~\eqref{eqn:del_T}--\eqref{eqn:del_U}, it is evident that
\begin{equation}\label{eqn:mod_oblique}
    \Delta \mathcal{O} = (1 - \cos^2 \alpha) \Big[\mathcal{O}_{\textnormal{SM}}(m_H) - \mathcal{O}_{\textnormal{SM}}(m_h)\Big].
\end{equation} 
Thus, for large $m_H$, $\alpha \sim 0,\,\pi$ is required, whereas large mixing angles are compatible with the EWPO constraint provided $m_H \sim m_h$.

Using the SM reference as $m_{h}^{\textnormal{ref}} = 125$\,GeV and $m_{t}^{\textnormal{ref}} = 172.5$\,GeV, the most recent global electroweak fit gives \cite{Haller:2018nnx}
\begin{equation}\label{eqn:STU_pars}
    \Delta S = 0.04 \pm 0.11, \quad \Delta T = 0.09 \pm 0.14, \quad \Delta U = -0.02 \pm 0.11,
\end{equation}
and the following correlation matrix
\begin{equation}\label{eqn:corr_mat}
    \rho_{ij} = 
    \begin{pmatrix}
        1 & 0.92 & -0.68 \\
          0.92 & 1 & -0.87 \\
          -0.68 & -0.87 & 1\\        
    \end{pmatrix}.
\end{equation}

To constrain the model parameter space from the EWPO, we use the following likelihood function \cite{Profumo:2014opa}
\begin{equation}
    \ln \like_{\textnormal{EWPO}}(\bm{\theta}) = -\frac{1}{2} \Delta \chi^2 = -\frac{1}{2} \sum_{i,\,j} (\Delta \mathcal{O}_i - \ovr{\Delta \mathcal{O}}_i) \left(\Sigma^2 \right)_{ij}^{-1} (\Delta \mathcal{O}_j - \ovr{\Delta \mathcal{O}}_j),
\end{equation}
where $\ovr{\Delta \mathcal{O}}_i$ denotes the central values for the shifts in Eq.~\eqref{eqn:STU_pars}, $\Sigma_{ij}^2 \equiv \sigma_i \rho_{ij} \sigma_j$ is the covariance matrix, $\rho_{ij}$ is the correlation matrix in Eq.~\eqref{eqn:corr_mat} and $\sigma_i$ are the associated errors in Eq.~\eqref{eqn:STU_pars}.

\subsection{Higgs searches at colliders}\label{subsec:HB_HS}
Due to a mixing between the interaction eigenstates $(\varphi, s)$, the coupling strengths between the mass eigenstates $(h, H)$ and SM particles are modified with respect to the SM expectation.~The effective squared couplings of $(h, H)$ to SM particles are \cite{Li:2014wia}
\begin{equation}\label{eqn:effC_tree}
    \left(\frac{g_{h\calX\ovr{\calX}}}{g_{h\calX\ovr{\calX}}^{\textnormal{SM}}}\right)^2 = \cos^2\alpha, \quad \left(\frac{g_{H\calX\ovr{\calX}}}{g_{H\calX\ovr{\calX}}^{\textnormal{SM}}}\right)^2 = \sin^2\alpha,
\end{equation}
where $\calX$ refers to a SM quark, lepton or gauge boson, and $g_{h\calX\ovr{\calX}}^{\textnormal{SM}}$ $(g_{H\calX\ovr{\calX}}^{\textnormal{SM}})$ are the coupling strengths for a SM-like Higgs boson with mass $m_h\,(m_H)$.~For the loop-induced processes, the effective squared couplings are given by \cite{Djouadi:2005gi}
\begin{equation}\label{eqn:effC_loop}
    \left(\frac{g_{h\calY\ovr{\calY}}}{g_{h\calY\ovr{\calY}}^{\textnormal{SM}}}\right)^2 = \frac{\Gamma_{h \rightarrow \calY\ovr{\calY}}}{\Gamma_{h \rightarrow \calY\ovr{\calY}}^{\textnormal{SM}}}  = \cos^2\alpha, \quad 
    \left(\frac{g_{H\calY\ovr{\calY}}}{g_{H\calY\ovr{\calY}}^{\textnormal{SM}}}\right)^2 = \frac{\Gamma_{H \rightarrow \calY\ovr{\calY}}}{\Gamma_{H \rightarrow \calY\ovr{\calY}}^{\textnormal{SM}}} = \sin^2\alpha,
\end{equation}
where $\calY \ovr{\calY} \in (\gamma \gamma, \, \gamma Z, \, gg, \, ggZ)$ and $\Gamma_{h \rightarrow \calY \ovr{\calY}}^{\textnormal{SM}}$ $(\Gamma_{H \rightarrow \calY \ovr{\calY}}^{\textnormal{SM}})$ are the decay rates for a SM-like Higgs boson with mass $m_h\,(m_H)$.~With modified branching ratios of $h/H$ into SM particles, the scalar sector of our model can be constrained using the direct Higgs searches performed at the lepton (e.g., LEP) and hadron (e.g., Tevatron, LHC) colliders. 

To constrain the model parameter space from the direct Higgs searches performed at the LEP, Tevatron and the LHC, we use the \HBver~\cite{Bechtle:2013wla} package.~From the model predictions for the two scalar masses, total decay widths, branching ratios, and effective squared couplings defined in Eqs.~\eqref{eqn:effC_tree} and \eqref{eqn:effC_loop}, \HB computes and compares the predicted signal rates for the search channels considered in multiple experimental analyses.~By comparing the predicted signal rates against the expected and observed cross-section limits from the direct Higgs searches, it determines whether or not a given parameter point is excluded at 95$\%$ C.L..

For the two physical scalars $(h, H)$, the signal strengths are given by \cite{Li:2014wia}
\begin{align}
    \mu_h = \frac{\Gamma_h^{\textnormal{SM}}\cos^4\alpha}{\Gamma_h^{\textnormal{SM}}\cos^2\alpha + \Gamma_{h\rightarrow\ovr{\psi}\psi} + \Gamma_{h \rightarrow HH}}, \label{eqn:h_mu} \\[1.5mm]
    \mu_H = \frac{\Gamma_H^{\textnormal{SM}}\sin^4\alpha}{\Gamma_H^{\textnormal{SM}}\sin^2\alpha + \Gamma_{H \rightarrow \ovr{\psi}\psi} + \Gamma_{H \rightarrow hh}}. \label{eqn:H_mu}
\end{align}
In the absence of invisible and cross Higgs decay modes, $\mu_h$ $(\mu_H)$ scales as $\cos^2\alpha$ $(\sin^2\alpha)$. However, when these decay modes are kinematically allowed, they suppress the $h/H$ signal strength with respect to the SM expectation.~Thus, the scalar sector of our model can also be constrained using the Higgs signal strength and mass measurements performed at the LHC. 

\begin{table}[t]
    \centering
	\begin{tabular}{cccc}
        \toprule
        Experiment & Channel & Obs. signal strength & Ref. \\ \midrule
        ATLAS & $h \rightarrow WW^*$          & $1.18_{-0.21}^{+0.24}$ & \cite{Aad:2015gba} \\[1.5mm]
        ATLAS & $h \rightarrow ZZ^*$          & $1.46_{-0.34}^{+0.40}$ & \cite{Aad:2015gba} \\[1.5mm]                
        ATLAS & $h \rightarrow \gamma\gamma$  & $1.17_{-0.26}^{+0.28}$ & \cite{Aad:2015gba} \\[1.5mm]                        
        ATLAS & $h \rightarrow \tau^+ \tau^-$ & $1.44_{-0.37}^{+0.42}$ & \cite{Aad:2015gba} \\[1.5mm]   
        ATLAS & $h \rightarrow b\ovr{b}$      & $0.63_{-0.37}^{+0.39}$ & \cite{Aad:2015gba} \\[1.5mm]
        CMS   & $h \rightarrow WW^*$          & $0.72_{-0.18}^{+0.20}$ & \cite{Chatrchyan:2013iaa} \\[1.5mm]
        CMS   & $h \rightarrow ZZ^*$          & $0.93_{-0.25}^{+0.29}$ & \cite{Chatrchyan:2013mxa} \\[1.5mm]                                                                                                                            
        CMS   & $h \rightarrow \gamma\gamma$  & $1.14_{-0.23}^{+0.26}$ & \cite{Khachatryan:2014ira} \\[1.5mm]        
        CMS   & $h \rightarrow \tau^+ \tau^-$ & $0.78_{-0.27}^{+0.27}$ & \cite{Chatrchyan:2014vua} \\[1.5mm]   
		CMS   & $h \rightarrow b\ovr{b}$      & $1.00_{-0.50}^{+0.50}$ & \cite{Chatrchyan:2014vua} \\
        \bottomrule
    \end{tabular}    
    \caption{A summary of Higgs boson signal strength measurements that are included in our analysis. For more details, see \texttt{Expt\_tables/latestresults-1.4.0-LHCinclusive/} directory of \HSver~\cite{Bechtle:2013xfa}.}
    \label{tab:exp_analysis}
\end{table}

To constrain the model parameter space from the Higgs signal strength and mass measurements, we use the \HSver~\cite{Bechtle:2013xfa} package. Assuming a Gaussian probability density function (p.d.f.)~for the two scalar masses, we compute a chi-square $\chi_{\textnormal{HS}}^2$ using the \emph{peak-centered} method.\footnote{A theoretical mass uncertainty of zero is assumed for both scalars as $m_h$ is fixed, whereas $m_H$ is a free model parameter.}~In this method, $\chi_{\textnormal{HS}}^2$ is evaluated by \emph{assigning}, for each signal (or peak) observed in multiple experimental analyses (see Table~\ref{tab:exp_analysis}), a combination of the two Higgs bosons from our model provided their masses lie within the experimental resolution of an analysis \cite{Stal:2013hwa}.~Following the assignment, a $\chi^2_\mu$ is evaluated by comparing the signal strength measurement for the peak to the model predicted signal strength. When a mass measurement is also available (e.g., from channels with a good mass-resolution such as the $h \rightarrow \gamma\gamma$ decay mode), a corresponding $\chi^2_m$ is also evaluated by comparing the model predicted and observed Higgs boson mass. Thus, the total $\chi_{\textnormal{HS}}^2$ is given by\footnote{For more details on the functional form of individual chi-squares, see Ref.~\cite{Bechtle:2013xfa}.}
\begin{equation}
	\chi_{\textnormal{HS}}^2 = \chi_\mu^2 + \chi^2_m = \chi^2_\mu + \sum_{i = 1}^2 \chi_{m_{i}}^2.
\end{equation}
In situations where more than one Higgs boson can contribute to a signal (as in our case), an optimal assignment of the Higgs bosons to the signals is achieved by minimising the overall $\chi^2_{\textnormal{HS}}$. The predicted signal strengths of the two scalars are added incoherently, assuming negligible interference effects. Finally, the computed $\chi_{\textnormal{HS}}^2$ is used to define a Higgs signal strength log-likelihood as
\begin{equation}\label{eqn:HS_chi}
    \ln \mathcal{L}_{\textnormal{HS}}(\bm{\theta}) = -\frac{1}{2} \chi_{\textnormal{HS}}^2.
\end{equation}
Thus, a large $\chi_{\textnormal{HS}}^2$ indicates a large deviation between the model predicted signal strength and the best-fit value for a fixed Higgs boson mass, and vice versa.

\section{Results}\label{sec:results}
We perform scans of our 7D model parameter space using \DE~\cite{Workgroup:2017htr} with \texttt{lambdajDE} = \texttt{true}, \texttt{NP} = 50,000 and \texttt{convthresh} = 10$^{-5}$.~To efficiently sample all parts of the parameter space (even the degenerate ones), we also run several targeted scans and combine the output chains to obtain high-quality profile likelihood (PL) plots.

We present our model results in the form of 1- and 2-dimensional PL plots.~For a model parameter $\theta_i$ where $i = 1,\ldots, 7$, a 1D PL $\like_{\textnormal{PL}} (\theta_i)$ is defined as
\begin{equation}\label{eqn:prof_like_1D}
    \like_{\textnormal{PL}} (\theta_i) \equiv \underset{\{\theta_j | \, j \, \neq \, i\}}{\textnormal{max}}\,\like (\bm{\theta}).
\end{equation}
Thus, $\like_{\textnormal{PL}} (\theta_i)$ is a function of $\theta_i$ only, i.e., all other parameters are \emph{profiled out}. Similarly, a 2D PL $\like_{\textnormal{PL}} (\theta_i, \theta_j)$ is defined as
\begin{equation}\label{eqn:prof_like_2D}
    \like_{\textnormal{PL}} (\theta_i, \theta_j) \equiv \underset{\{\theta_k | \, k \, \neq \, i, \, j \}}{\textnormal{max}}\,\like (\bm{\theta}).
\end{equation}
Thus, $\like_{\textnormal{PL}} (\theta_i, \theta_j)$ is a function of $\theta_i$ and $\theta_j$ only. Using Eqs.~\eqref{eqn:prof_like_1D} and \eqref{eqn:prof_like_2D}, we can define a PL ratio \cite{Cowan:2010js} as  
\begin{equation}\label{eqn:prof_ratio}
    \Lambda (\theta_i) = \frac{\like_{\textnormal{PL}} (\theta_i)}{\like (\hat{\bm{\theta}})}, \quad \Lambda (\theta_i, \theta_j) = \frac{\like_{\textnormal{PL}} (\theta_i, \theta_j)}{\like (\hat{\bm{\theta}})},
\end{equation}
where $\hat{\bm{\theta}} \equiv (\hat{\theta}_1, \ldots, \hat{\theta}_7)$ is the best-fit point, i.e., a parameter point that maximises the total likelihood function $\like (\bm{\theta})$.~Using Wilks' theorem \cite{Wilks:1938dza}, Eq.~\eqref{eqn:prof_ratio} can be used to construct $1\sigma$ $(2\sigma)$ contours corresponding to $\sim 68.3\% \, (95.4\%)$\,C.L.~regions. 

In the following subsections, we present our model results in the form of 1D and 2D PL plots. These are generated using the \textsf{pippi\_v2.0} \cite{Scott:2012qh} package. 

\subsection{EWBG only}\label{subsec:EWBGonly}
We start by finding regions in the model parameter space where a successful EWBG is potentially viable.~This is achieved by performing a 7D scan of the model using \emph{only} the $v_c/T_c$ log-likelihood, i.e.,
\begin{equation}
    \ln \like (\bm{\theta}) = \ln \like_{v_c/T_c} (\bm{\theta}),
\end{equation}
where $\ln \like_{v_c/T_c} (\bm{\theta})$ is defined in subsection~\ref{subsec:vcTc}.~The resulting 2D PL plots are shown in Fig.~\ref{fig:2D_prof_like_vcTc}.~In the dark blue regions where the PL ratio $\Lambda \equiv \like/\like_{\textnormal{max}} = 1$, the dimensionless parameter $v_c/T_c \geq 0.6$ and a successful EWBG can be viable.~To understand the results in more detail, we go over each panel in Fig.~\ref{fig:2D_prof_like_vcTc} one-by-one.
\begin{enumerate}
    \item $(m_H, s_0)$ plane: For $m_H \lesssim 1.3$\,TeV, all values of $s_0$ and some combination of 5 profiled out parameters (namely $\mu_3$, $\lambda_S$, $\alpha$, $m_\psi$ and $g_S$) give $v_c/T_c \geq 0.6$ and maximise the $v_c/T_c$ log-likelihood, thus $\Lambda = 1$ everywhere.~Due to the dependence of $s_0$ in Eq.~\eqref{eqn:lPS}, large values of $|s_0|$ should lead to runaway directions, $\lambda_{\pp S} \leq -2 \sqrt{\lambda_\pp \lambda_S}$, and/or non-perturbative coupling, $|\lambda_{\pp S}| \geq 4\pi$.~With $\alpha  = \pi/2$, a large contribution from $m_H$ to $\lambda_{\pp S}$ can be suppressed.~However, this choice of $\alpha$ makes $\lambda_\Phi$ in Eq.~\eqref{eqn:lP} non-perturbative as its contribution appears as $m_H^2 \sin^2 \alpha$.~Ultimately, the solution is to choose a small value for $\lambda_S$ as its contribution in Eq.~\eqref{eqn:lPS} appears as $-\lambda_S s_0^2$. In addition, small values of $\mu_3$ can also help in keeping $|\lambda_{\pp S}| < 4\pi$.~Thus, for $m_H \lesssim 1.3$\,TeV, large values of $|s_0|$ can facilitate EWBG. 

    For $m_H \gtrsim 1.3$\,TeV and $|s_0| \gtrsim 50$\,GeV, the white region ($\Lambda = 0$) is disfavoured as it leads to $|\lambda_{\pp S}| \geq 4\pi$. This is expected as the contribution from $m_H$ in Eq.~\eqref{eqn:lPS} is dominant at large values.~With large $|s_0|$, no choice of $\mu_3$, $\lambda_S$ and $\alpha$ can keep $|\lambda_{\pp S}| < 4\pi$. In fact, the requirement $|\lambda_{\pp S}| < 4\pi$ translates into an upper limit on $m_H$ as a function of $s_0$, $\mu_3$, $\lambda_S$ and $\alpha$. Using Eq.~\eqref{eqn:lPS}, we get 
        \begin{equation}\label{eqn:pert_lPS}
      \frac{v_0}{s_0} (m_H^2 - m_h^2) \sin 2\alpha + 4 (m_h^2 \sin^2 \alpha + m_H^2 \cos^2 \alpha + \mu_3 s_0 - 2 \lambda_S s_0^2) < 8 \pi v_0^2.
    \end{equation}
    For a fixed $m_H$ and $s_0$, Eq.~\eqref{eqn:pert_lPS} has 3 degrees of freedom.~As $\mu_3$, $\lambda_S$ and $\alpha$ are profiled over, it is non-trivial to predict the exact shape of the upper limit on $m_H$ as a function of $s_0$. The upper limit also weakens as $|s_0|$ increases. The net result is that for $m_H \gtrsim 5$\,TeV, $|s_0| \lesssim 50$\,GeV is required to facilitate EWBG. 
    
    \item $(m_H, \alpha)$ plane: Similar to the $(m_H, s_0)$ plane for $m_H \lesssim 1.3$\,TeV, some combination of the profiled out parameters gives $v_c/T_c \geq 0.6$ for all values of $\alpha$. However, when $m_H \gtrsim 1.3$\,TeV and $\alpha \neq 0, \, \pi$, the Higgs quartic coupling $\lambda_\pp$ in Eq.~\eqref{eqn:lP} becomes non-perturbative.~In fact, the requirement $|\lambda_\pp| < 4\pi$ translates into the following upper limit on $m_H$ as a function of $\alpha$
    \begin{equation}\label{eqn:lP_non_pert}
        m_H^2 \sin^2\alpha < 8\pi v_0^2 - m_h^2 \cos^2 \alpha.
    \end{equation}
    When $\alpha = 0, \, \pi$, the above condition is satisfied for all values of $m_H$.~Thus, a successful EWBG can be viable at large values of $m_H$.~On the other hand, when $\alpha = \pi/2$, Eq.~\eqref{eqn:lP_non_pert} imposes the strongest upper limit on $m_H$, namely $m_H \lesssim 1.23$\,TeV. As $\alpha \rightarrow 0, \, \pi$, the upper limit on $m_H$ becomes weaker, as is evident from the plot.
    
    \item $(m_H, \mu_3)$ and $(m_H, \lambda_S)$ planes: In these two planes, all possible combinations of $(m_H, \mu_3)$, $(m_H, \lambda_S)$ and profiled out parameters give $v_c/T_c \geq 0.6$.~Thus, the PL ratio is roughly flat and equal to 1 everywhere; hence, we do not show these planes in Fig.~\ref{fig:2D_prof_like_vcTc}. In fact, the $v_c/T_c$ likelihood is weakly dependent on $\mu_3$ and $\lambda_S$ as expected from Eq.~\eqref{eqn:lPS}. For instance, at large values of $\mu_3$ or $\lambda_S$ which would give $|\lambda_{\pp S}| \geq 4\pi$ or $\lambda_{\pp S} \leq -2 \sqrt{\lambda_{\pp} \lambda_S}$, small values of $s_0$ can be chosen to avoid such situations. 
    
    \item $(m_H, m_\psi)$ plane: For $m_H \lesssim 5$\,TeV, all values of $m_\psi$ give $v_c/T_c \geq 0.6$. As $m_\psi$ does not appear directly in Eqs.~\eqref{eqn:lP} and \eqref{eqn:lPS}, the $v_c/T_c$ likelihood is weakly dependent on $m_\psi$. This is expected as the contribution from $m_\psi$ to the effective potential appears only at 1-loop order.
    
	\begin{figure}[t]
	    \centering	    
	    
	    \includegraphics[scale=0.6]{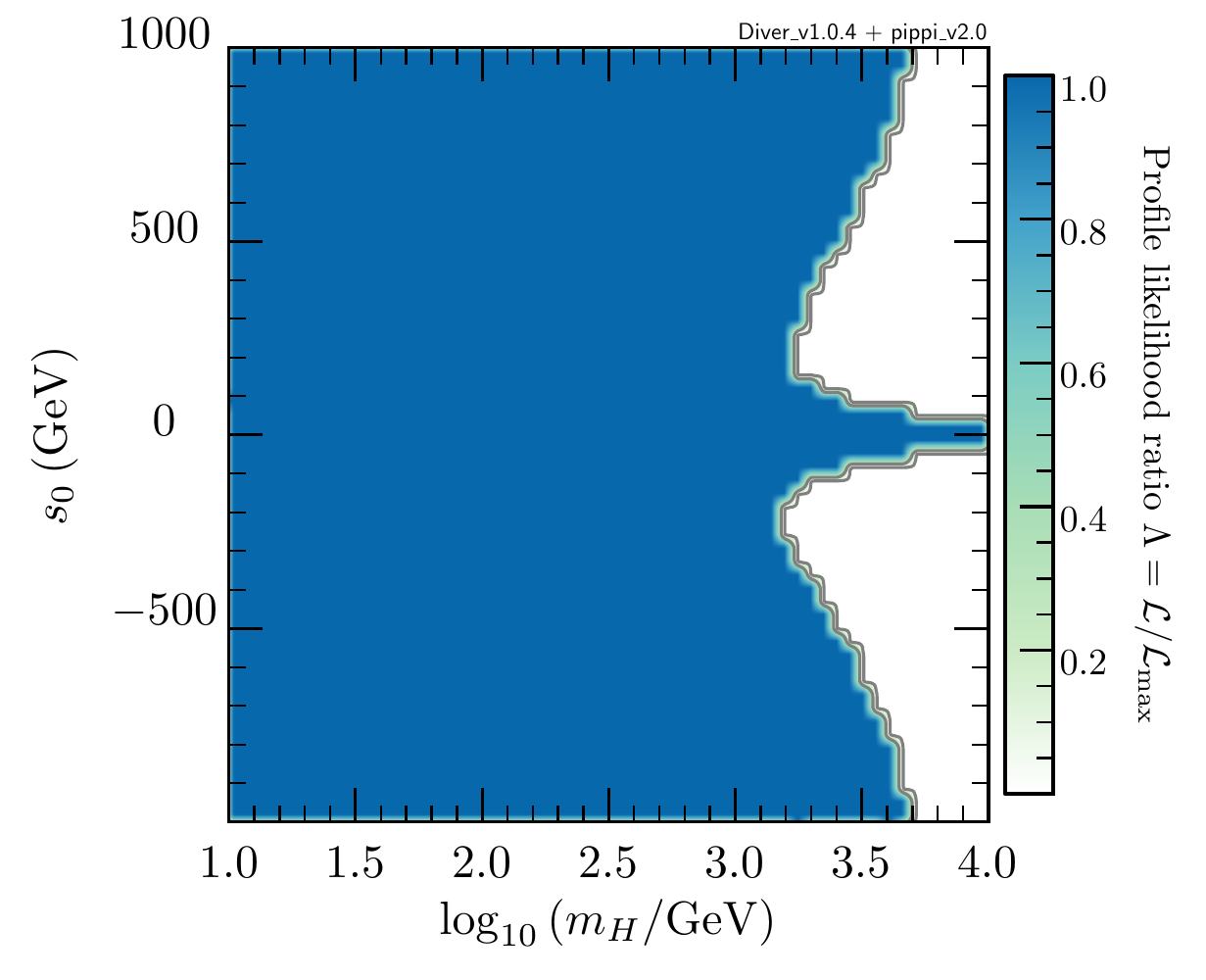}
	    \includegraphics[scale=0.6]{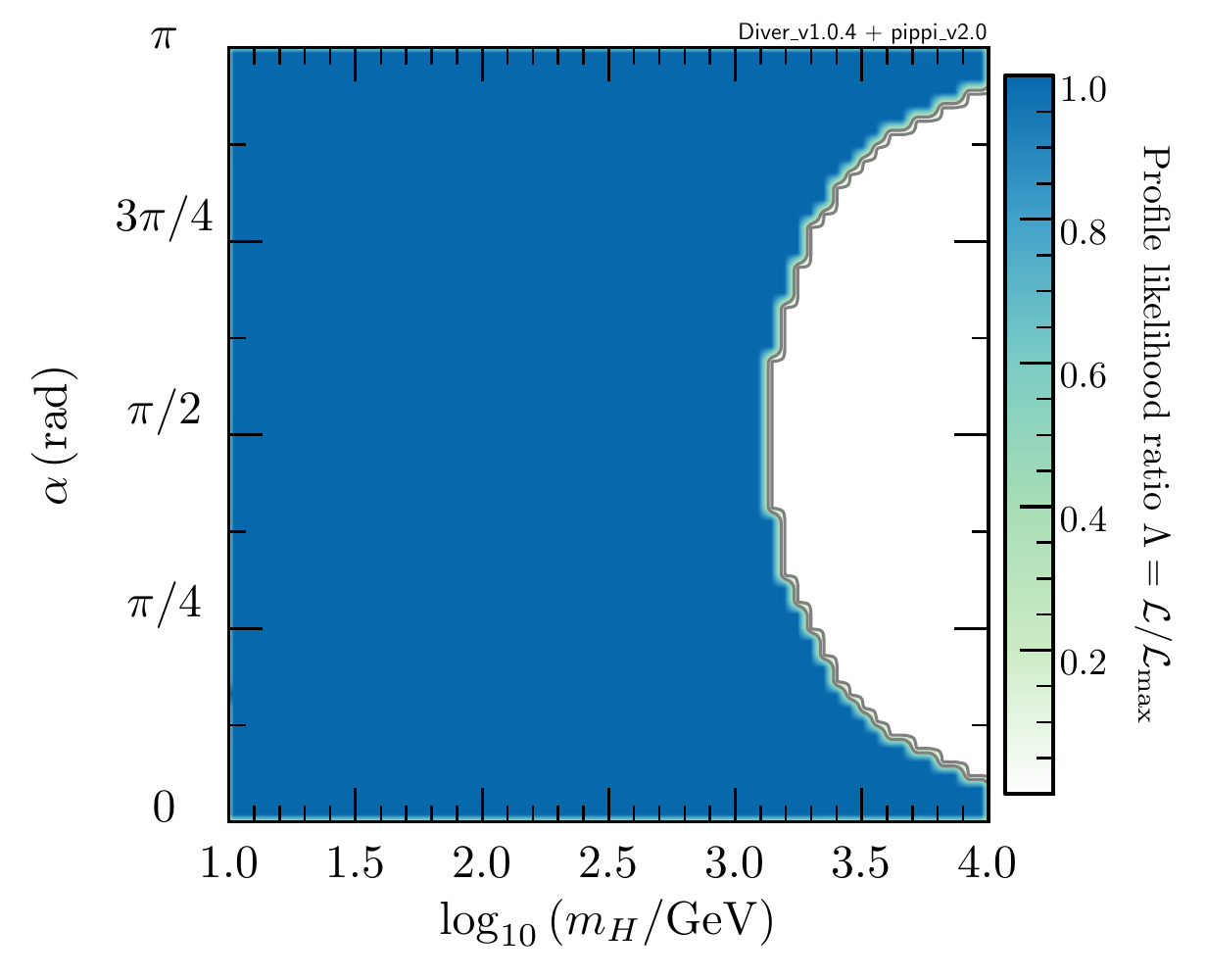}    
		
	    \includegraphics[scale=0.6]{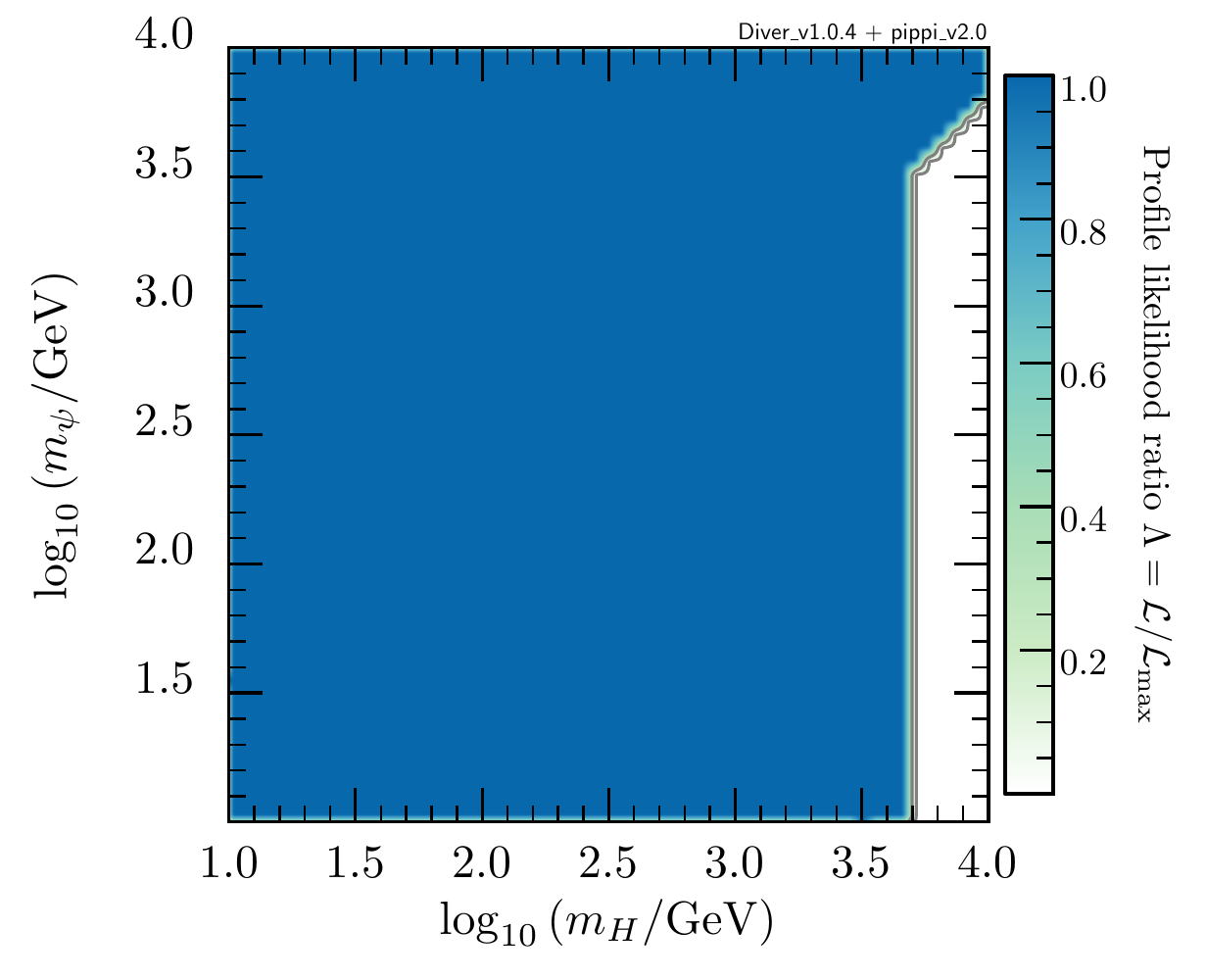}
	    \includegraphics[scale=0.6]{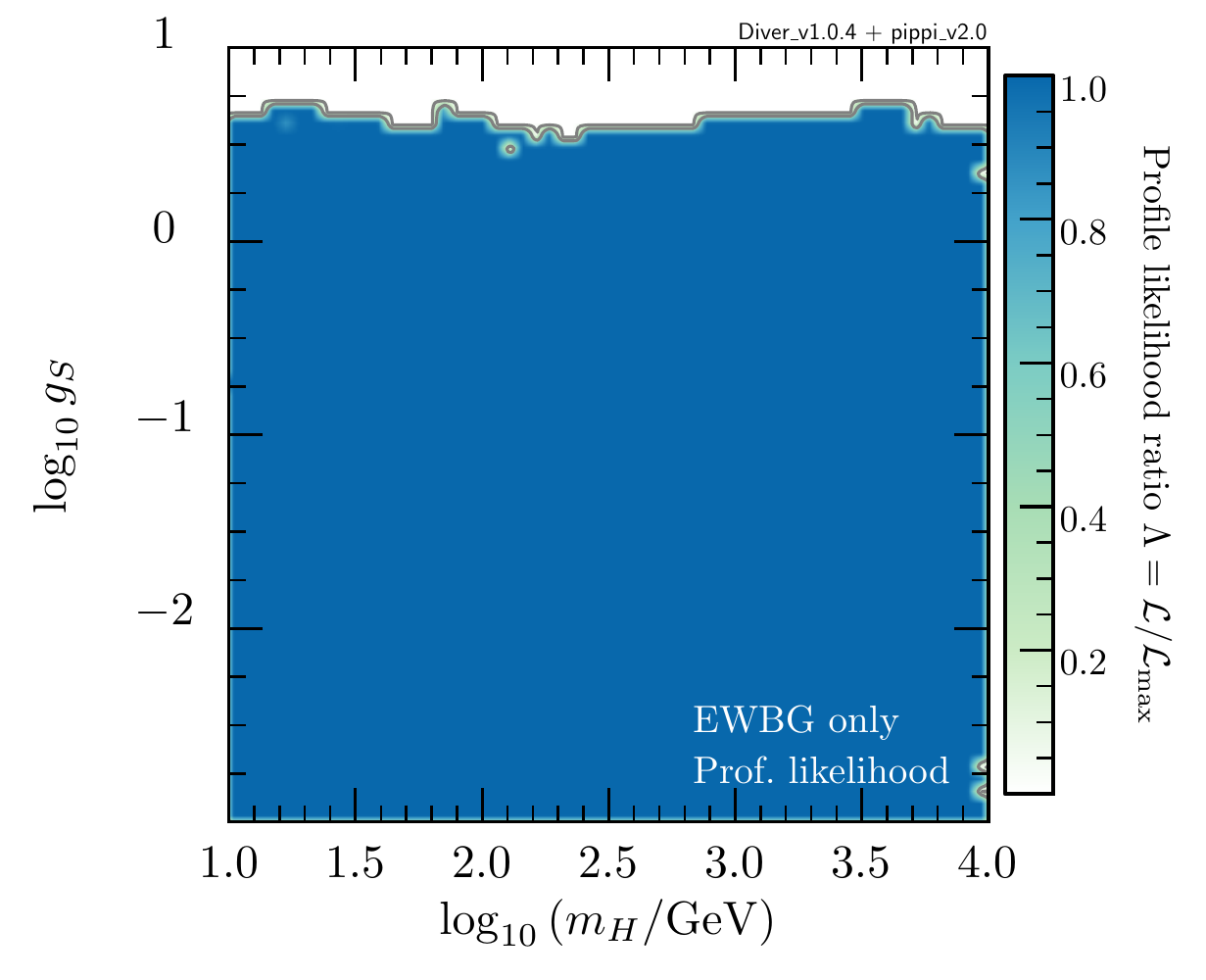}    
	    
	\caption{2D profile likelihood (PL) plots from a 7D scan of our model using \emph{only} the electroweak baryogenesis (EWBG) constraint.~The contour lines mark out the $1\sigma$\,(68.3$\%$)  and $2\sigma$\,(95.4$\%$) C.L.~regions.~In regions where $\Lambda \equiv \like/\like_{\textnormal{max}} = 1$, a successful EWBG can be viable as $v_c/T_c \geq 0.6$ (see text for more details).~The parameter planes $(m_H, \mu_3)$ and $(m_H, \lambda_S)$ are not shown as they are unconstrained by the EWBG constraint.}
	    \label{fig:2D_prof_like_vcTc}
	\end{figure}

    For $m_H \gtrsim 5$\,TeV and $m_\psi \lesssim 3.2$\,TeV, no combination of the profiled out parameters can keep $|\lambda_{\pp S}| < 4\pi$.~On the other hand, when $m_\psi \gtrsim 3.2$\,TeV, one can arrange for a cancellation of large quantum corrections to obtain perturbative couplings, although all such solutions carry some degree of extra tuning. 
    
    \item $(m_H, g_S)$ plane: For $g_S \lesssim 5.62$, all values of $m_H$ and profiled out parameters give $v_c/T_c \geq 0.6$, and maximise the $v_c/T_c$ likelihood.~However, values of $g_S > 5.62$ lead to runaway directions in the potential as the contribution from $g_S$ in the 1-loop corrections become large.
\end{enumerate}
In summary, it is not difficult to facilitate a successful EWBG in our model.~For any specific model parameter, usually some combination of the remaining parameters give viable points even if the parameter in question causes problems.~For instance, large values of $m_H$ generally push up the EWSB minimum and cause it to not become the global minimum at $T = 0$. However, this effect can be counteracted by choosing a large value for $m_\psi$ which gives a large negative contribution to the effective potential.~One exception is $g_S > 5.62$ which always generates runaway directions in the effective potential.~For the remaining model parameters, namely $(m_H, s_0, \mu_3, \lambda_S, \alpha, m_\psi)$, the 1D PL ratio $\Lambda$ is roughly flat and equal to 1 for all parameter values.~Thus, we do not show the 1D PL plots for our model parameters. 


%
%
    
\subsection{Global fit}
With some intuition on the choice of free model parameters that can facilitate a successful EWBG, we present results from a global fit of our model using the total log-likelihood function in Eq.~\eqref{eqn:tot_like}. In practice, we consider two scenarios in which the fermion DM accounts for either a small fraction $(f_\textnormal{rel} \leq 1)$ or all $(f_{\textnormal{rel}} = 1)$ of the observed DM abundance. In the former case, we use a relic density likelihood that is one-sided Gaussian, whereas in the latter, we use a Gaussian likelihood. For more details, see subsection~\ref{subsec:relic}.

\subsubsection{Scenario I: $f_{\textnormal{rel}} \leq 1$}\label{subsec:freleqone}
The resulting 2D PL plots from our 7D scans are shown in Fig.~\ref{fig:2D_prof_like_all}.~For $m_H \lesssim m_h/2 = 62.6$\,GeV, the parameter planes are ruled out by the observed Higgs signal strength measurements, EWPO and direct Higgs searches performed at the LEP experiment.~As the decay channel $h \rightarrow H H$ is kinematically allowed and dominant in this region for all values of the mixing angle $\alpha$, it reduces the SM-like Higgs signal strength $\mu_h$ with respect to SM expectation, see Eq.~\eqref{eqn:h_mu}.~This translates into a large $\chi^2_\mu$ in Eq.~\eqref{eqn:HS_chi} and is thus disfavoured. 

To understand the remaining set of results in more detail, we go over each panel in Fig.~\ref{fig:2D_prof_like_all} one-by-one.
\begin{enumerate}
	\item $(m_H, s_0)$ plane: For $m_H \gtrsim 4$\,TeV, the parameter planes are ruled out by the EWBG constraint as they either lead to runaway directions, $\lambda_{\pp S} \leq -2\sqrt{\lambda_\pp \lambda_S}$, or non-perturbative couplings, $|\lambda_{\pp}|, |\lambda_{\pp S}| \geq 4\pi$.~Although, some combinations of the profiled out parameters can give a successful EWBG at large values of $m_H$ (see Fig.~\ref{fig:2D_prof_like_vcTc}), they are often not compatible with the remaining constraints. This is especially true for the EWPO constraint which only depends on $m_H$ and $\alpha$.~For large $m_H$, $\alpha \simeq 0, \pi$ is required in order to satisfy the EWPO constraint.

	\begin{figure}[t]
	    \centering
	    
	    \includegraphics[scale=0.6]{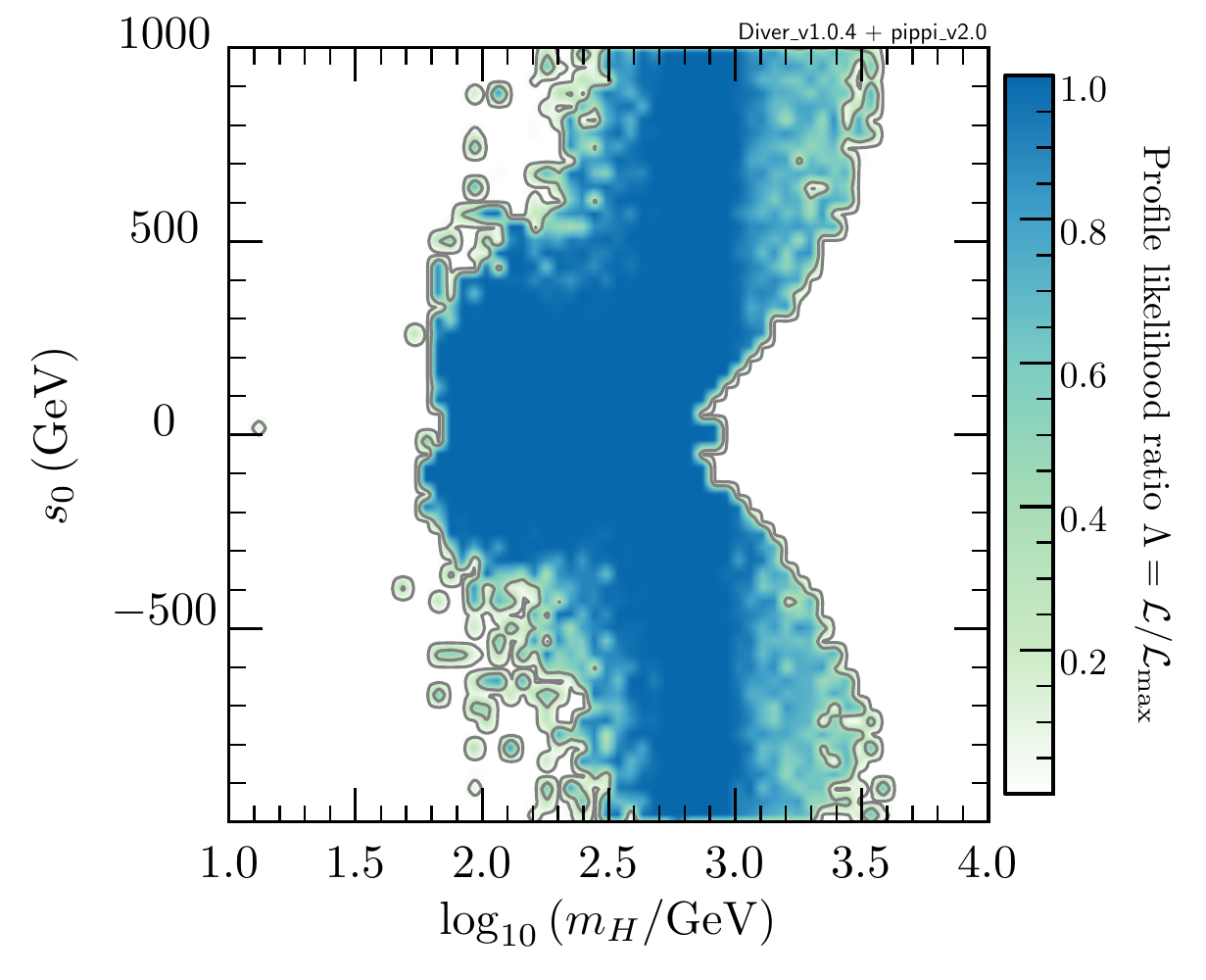} 
	    \includegraphics[scale=0.6]{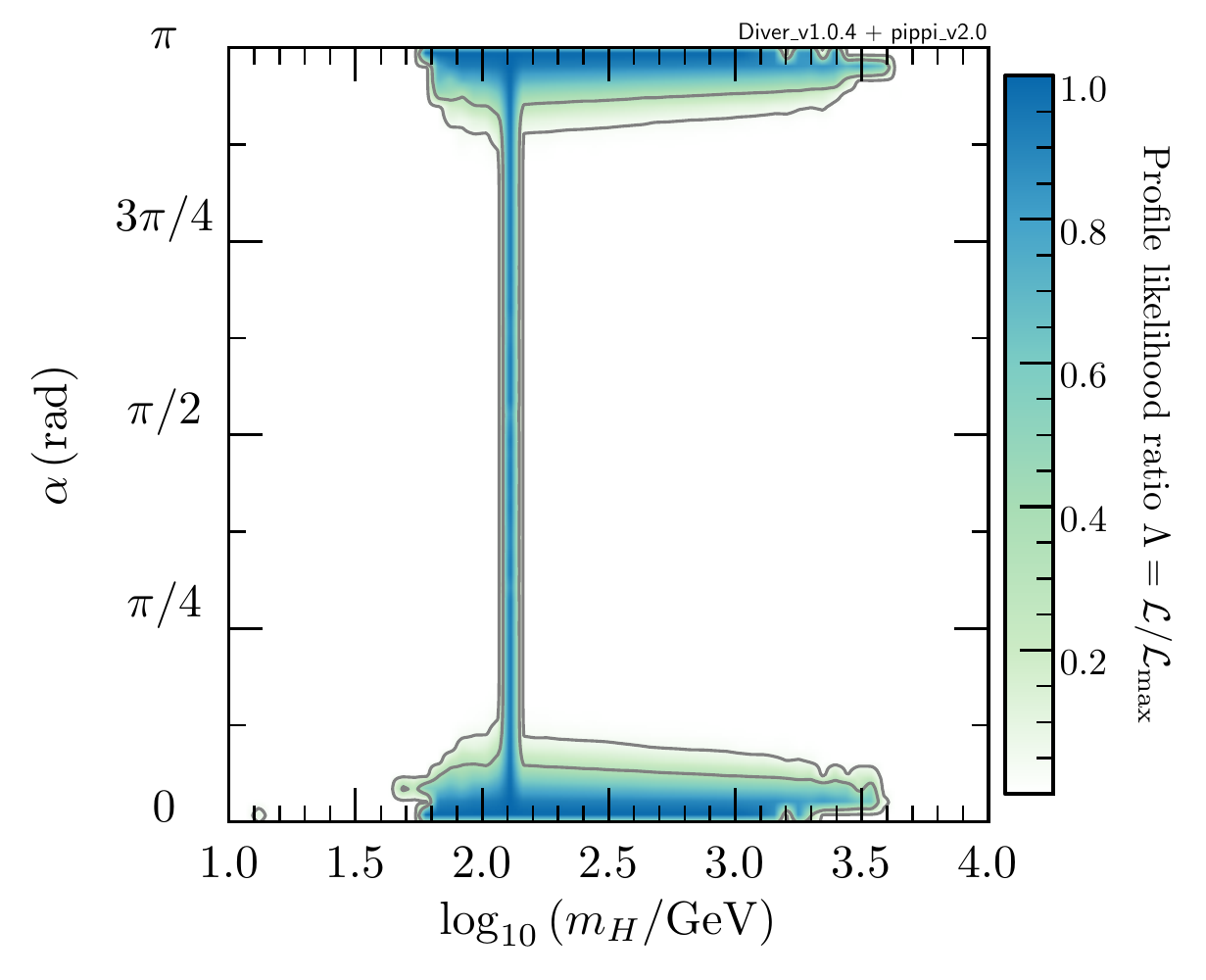}     
	
	    \includegraphics[scale=0.6]{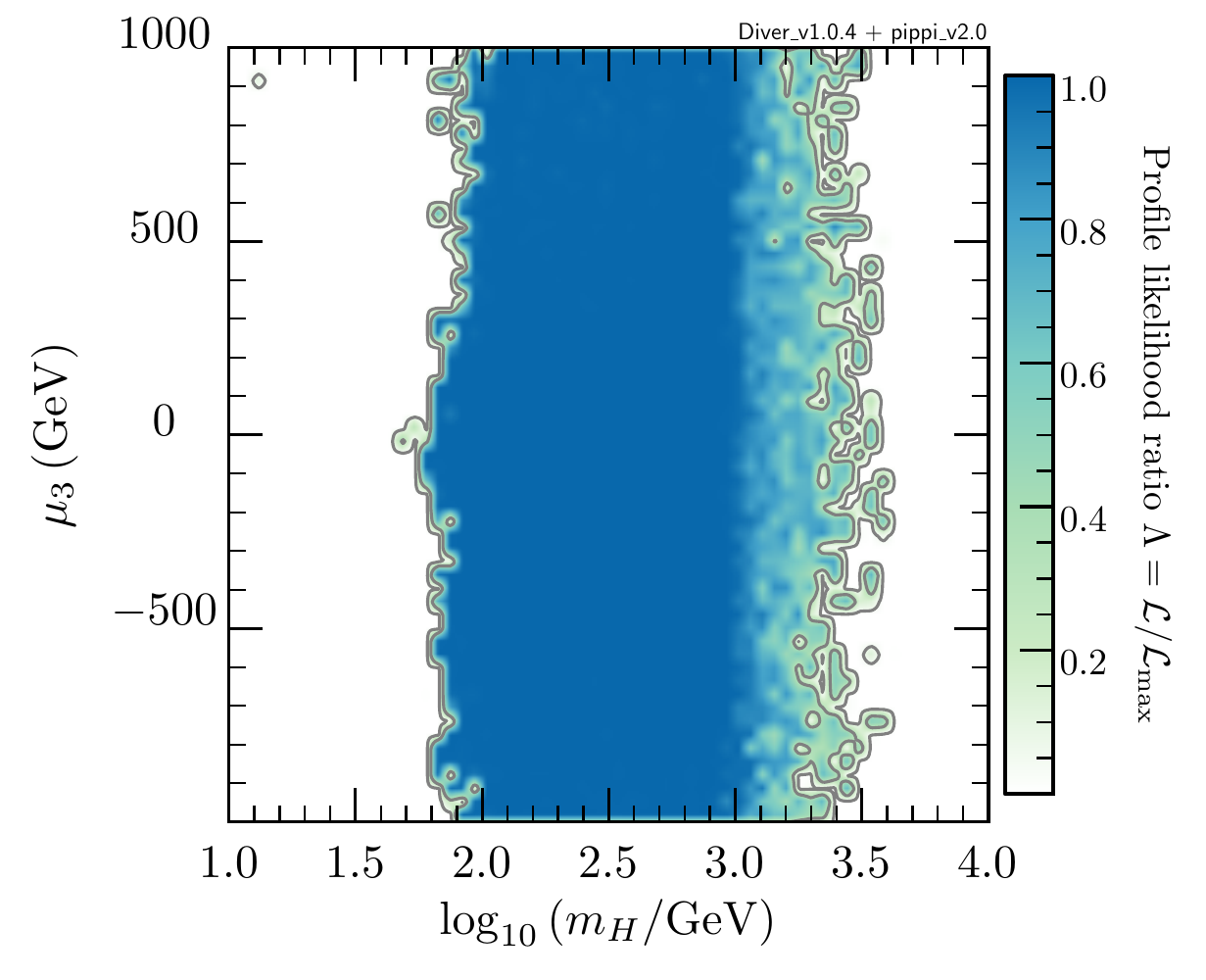} 
	    \includegraphics[scale=0.6]{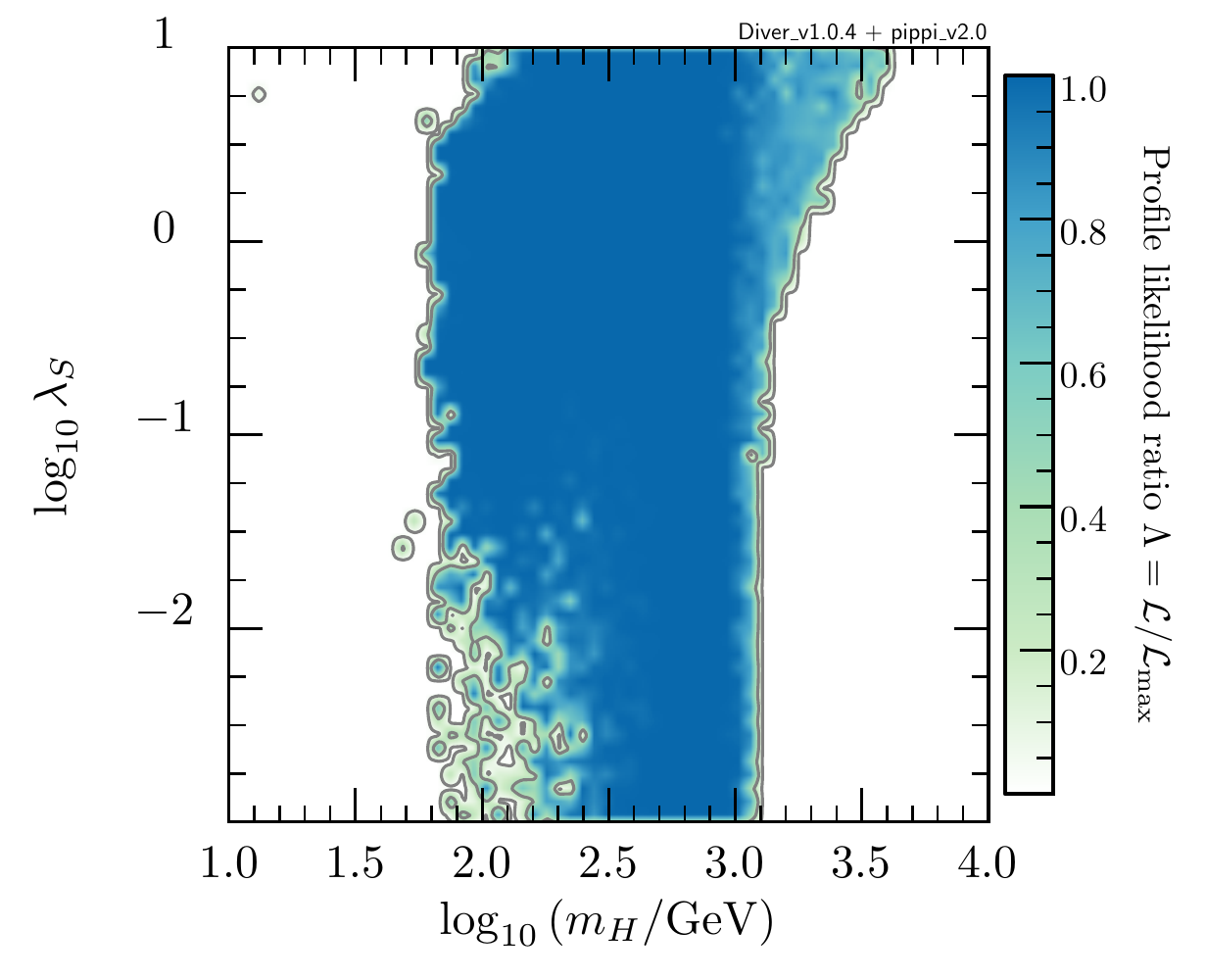}     
	
	    \includegraphics[scale=0.6]{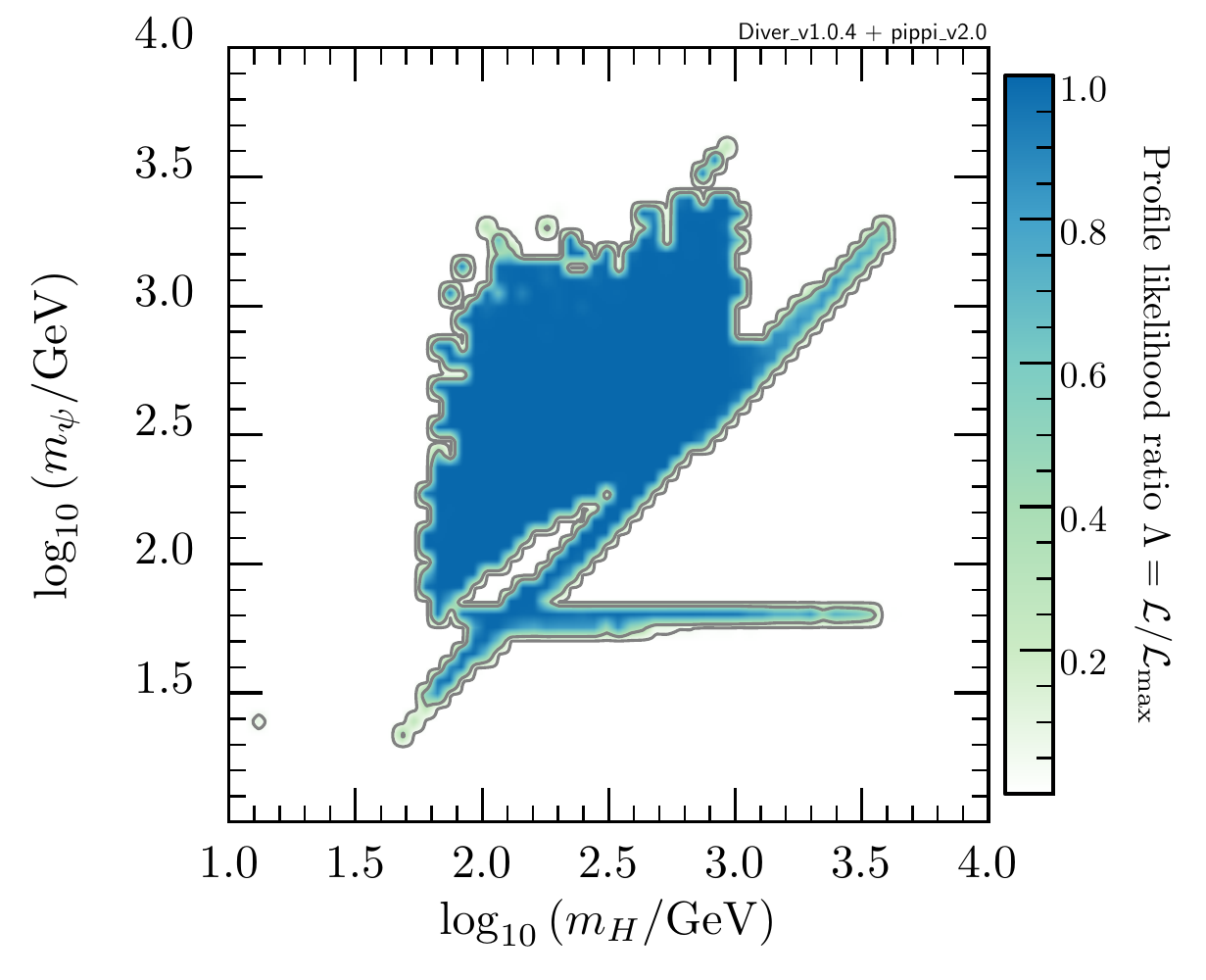} 
	    \includegraphics[scale=0.6]{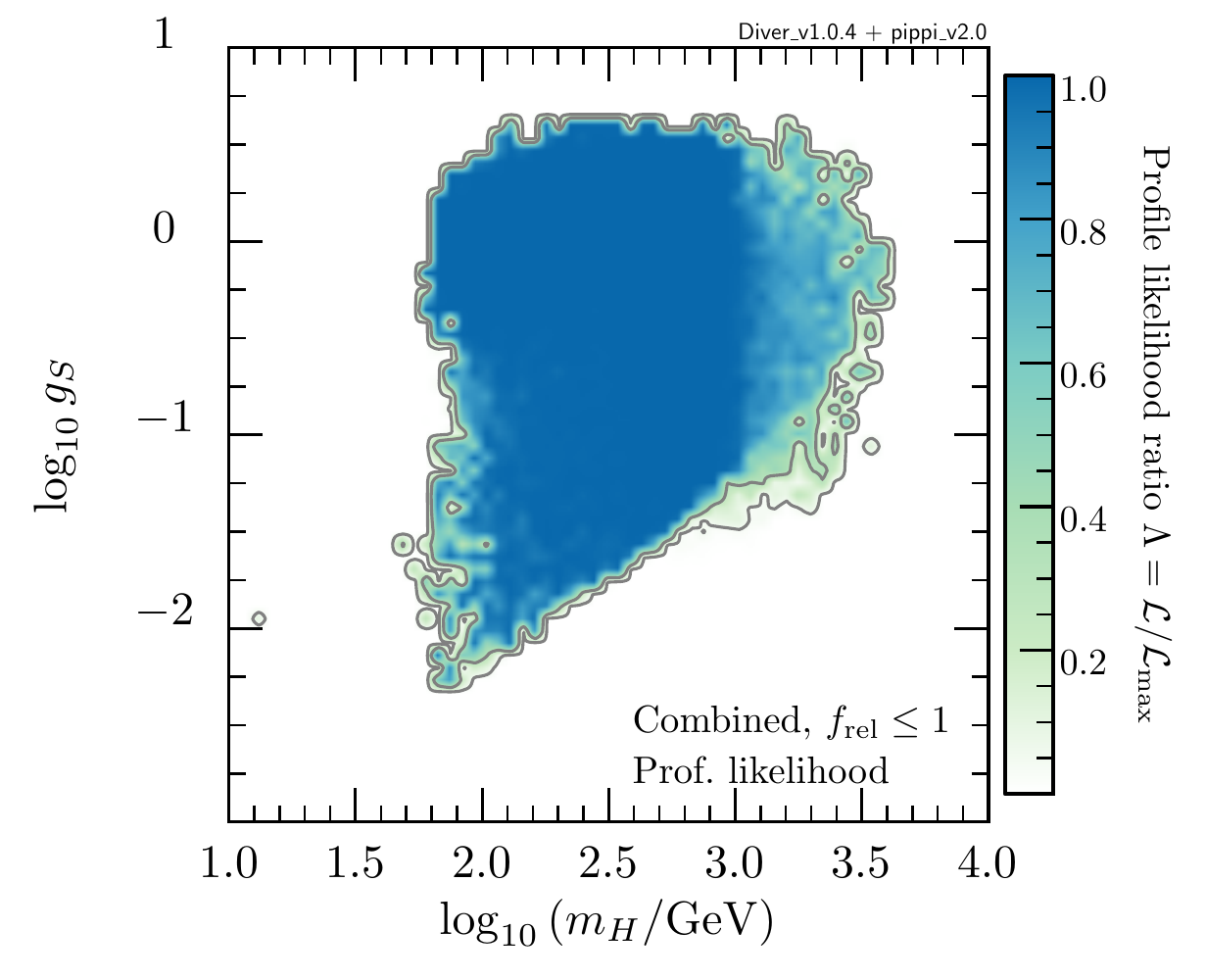}
	
	    \caption{2D PL plots from a global fit of our model assuming $f_{\textnormal{rel}} \equiv \Omega_\psi/\Omega_{\textnormal{DM}} \leq 1$. The contour lines mark out the $1\sigma$\,(68.3$\%$)  and $2\sigma$\,(95.4$\%$) C.L.~regions.}
	    \label{fig:2D_prof_like_all}
	\end{figure}
	
	\item $(m_H, \alpha)$ plane: We see that the model is allowed by all constraints for a range of low and high $m_H$ values provided $\alpha \simeq 0, \pi$.~This is expected as the second scalar $H$ is decoupled in this regime and gives no new contribution to the observed Higgs signal strengths. However, when $m_H \simeq m_h$, the two scalars are indistinguishable from the point of direct Higgs searches and Higgs signal strength measurements. As is evident in Eqs.~\eqref{eqn:G_coupling} and \eqref{eqn:mod_oblique}, direct detection and EWPO constraints respectively are also relaxed in this regime. The net result is that all values of $\alpha$ are allowed when $m_H \simeq m_h$.

	\item $(m_H, \mu_3)$ and $(m_H,\lambda_S)$ planes: These parameter planes are mostly unconstrained by our global fit except for $m_H \lesssim m_h/2$ (excluded by the Higgs signal strength measurements) and $m_H \gtrsim 4$\,TeV (ruled out by the EWBG constraint). For $m_H \gtrsim 1.3$\,TeV, large values of $\lambda_S$ are required to facilitate a successful EWBG. 
	
	\item $(m_H, m_\psi)$ plane: For $m_\psi \lesssim m_h/2$, the fermion DM can only annihilate into light SM quarks, thereby giving $f_{\textrm{rel}} > 1$.~On the other hand, $m_\psi \gtrsim m_h/2$ is constrained by the DM relic density and XENON1T limits. When $m_\psi \simeq m_h/2$, all values of $m_H$ up to $\sim 4$\,TeV are allowed by the \emph{Planck} measured relic density and XENON1T limits; this region appears in the plot as a horizontal band. In this band, small values of $g_S$ can yield a fermion DM relic density and DM-nucleon cross-section that is compatible with the \emph{Planck} measured value and XENON1T limit respectively. 
	
	For $m_\psi \in [m_h/2, m_H/2]$, the region is disfavoured by either the \emph{Planck} measured relic density or XENON1T limit.~This is generally expected from an incompatibility between small values of $g_S$ which are favoured by the XENON1T limit (as it gives a small DM-nucleon cross-section $\sigma_{\textnormal{SI}}^{\psi \calN})$ but disfavoured by the relic density constraint (as it gives $f_{\textnormal{rel}} > 1$) and vice versa.
	    
	The diagonal band corresponds to the second resonance $m_\psi \simeq m_H/2$. Similar to the first resonance $m_\psi \simeq m_h/2$, all  points in this band are allowed by the relic density and XENON1T limits. As $g_S$ is profiled over, small values of $g_S$ can easily give $f_{\textnormal{rel}} \leq 1$ and $\sigma_{\textnormal{SI}}^{\textnormal{eff}} \leq \sigma_{\textnormal{XENON1T}}$. On the other hand, when $m_H \gtrsim 4$\,TeV, parameter points are disfavoured by the EWPO and EWBG constraints.~For $m_\psi \gtrsim 3.2$\,TeV, the region is robustly excluded by the combined constraints.

	\item $(m_H, g_S)$ plane: In this plane, a lower limit on $g_S$ comes from the DM relic density constraint as smaller values of $g_S$ lead to an overabundance of the fermion DM in the universe today.~This lower limit becomes weaker as $m_H$ increases.~For $m_H \gtrsim 4$\,TeV, the coupling $\lambda_{\pp S}$ becomes non-perturbative, thus this region is disfavoured.~Similarly, values of $g_S \gtrsim 3.2$ are disfavoured by the EWBG constraint as they lead to runaway directions in the potential, see Fig.~\ref{fig:2D_prof_like_vcTc}.  
\end{enumerate}

\begin{figure}[t]
    \centering  
    
    \includegraphics[scale=0.45]{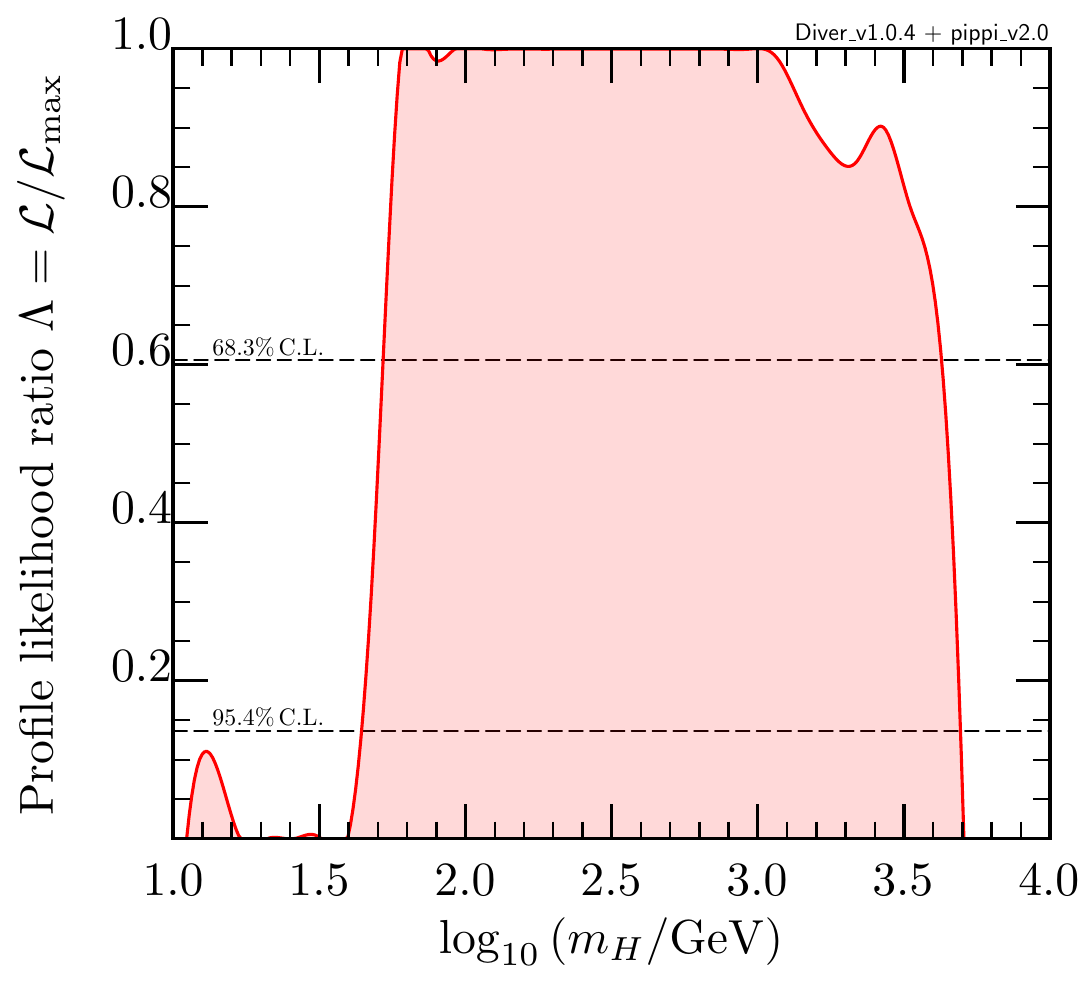}
    \includegraphics[scale=0.45]{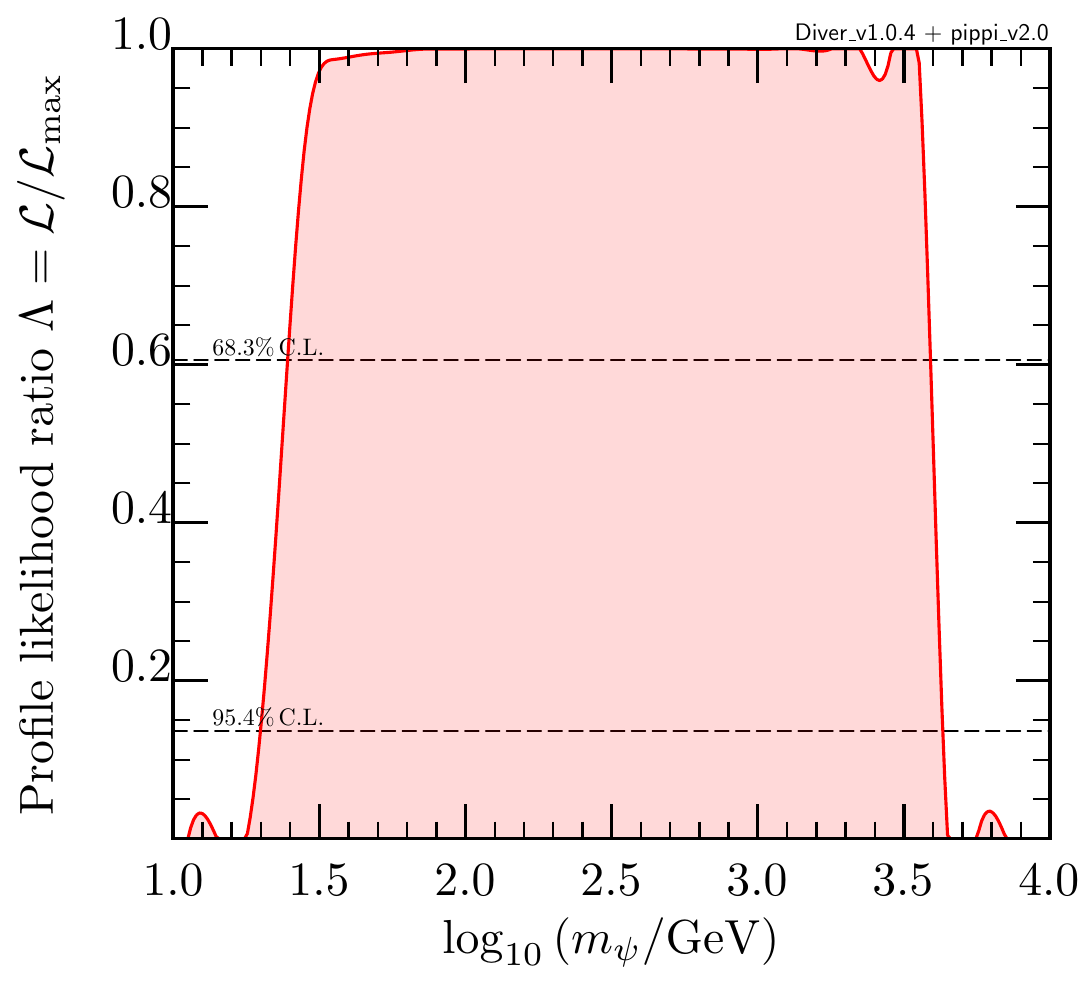} 
    \includegraphics[scale=0.45]{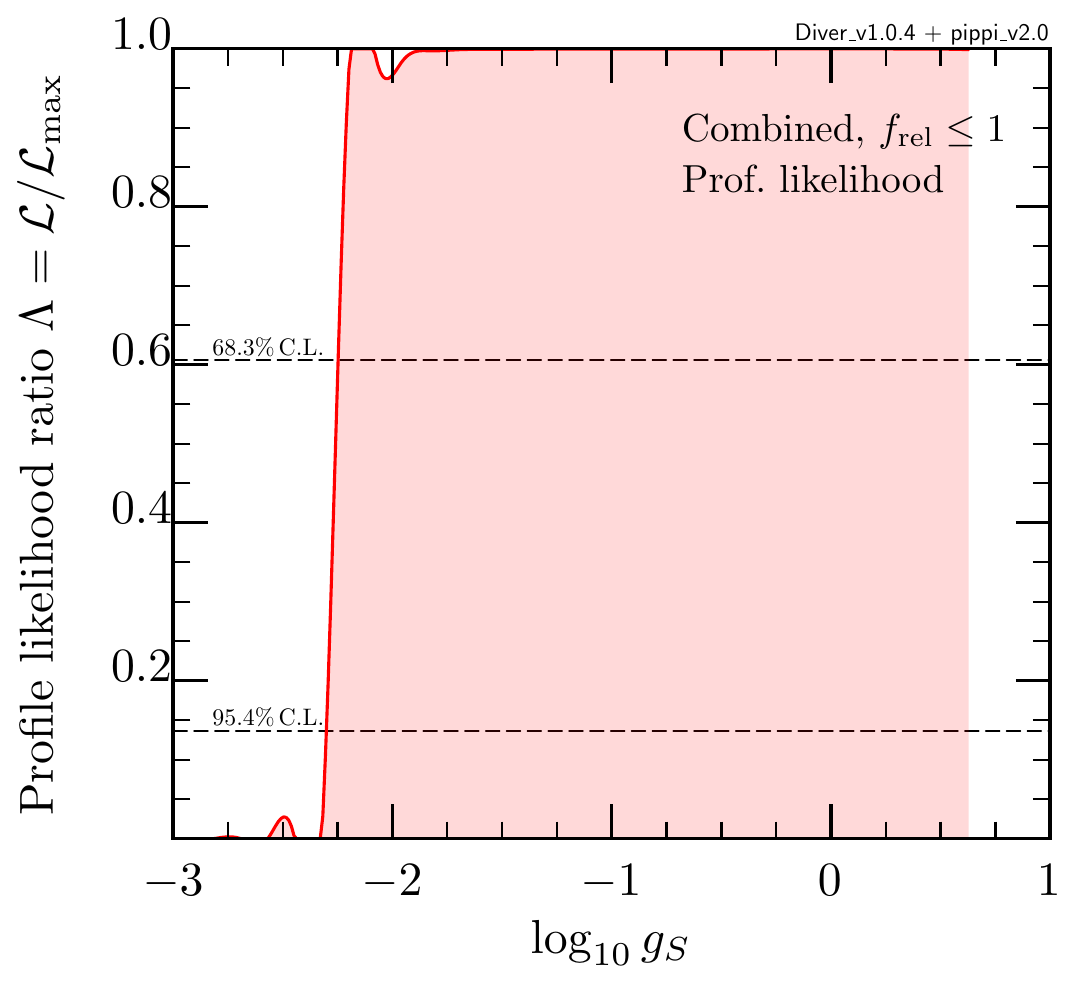}	
    
    \caption{1D PL plots for $m_H$ (left), $m_\psi$ (center) and $g_S$ (right) assuming $f_{\textnormal{rel}} \leq 1$. The respective plots for $s_0$, $\mu_3$ and $\lambda_S$ are not shown as they are unconstrained by our global fit.}
    \label{fig:1D_prof_like_all}
\end{figure}

In Fig.~\ref{fig:1D_prof_like_all}, we show the 1D PL plots for the parameters $m_H$, $m_\psi$ and $g_S$.\footnote{For the remaining parameters, we find that the PL ratio $\Lambda$ is roughly flat and equal to 1 at all values. In other words, the parameters $s_0$, $\mu_3$, $\lambda_S$ and $\alpha$ are unconstrained by our global fit.}~From these plots, it is evident that the combined constraints impose an upper \emph{and} lower limit on $m_H$, $m_\psi$ and $g_S$, namely
\begin{equation}
	m_h/2 \lesssim m_H \lesssim 5\,\mathrm{TeV}, \quad 32\,\mathrm{GeV} \lesssim m_\psi \lesssim 3.2\,\mathrm{TeV}, \quad 5.6 \times 10^{-3} \lesssim g_S \lesssim 3.5.
\end{equation}
These limits are based on our chosen ranges and priors for the free model parameters (as summarised in Table~\ref{tab:par_ranges}).~For instance, our lower limit on $m_H$ can be softened by reducing the branching ratio $\mathcal{BR}(h \rightarrow HH)$ when $\alpha = 0$, $\pi$. In these cases, the reduced branching ratio can give a better fit to the observed signal strength measurements for a SM-like Higgs boson $h$. For non-zero mixing angles, however, this part of the parameter space is strongly constrained by the direct Higgs searches performed at the LEP experiment.

\begin{figure}[t]
	\centering
	
    \includegraphics[scale=0.6]{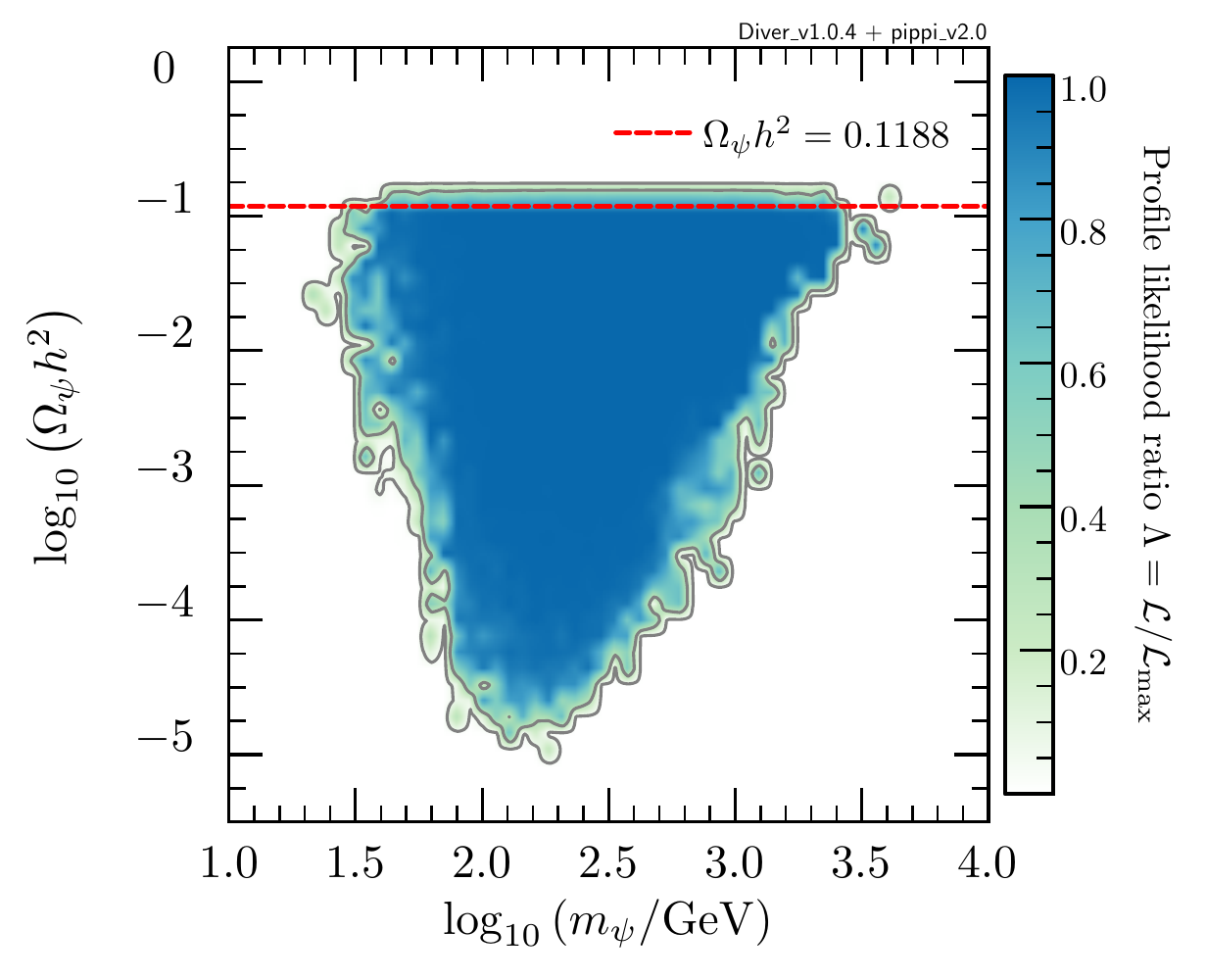}
    \includegraphics[scale=0.6]{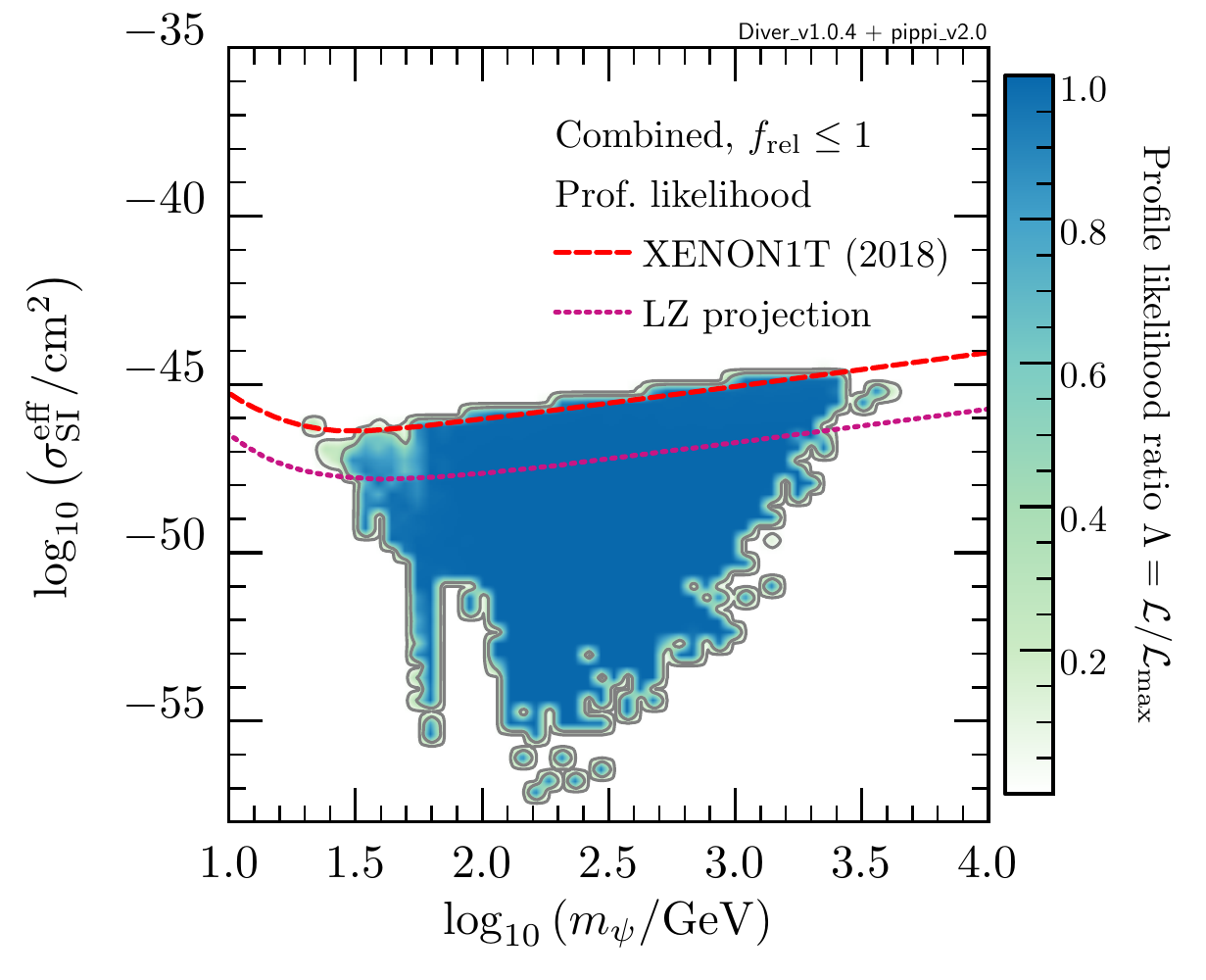}

    \caption{\emph{Left panel}: Fermion DM relic density vs.~the fermion DM mass. The red dashed curve corresponds to the \emph{Planck} measured value \cite{Ade:2015xua}.~\emph{Right panel}: Effective SI DM-nucleon cross-section vs.~the fermion DM mass.~The red dashed curve shows the current 90$\%$ C.L.~upper limit from the XENON1T (2018) \cite{Aprile:2018dbl}, whereas the violet dotted curve shows the projected sensitivity of the LUX-ZEPLIN (LZ) \cite{Akerib:2018lyp} experiment.}
    \label{fig:RD_and_DD_leq}
\end{figure}

In Fig.~\ref{fig:RD_and_DD_leq}, we show the key observables such as the fermion DM relic density (left panel) and effective SI DM-nucleon cross-section (right panel).~These can be compared against the \emph{Planck} measured value and XENON1T limit.~It is evident that all of the sampled points satisfy $f_{\textnormal{rel}} \leq 1$ and $\sigma_{\textnormal{SI}}^{\textnormal{eff}} \leq \sigma_{\textnormal{XENON1T}}$. We also show the projected sensitivity of the LUX-ZEPLIN (LZ) \cite{Akerib:2018lyp} experiment. Intriguingly, the LZ experiment will probe 2 orders of magnitude smaller DM-nucleon cross sections than the XENON1T experiment. Due to the two resonances $m_\psi \simeq m_{h,H}/2$ and the ability to profile over $\alpha$, the direct detection cross-section in our model can be significantly suppressed to avoid bounds from current and future direct search experiments. 

\subsubsection{Scenario II: $f_{\textnormal{rel}} = 1$}

\begin{figure}[t]
    \centering
    
	\includegraphics[scale=0.45]{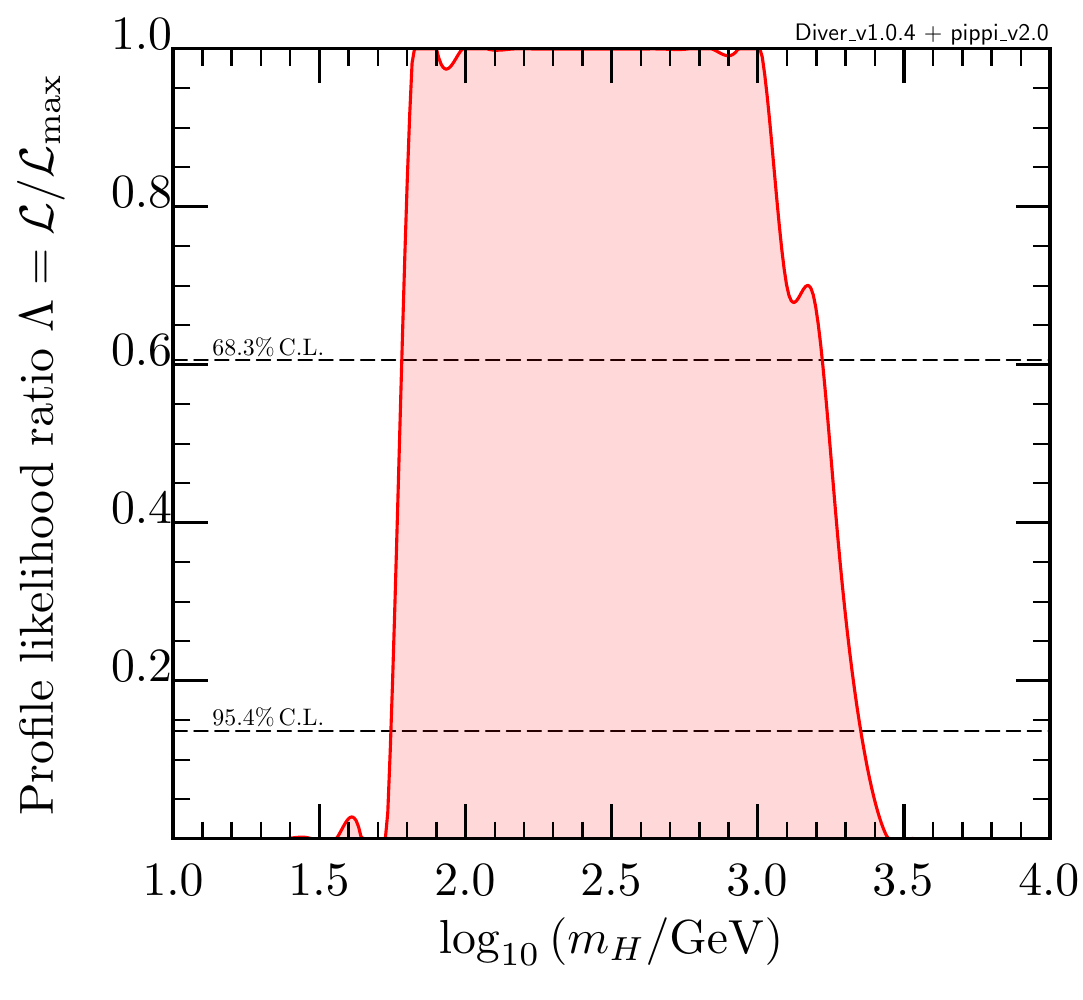}
    \includegraphics[scale=0.45]{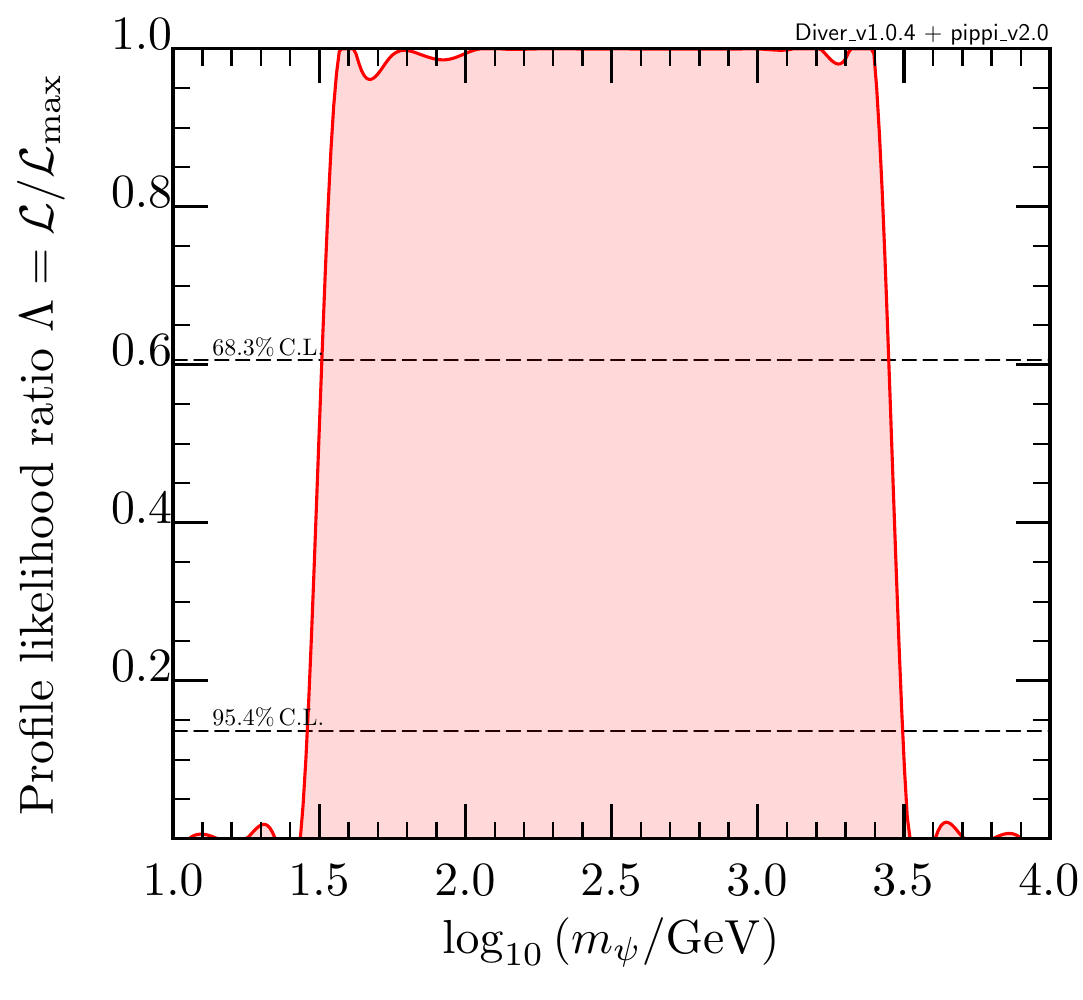}
    \includegraphics[scale=0.45]{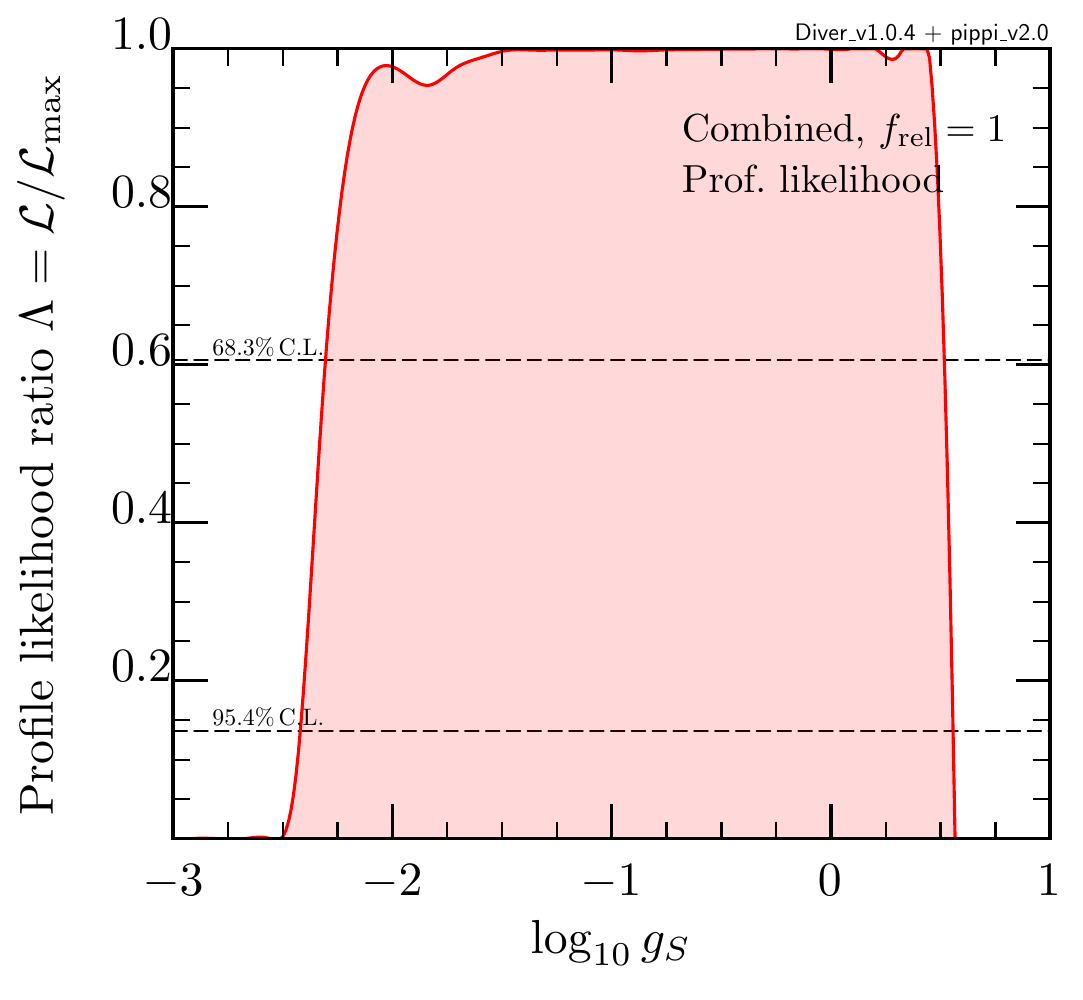}	
    
    \caption{Same as Fig.~\ref{fig:1D_prof_like_all} except for the $f_{\textnormal{rel}} = 1$ case.}
    \label{fig:1D_prof_like_all_eq}
\end{figure}

In this subsection, we present results from our global fit assuming $f_{\textnormal{rel}} = 1$.~The only difference with respect to the $f_{\textnormal{rel}} \leq 1$ case is our use of a Gaussian likelihood function for the \emph{Planck} measured DM relic density. In this case, not only small values of $g_S$ are disfavoured by the relic density constraint (as they give $f_{\textnormal{rel}} > 1$), large values of $g_S$ are also disfavoured (as they give $f_{\textnormal{rel}} \ll 1)$.

\begin{figure}[t]
    \centering
    
    \includegraphics[scale=0.6]{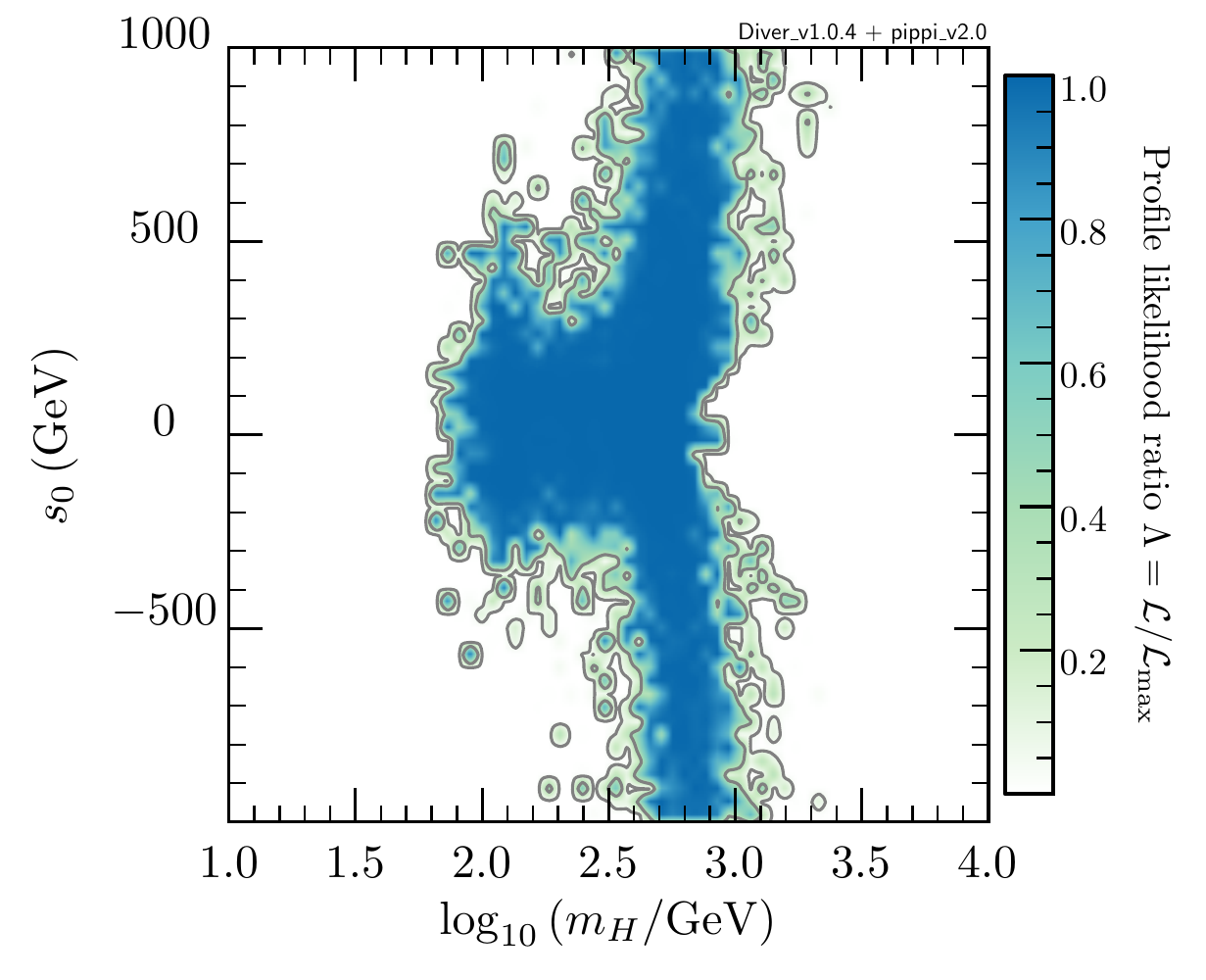} 
    \includegraphics[scale=0.6]{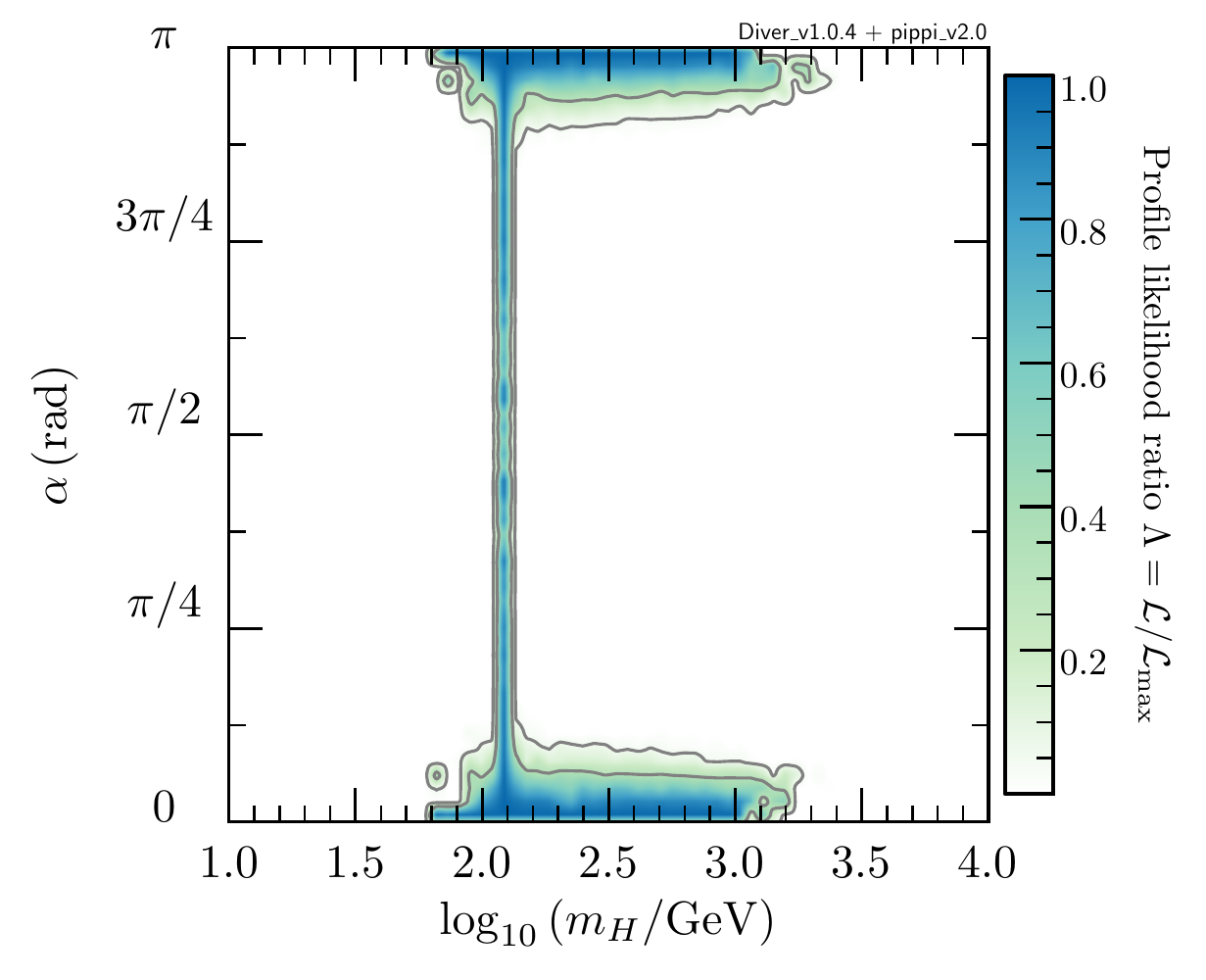}     

    \includegraphics[scale=0.6]{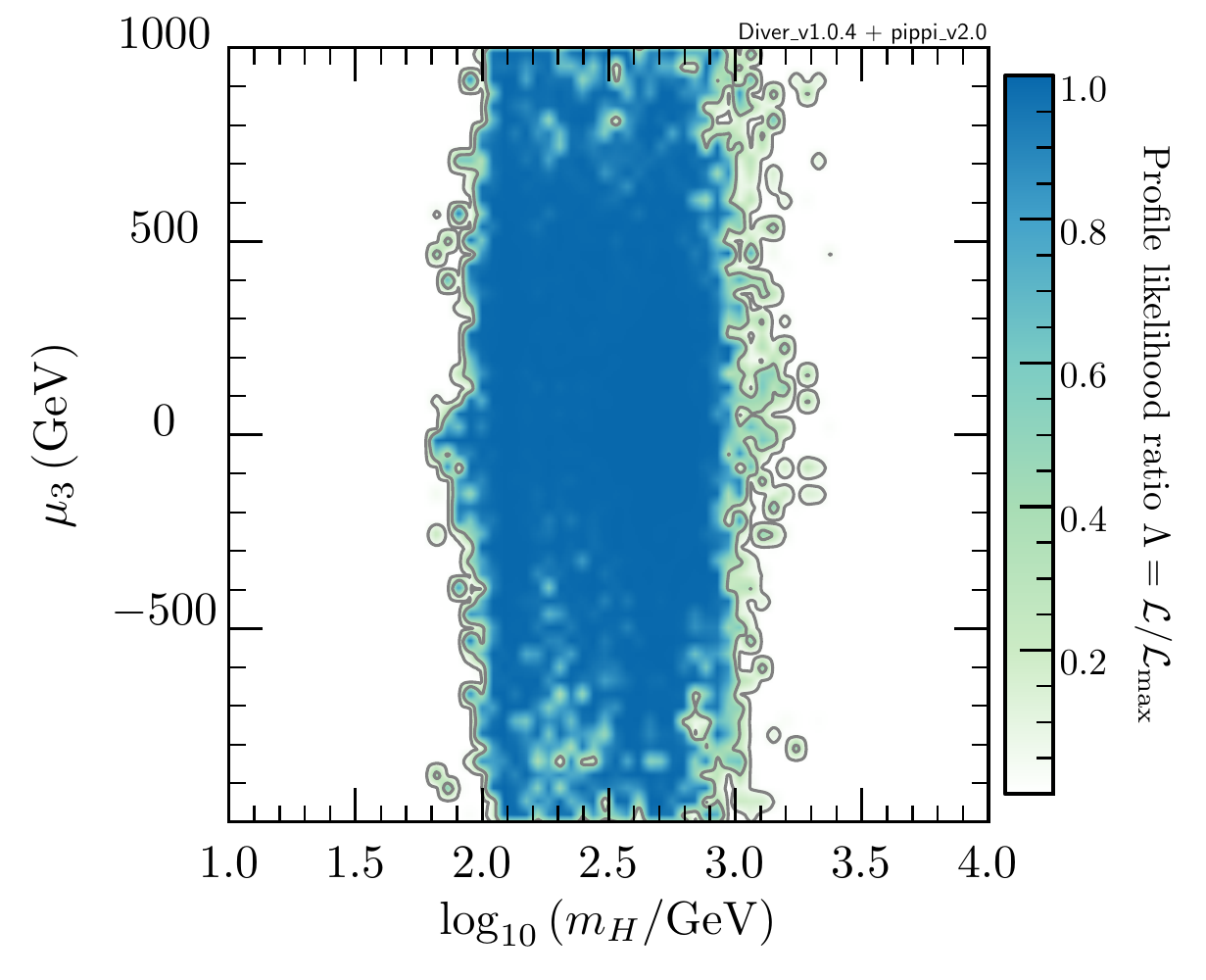}
	\includegraphics[scale=0.6]{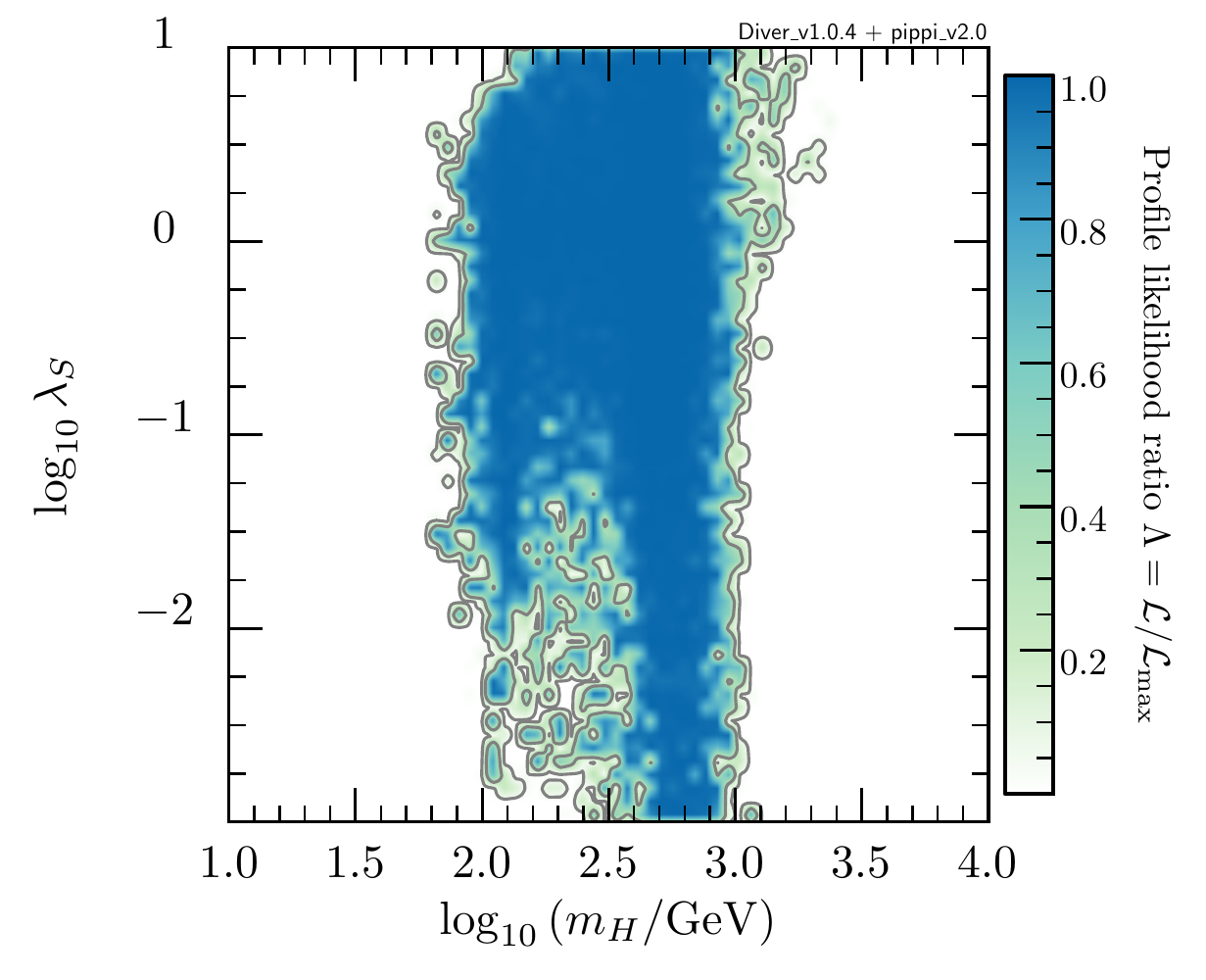}	         

    \includegraphics[scale=0.6]{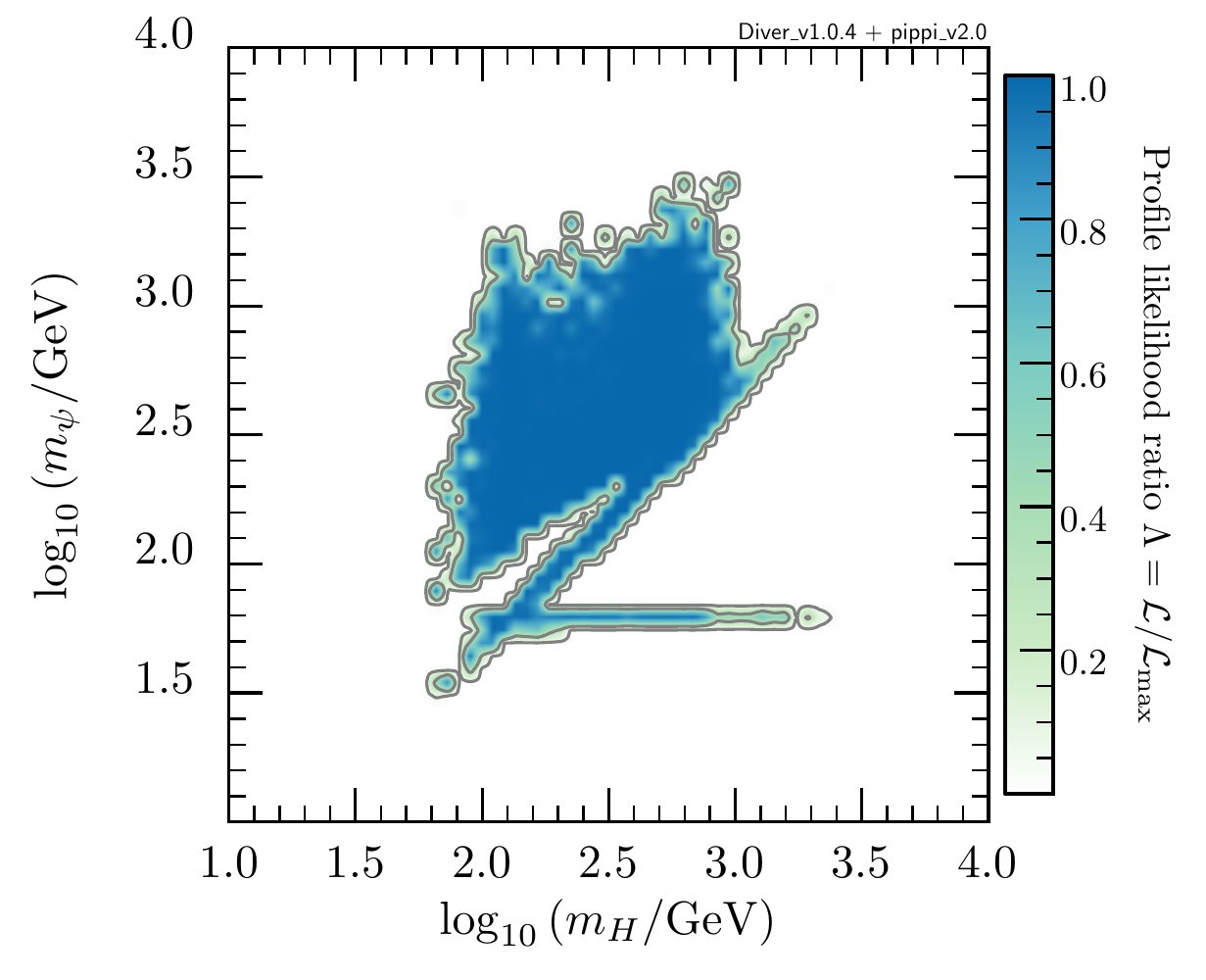}
    \includegraphics[scale=0.6]{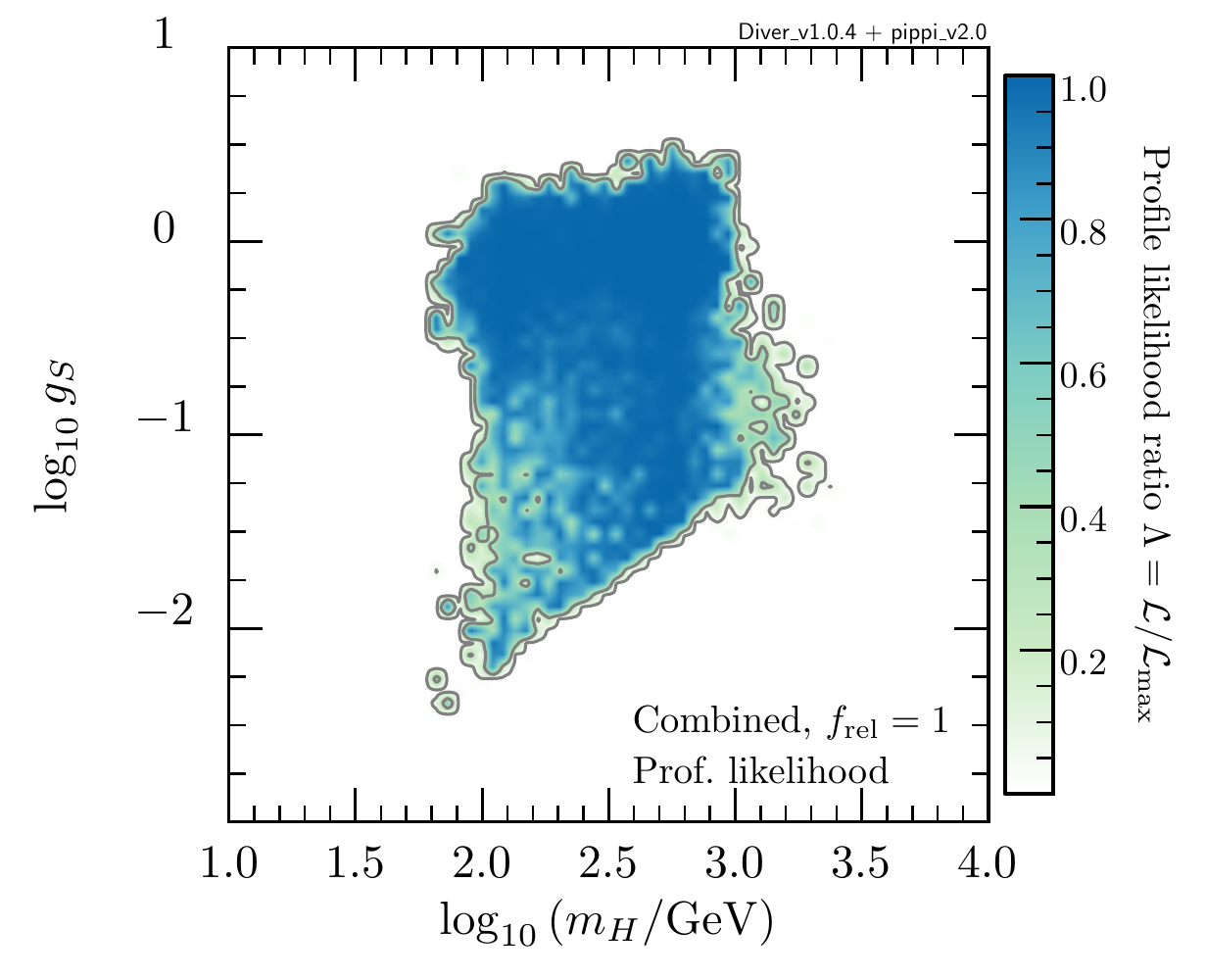}

    \caption{Same as Fig.~\ref{fig:2D_prof_like_all} except for the $f_{\textnormal{rel}} = 1$ case.}
    \label{fig:2D_prof_like_all_eq}
\end{figure}

%

The resulting 1D and 2D PL plots from our 7D scans are shown in Figs.~\ref{fig:1D_prof_like_all_eq} and \ref{fig:2D_prof_like_all_eq} respectively.~In comparison with Figs.~\ref{fig:1D_prof_like_all} and \ref{fig:2D_prof_like_all} respectively, the shape of the 1 and 2$\sigma$ C.L. contours is mostly similar; thus we refer to subsection~\ref{subsec:freleqone} to avoid repetition.~However, the allowed parameter space is significantly smaller. This is expected as the allowed region not only has to reproduce the observed DM abundance, it also has to yield a successful EWBG.

In general, we find that our model can easily satisfy all constraints provided $\alpha \simeq 0$, $\pi$.~This is expected as the new scalar $H$ is decoupled in this regime and provides no new contribution to the observed Higgs signal strength measurements. Thus, the allowed final states from the fermion DM annihilation are $hh$, $HH$ and $hH$, which gives $f_{\textnormal{rel}} = 1$.

An important point to note is that in the $\mathbb{Z}_2$ symmetric case \cite{Beniwal:2017eik}, the scalar Higgs portal coupling \emph{cannot} simultaneously explain the observed DM abundance and matter-antimatter asymmetry. This is expected as large values of the portal coupling are required to yield a successful EWBG, whereas small values are required to satisfy the direct detection limits. In contrast, our model contains additional couplings (e.g., $\mu_3$ and $\mu_{\pp S}$) between the new scalar $S$ and SM Higgs boson; these couplings can aid in generating a strong first-order EWPT. As $\mu_3$ and $\mu_{\pp S}$ does not significantly affect the phenomenology of the fermion DM (which is mostly determined by $g_S$ and $\alpha$), the fermion DM can easily saturate the observed DM abundance. These two features together allow the model to \emph{simultaneously} explain the observed DM abundance and matter-antimatter asymmetry.


\subsection{Gravitational wave signals}
The computation of gravitational wave (GW) signals requires a detailed study of the dynamics of the phase transition (PT). Luckily, the analysis of bubble nucleation is to some extent generic, and the steps required are always similar, albeit using a different scalar potential.~In our model, the main difficulty is that the transition always involves both scalar fields, and finding a correct tunnelling path in the 2D field space is always necessary.~We tackle this problem in the same way as in Ref.~\cite{Beniwal:2017eik}.~In particular, we use the method described there to find the appropriate tunnelling path and the bubble solutions which drive the transition in each case. The main drawback of this calculation is that it is computationally expensive when compared to all other constraints discussed in section~\ref{sec:const_like}. Thus, we first identify interesting points in the model parameter space from our global fit, and check the detailed PT dynamics and GW signals afterwards.

For each viable point from our global fit assuming $f_{\textnormal{rel}} \leq 1$, we compute the tunnelling path and corresponding action as in Ref.~\cite{Beniwal:2017eik}. Next, we compute the fraction of volume converted to the true vacuum \cite{Ellis:2018mja} to accurately calculate the bubble percolation temperature $T_p$.~This allows us to identify cases in which too strong supercooling renders percolation impossible; as temperature drops, the false vacuum energy dominates the expansion of the universe and an inflationary phase begins~\cite{Ellis:2018mja, Turner:1992tz, Guth:1982pn}. We find that a significant number of interesting points are excluded as the decay is too suppressed and the transition never successfully finishes. This happens because the extended parameter space with respect to the simple scalar potential in Ref.~\cite{Beniwal:2017eik} allows for a formation of a large tree-level barrier which can persist even at $T = 0$ and suppress the vacuum decay probability. In our 7D scans, we only use the approximation involving the critical temperature $T_c$ at which the symmetric and EWSB minima are degenerate. Such points are perfectly valid and predict $v_c/T_c>1$ as required for a successful EWBG. However, after a more detailed analysis, we find that roughly $50\%$ of all points remain viable and their GW spectra have large enough amplitudes to be shown in our plots. 

For the viable parameter points, we calculate the ratio of the released latent heat to the energy density of the plasma background, $\alpha_{\rm GW}$\footnote{Not to be confused with the mixing angle $\alpha$.}, and the size of bubbles carrying the most energy at percolation $R_{\rm MAX}$, which we then convert to the more familiar inverse time of the phase transition $\beta/H = (8\pi)^{1/3} v_w/(HR_{\rm MAX})$ \cite{Ellis:2018mja, Enqvist:1991xw}. These two parameters are essential for computing the GW spectra~\cite{Grojean:2006bp, Caprini:2015zlo}. We also assume that the speed of bubble walls is close to the speed of light $(v_w/c \approx 1)$ which is valid for the very strong first-order PTs that we are mostly interested in.

\begin{figure}[t]
    \centering
    \includegraphics[scale=0.9]{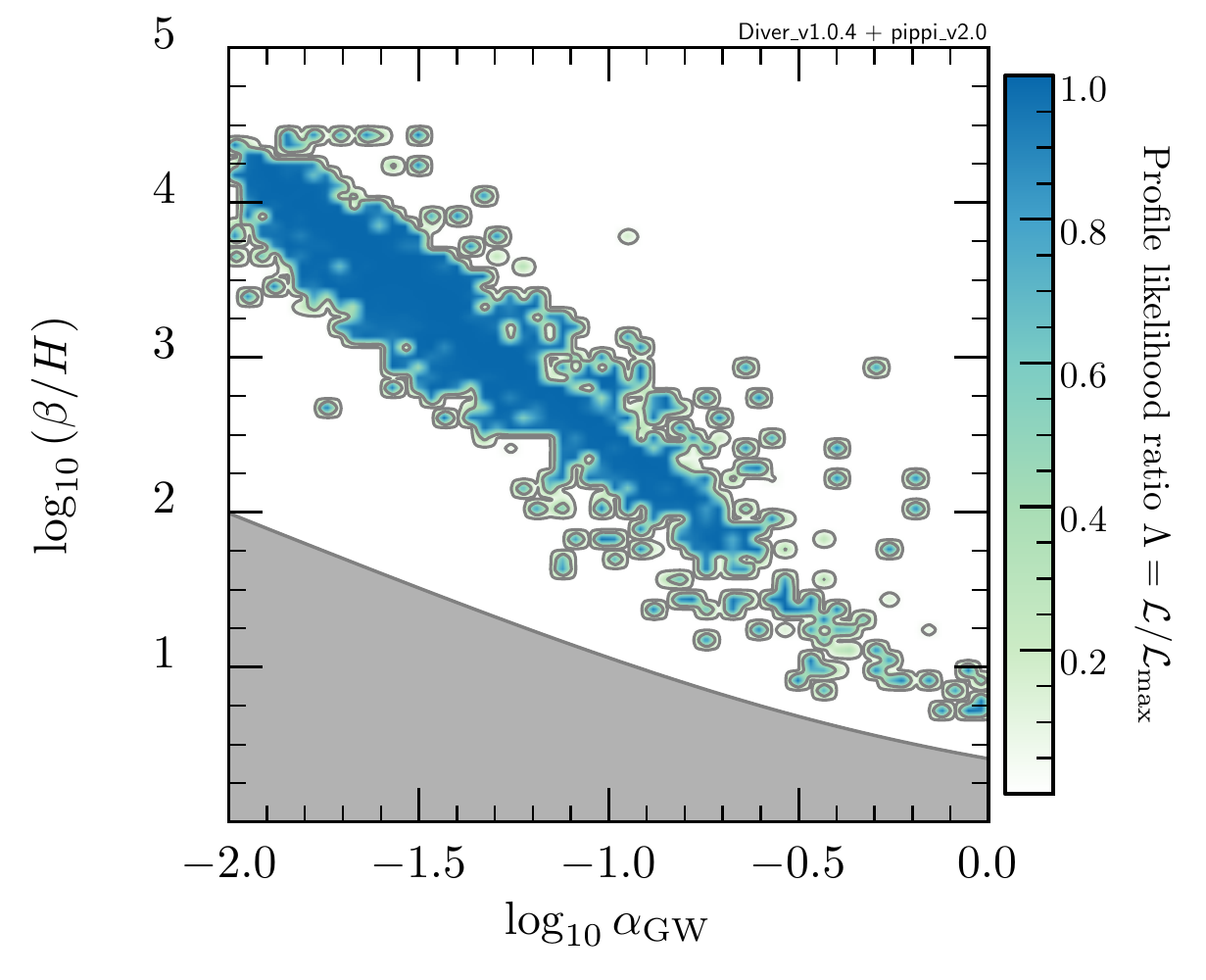}
    
    \caption{Viable points from our global fit assuming $f_{\textnormal{rel}} \leq 1$ in terms of the parameters $\alpha_{\rm GW}$ and $\beta/H$. The grey region shows where the sound waves last more than a Hubble time (assuming $v_w/c \approx 1$ which results in the largest allowed area) and reliably produce a GW signal.}
    \label{fig:AlphaBetaPlot}    
\end{figure}

Our calculation of the GW spectrum is based on Ref.~\cite{Ellis:2018mja}. In particular, we do not include the signal contribution from collisions of bubbles~\cite{Kamionkowski:1993fg, Huber:2008hg,Jinno:2016vai,Jinno:2017fby} as the bubbles reach equilibrium with the surrounding plasma and most of the energy is pumped into fluid shells around them~\cite{Bodeker:2017cim}.~The two remaining sources are sound waves in the plasma~\cite{Hindmarsh:2013xza, Hindmarsh:2015qta, Hindmarsh:2016lnk, Hindmarsh:2017gnf} and turbulence~\cite{Caprini:2009yp, Kosowsky:2001xp, Gogoberidze:2007an, Niksa:2018ofa} ensuing after the sound waves period.~We also check the condition for the sound waves to last more than one Hubble time which was assumed to hold while obtaining the GW spectra in references above.~We show this criterion in the $(\alpha_{\rm GW},\,\beta/H)$ plane (see Refs.~\cite{Ellis:2018mja,Hindmarsh:2017gnf} for more details) along with our results in Fig.~\ref{fig:AlphaBetaPlot}, assuming $v_w/c \approx 1$ which results in the largest allowed parameter space. We find that no parameter points are consistent with this criterion. This implies that the standard formula for sound wave spectra~\cite{Caprini:2015zlo} is probably overestimating the true signal. On the other hand, the turbulence signal will be stronger than the standard estimate as the turbulent motion begins more promptly after the PT.

\begin{figure}
    \centering
	{\Large Sound Waves} \\
	\includegraphics[scale=0.83]{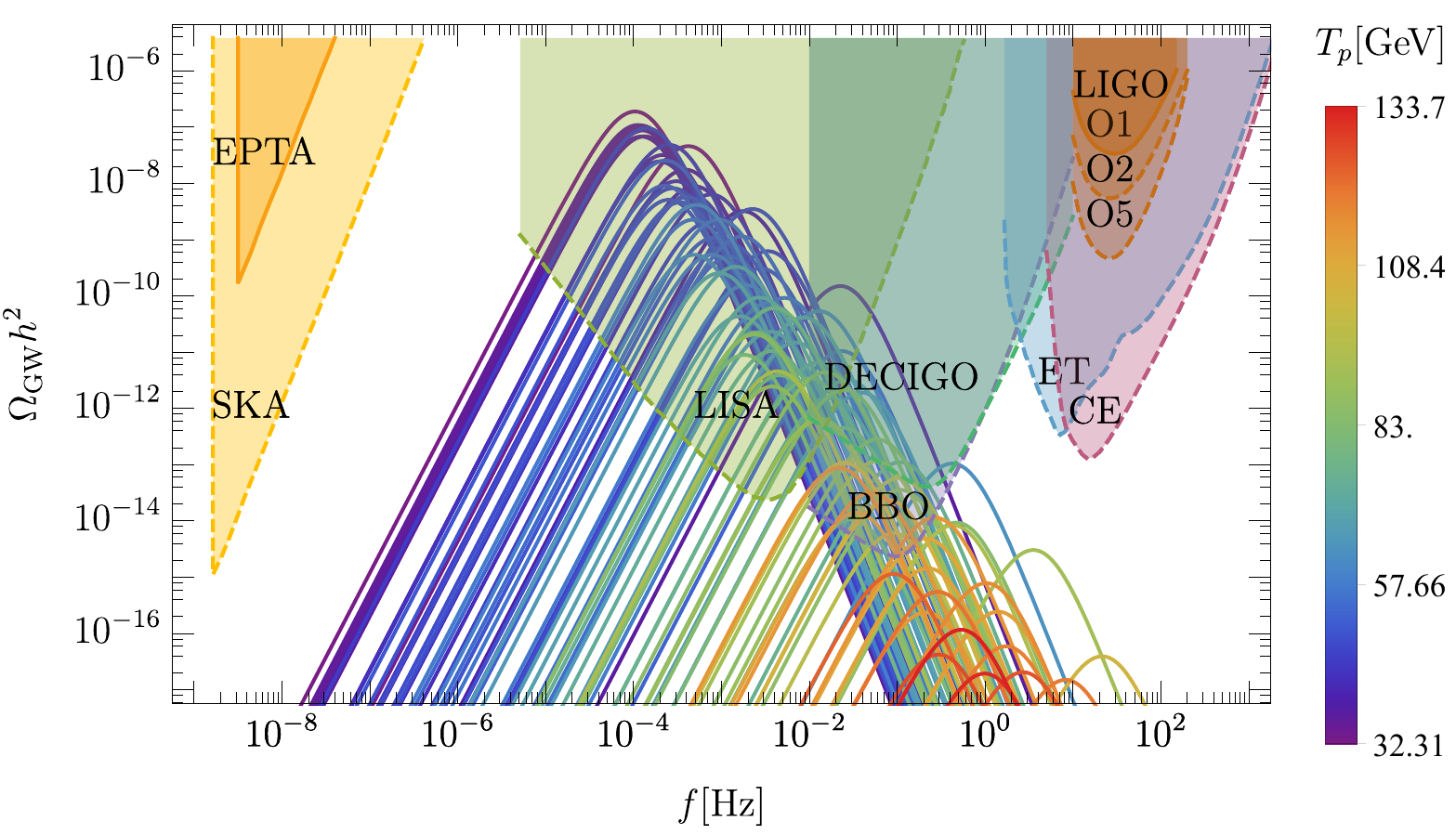} 
	\\[3mm]
	{\Large Turbulence} \\
  	\includegraphics[scale=0.83]{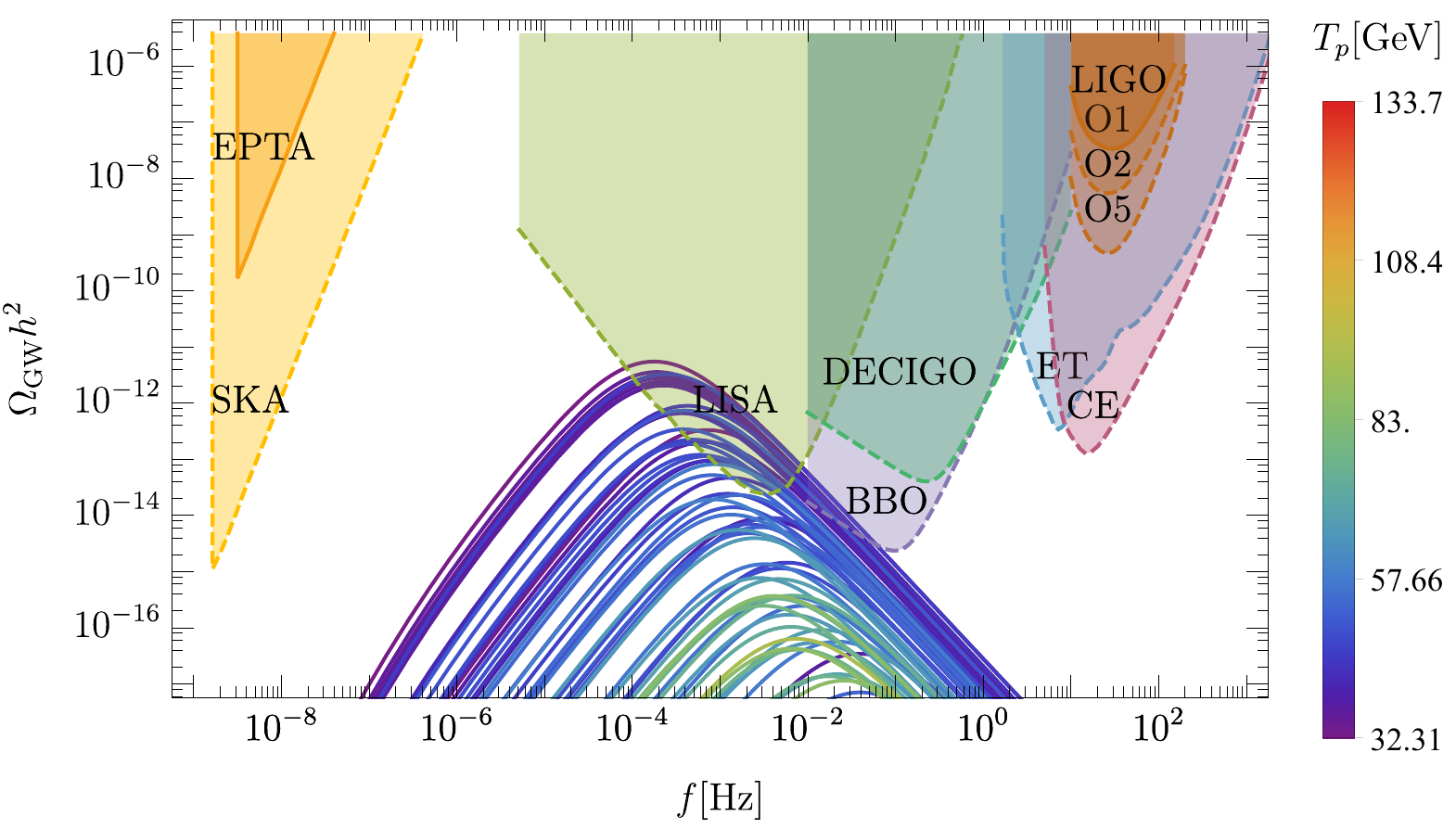}
  	
    \caption{Gravitational wave (GW) spectra of viable points as sourced by sound waves (top panel) and turbulence (bottom panel) along with their dependence on the percolation temperature $T_p$. Current sensitivity bands of LIGO and EPTA, as well as detection prospects of LISA, DECIGO, BBO and SKA are also shown for comparison (see text for more details).}
    \label{fig:GWresTp}
\end{figure}

In Fig.~\ref{fig:GWresTp}, we show the resulting GW spectra of viable points as sourced by sound waves (top panel) and turbulence (bottom panel), and their dependence on the percolation temperature $T_p$. As the condition to reliably generate a GW signal from sound waves is not fulfilled, a dedicated numerical simulation would be necessary to ascertain the spectral shape. We expect that the final results will lie somewhere between these two figures. In these figures, we discarded points with almost identical GW parameters to avoid plotting many overlapping results. Thus, we only show 100 representative lines out of 10,000 GW spectra as computed from our results.~We also show the detection prospects of Laser Interferometer Space Antenna (LISA) \cite{Bartolo:2016ami} (assuming the most optimistic A5M5 configuration), Deci-hertz Interferometer Gravitational-wave Observatory (DECIGO) and Big Bang Observer (BBO) \cite{Yagi:2011wg}. The current and future sensitivity bands of LIGO \cite{TheLIGOScientific:2014jea,TheLIGOScientific:2016wyq,Thrane:2013oya}, the European Pulsar Timing Array (EPTA) \cite{vanHaasteren:2011ni}, the Square Kilometre Array~(SKA) \cite{Janssen:2014dka}, Cosmic Explorer (CE)~\cite{Evans:2016mbw} and the Einstein Telescope (ET)~\cite{Punturo:2010zz,Hild:2010id} fall in a different frequency range than that of the viable points.~Thus, these experiments give no hope for detection of any of our results.

In summary, we find that the GW spectra of viable points that are interesting from the point of view of baryogenesis can lie within reach of future GW experiments such as LISA, DECIGO and BBO.
However, the uncertainty of the sound wave spectrum can have a dramatic impact on the results.~In the overly optimistic case of the standard sound wave signal, roughly $15 \%$ of all of our viable points would be detectable by LISA while for the most pessimistic turbulence-only spectrum, this number falls below half a percent.
We also confirm that pulsar timing arrays and terrestrial experiments (e.g., LIGO) are not sensitive to frequencies that result in a GW signal from a strong EWPT. 
Notably, our results are qualitatively very similar to the $\mathbb{Z}_2$ symmetric ones in Ref.~\cite{Ellis:2018mja}, despite our general non-$\mathbb{Z}_2$ symmetric potential. This leads us to believe that our current knowledge of the Higgs boson properties (most notably, a constraint on the mixing angle $\alpha$) is enough to significantly constrain viable potentials, and bring them closer to the $\mathbb{Z}_2$ symmetric case. 

\section{Conclusions}\label{sec:conclusions}
In this paper, we have performed the most comprehensive and up-to-date study of the extended scalar singlet model with a fermionic DM candidate.~After performing a 7D scan of the model using \emph{only} the EWBG constraint, we found regions in the model parameter space that can facilitate a successful EWBG. From our 1D PL plots, we showed that a successful EWBG can be viable in all parts of the model parameter space provided $g_S \lesssim 5.62$.

After building intuition from the EWBG only results, we performed a global fit of our model using the constraints from the \emph{Planck} measured DM relic density, direct detection limits from the XENON1T (2018) experiment, electroweak precision observables (EWPO) and Higgs searches at colliders.~This allowed us to constrain parts of the 7D model parameter space.~In particular, our global fit placed an upper \emph{and} lower limit on $m_H$, $m_\psi$ and $g_S$, namely $m_h/2 \lesssim m_H \lesssim 5$\,TeV, $32\,\mathrm{GeV} \lesssim m_\psi \lesssim 3.2$\,TeV and $5.6 \times 10^{-3} \lesssim g_S \lesssim 3.5$. Moreover, we confirmed that our model can \emph{simultaneously} yield a strong first-order phase transition and saturate the observed DM abundance. This is an important feature which is missing in the $\mathbb{Z}_2$ symmetric case. 

From the viable points that satisfied all of the available constraints, we computed the GW spectra, and checked the discovery prospects of the model at current and future GW experiments. In doing so, we found that the GW spectra of viable points can be within reach of future GW experiments such as LISA, DECIGO and BBO. We checked that the condition for sound waves to be a long-lasting source of GWs is not satisfied for any of our results.~This implies that the standard sound wave spectrum used in the literature likely overestimates the true signal, whereas the turbulence signal can be stronger than the standard prediction as the turbulence sets in quicker after the end of the phase transition. Unfortunately, the overall result will still likely be a reduction of the overall spectrum, thereby reducing the discovery prospects. Specifically, in our results we find that $15 \%$ of our viable points would be within the reach of LISA if the final spectrum was close to the standard sound wave prediction. However, this number falls down to less than half a percent in the most pessimistic case of only a turbulence-sourced GW signal. 


\acknowledgments{We thank Philip Bechtle, Malcolm Fairbairn, Jos\'{e} M.~No, Chris Rogan and Pat Scott for helpful discussions.~AB thanks Sean Crosby from the CoEPP Research Computing for providing computing assistance and allocating resources.~This work was supported by the Swedish Research Council (contract 621-2014-5772), ARC Centre of Excellence for Particle Physics at the Terascale (CoEPP) (CE110001104) and the Centre for the Subatomic Structure of Matter (CSSM). This work also made use of the supercomputing resources provided by the Phoenix HPC service at the University of Adelaide.~ML was supported in part by the Polish MNiSW grant IP2015 043174 and STFC grant number ST/L000326/1.~AB was supported by the Australian Postgraduate Award (APA). MW is supported by the Australian Research Council Future Fellowship FT140100244.~Feynman diagrams were drawn using the \texttt{TikZ-Feynman\_v1.1.0} package \cite{Ellis:2016jkw}.}

\appendix

\section{Tree-level scalar potential}\label{app:tree-potential}
The tree-level scalar potential in Eq.~\eqref{eqn:pot-part} expands to 
\begin{equation}\label{eqn:pot-expand}
    V_{\textnormal{tree}} = - \mu_\pp^2 \PP \pp + \lambda_\pp (\PP \pp)^2 - \frac{1}{2} \mu_S^2 S^2 - \frac{1}{3} \mu_3 S^3 + \frac{1}{4} \lambda_S S^4 + \mu_{\pp S} \PP \pp S + \frac{1}{2} \lambda_{\pp S} \PP \pp S^2. 
\end{equation}
With the following definitions
\begin{equation*}
    \pp = 
    \begin{pmatrix}
        G^+ \\
        \frac{1}{\sqrt{2}} (\phi + i G^0)
    \end{pmatrix}, \quad 
    \PP =
    \begin{pmatrix}
        G^-, & \frac{1}{\sqrt{2}} (\phi - i G^0)
    \end{pmatrix},
\end{equation*}
where $G^- \equiv (G^+)^*$, the potential in Eq.~\eqref{eqn:pot-expand} depends on 2 complex ($G^+$, $G^-$) and 3 real ($G^0$, $\phi$, $S$) scalar fields. 

After EWSB, the $\phi$ and $S$ fields acquire their VEVs in Eq.~\eqref{eqn:vevs}. Thus, the following partial derivatives
\begin{equation*}
    \frac{\partial V_{\textnormal{tree}}}{\partial G^0}, \, \frac{\partial V_{\textnormal{tree}}}{\partial G^-}, \, \frac{\partial V_{\textnormal{tree}}}{\partial G^+}, \, \frac{\partial V_{\textnormal{tree}}}{\partial \phi}, \, \frac{\partial V_{\textnormal{tree}}}{\partial S},
\end{equation*}
must vanish at the EWSB minimum $(\left.\langle \phi \rangle\right|_{T=0}, \left.\langle S \rangle\right|_{T=0}) = (v_0, s_0)$. This gives
\begin{align*}
    0 &= \left.\frac{\partial V_{\textnormal{tree}}}{\partial G^0}\right|_{(v_0, \, s_0)} = \left.\frac{\partial V_{\textnormal{tree}}}{\partial G^-}\right|_{(v_0, \, s_0)} = \left.\frac{\partial V_{\textnormal{tree}}}{\partial G^+}\right|_{(v_0, \, s_0)}, \\[1.5mm]
    0 &= \left. \frac{\partial V_{\textnormal{tree}}}{\partial \phi}\right|_{(v_0,\,s_0)} = - \mu_\pp^2 v_0 + \lambda_\pp v_0^3 + \mu_{\pp S} s_0 v_0 + \frac{1}{2} \lambda_{\pp S} v_0 s_0^2, \\[1.5mm]
    0 &= \left.\frac{\partial V_{\textnormal{tree}}}{\partial S}\right|_{(v_0,\,s_0)} = - \mu_S^2 s_0 - \mu_3 s_0^2 + \lambda_S s_0^3 + \frac{1}{2} \mu_{\pp S} v_0^2 + \frac{1}{2} \lambda_{\pp S} s_0 v_0^2. 
\end{align*}
A simple rearrangement gives us the following EWSB conditions
\begin{align}
    \mu_\pp^2 &= \lambda_\pp v_0^2 + \mu_{\pp S} s_0 + \frac{1}{2} \lambda_{\pp S} s_0^2, \label{eqn:mu_pp} \\
    \mu_{S}^2 &= - \mu_3 s_0 + \lambda_S s_0^2 + \frac{\mu_{\pp S} v_0^2}{2 s_0} + \frac{1}{2} \lambda_{\pp S} v_0^2. \label{eqn:mu_s} 
\end{align}
Now, we compute the second-order partial derivatives at the EWSB minimum.~The only non-zero ones are given by
\begin{align*}
    \left. \frac{\partial^2 V_{\textnormal{tree}}}{\partial G^0 \,\partial G^0}\right|_{(v_0, \, s_0)} = \left. \frac{\partial^2 V_{\textnormal{tree}}}{\partial G^- \,\partial G^+}\right|_{(v_0, \, s_0)} &= \left. \frac{\partial^2 V_{\textnormal{tree}}}{\partial G^+ \,\partial G^-}\right|_{(v_0, \, s_0)} = -\mu_\pp^2 + \lambda_\pp v_0^2 + \mu_{\pp S} s_0 + \frac{1}{2} \lambda_{\pp S} s_0^2, \\[1.5mm]
    \left. \frac{\partial^2 V_{\textnormal{tree}}}{\partial \phi^2} \right|_{(v_0, \, s_0)}  &= -\mu_\pp^2 + 3 \lambda_\pp v_0^2 + \mu_{\pp S} s_0 + \frac{1}{2} \lambda_{\pp S} s_0^2, \\[1.5mm]
    \left. \frac{\partial^2 V_{\textnormal{tree}}}{\partial S^2} \right|_{(v_0, \, s_0)} &= - \mu_S^2 - 2 \mu_3 s_0 + 3 \lambda_S s_0^2 + \frac{1}{2} \lambda_{\pp S} v_0^2, \\[1.5mm]
    \left. \frac{\partial^2 V_{\textnormal{tree}}}{\partial \phi \,\partial S} \right|_{(v_0, \, s_0)} = \left. \frac{\partial^2 V_{\textnormal{tree}}}{\partial S \,\partial \phi} \right|_{(v_0, \, s_0)} &= \mu_{\pp S} v_0 + \lambda_{\pp S} v_0 s_0. 
\end{align*}
Using Eqs.~\eqref{eqn:mu_pp} and \eqref{eqn:mu_s}, these expressions can be simplified to
\begin{align}
    \left. \frac{\partial^2 V_{\textnormal{tree}}}{\partial G^0 \,\partial G^0}\right|_{(v_0, \, s_0)} = \left. \frac{\partial^2 V_{\textnormal{tree}}}{\partial G^- \,\partial G^+}\right|_{(v_0, \, s_0)} &= \left. \frac{\partial^2 V_{\textnormal{tree}}}{\partial G^+ \,\partial G^-}\right|_{(v_0, \, s_0)} = 0,  \\[1.5mm]
    \left. \frac{\partial^2 V_{\textnormal{tree}}}{\partial \phi^2} \right|_{(v_0, \, s_0)}  &= 2 \lambda_\pp v_0^2, \label{eqn:second-der-1} \\[1.5mm]
    \left. \frac{\partial^2 V_{\textnormal{tree}}}{\partial S^2} \right|_{(v_0, \, s_0)} &= - \mu_3 s_0 + 2 \lambda_S s_0^2  - \frac{\mu_{\pp S} v_0^2}{2 s_0}, \\[1.5mm] 
    \left. \frac{\partial^2 V_{\textnormal{tree}}}{\partial \phi \,\partial S} \right|_{(v_0, \, s_0)} = \left. \frac{\partial^2 V_{\textnormal{tree}}}{\partial S \,\partial \phi} \right|_{(v_0, \, s_0)} &= \mu_{\pp S} v_0 + \lambda_{\pp S} v_0 s_0. \label{eqn:second-der-4}
\end{align}
After EWSB, the $\phi$ and $S$ fields can be expanded as
\begin{equation}\label{eqn:redef}
    \phi = v_0 + \varphi, \quad S = s_0 + s.
\end{equation}
As $\partial V_{\textnormal{tree}}/\partial \phi = \partial V_{\textnormal{tree}}/\partial \varphi$ and $\partial V_{\textnormal{tree}}/\partial S = \partial V_{\textnormal{tree}}/\partial s$, a mass-term for the real scalar fields $\mathcal{A}^T = (\varphi, s)$ is 
\begin{equation}\label{eqn:phismassmatrix}
    \lagr_{\textnormal{mass--term}} = -\frac{1}{2} \mathcal{A}^T \calM^2 \mathcal{A},
\end{equation}
where 
\begin{equation}\label{eqn:mass-mat}
    \calM^2 =    
    \begin{pmatrix}
        \calM_{\varphi \varphi}^2 & \calM_{\varphi s}^2 \\[1.5mm]
        \calM_{s \varphi}^2 & \calM_{s s}^2
    \end{pmatrix}
    \equiv 
    \begin{pmatrix}
        \left.\frac{\partial^2 V_{\textnormal{tree}}}{\partial \varphi^2}\right|_{(v_0, \, s_0)} & \left.\frac{\partial^2 V_{\textnormal{tree}}}{\partial \varphi \,\partial s}\right|_{(v_0, \, s_0)} \\[5mm]
        \left.\frac{\partial^2 V_{\textnormal{tree}}}{\partial s \,\partial \varphi}\right|_{(v_0, \, s_0)} & \left.\frac{\partial^2 V_{\textnormal{tree}}}{\partial s^2}\right|_{(v_0, \, s_0)} 
    \end{pmatrix}
\end{equation}
is the squared mass matrix. Using Eqs.~\eqref{eqn:second-der-1}--\eqref{eqn:second-der-4}, the matrix elements are given by
\begin{equation}\label{eqn:mass-matrix}
    \calM_{\varphi \varphi}^2 = 2 \lambda_\pp v_0^2, \quad \calM_{ss}^2 = - \mu_3 s_0 + 2 \lambda_S s_0^2 - \frac{\mu_{\pp S} v_0^2}{2 s_0}, \quad \calM_{\varphi s}^2 = \calM_{s \varphi}^2 = \mu_{\pp S} v_0 + \lambda_{\pp S} v_0 s_0. 
\end{equation}

For the EWSB minimum to be a stable (i.e., not a saddle point) solution of Eq.~\eqref{eqn:pot-expand}, the symmetric $5 \times 5$ Hessian matrix $\mathcal{H}$ must be positive-definite. At the EWSB minimum, it is given by
\begin{equation*}
    \left.\mathcal{H}\right|_{(v_0, \, s_0)} = 
    \bordermatrix{
    & G^0 & G^- & G^+ & \phi & S \cr 
    G^0 & 0 & 0 & 0 & 0 & 0 & \cr
    G^- & 0 & 0 & 0 & 0 & 0 & \cr
    G^+ & 0 & 0 & 0 & 0 & 0 & \cr
    \phi & 0 & 0 & 0 & 2 \lambda_\pp v_0^2 & \mu_{\pp S} v_0 + \lambda_{\pp S} v_0 s_0 \cr    
    S & 0 & 0 & 0 & \mu_{\pp S} v_0 + \lambda_{\pp S} v_0 s_0 & - \mu_3 s_0 + 2 \lambda_S s_0^2  - \frac{\mu_{\pp S} v_0^2}{2 s_0} \cr    
    }.
\end{equation*}
The above matrix is guaranteed to be positive-definite if the determinant (eigenvalue) of the $2 \times 2$ sub-matrix is non-zero (positive). These two requirements give
\begin{equation}
    \lambda_\pp > 0, \quad 2 \lambda_S s_0^2 - \left(\mu_3 s_0  + \frac{\mu_{\pp S} v_0^2}{2 s_0} \right) > 0. 
\end{equation}

To study the bounds of the tree-level scalar potential, Eq.~\eqref{eqn:pot-expand} can be expressed in terms of the $\phi$ and $S$ fields as
\begin{equation}\label{eqn:bound-pot}
    V_{\textnormal{tree}} \simeq - \frac{1}{2} \mu_\pp^2 \phi^2 + \frac{1}{4} \lambda_\pp \phi^4 - \frac{1}{2} \mu_S^2 S^2 - \frac{1}{3} \mu_3 S^3 + \frac{1}{4} \lambda_S S^4 + \frac{1}{2} \mu_{\pp S} \phi^2 S + \frac{1}{4} \lambda_{\pp S} \phi^2 S^2. 
\end{equation}
Depending on the chosen direction in the $(\phi, S)$ plane, three scenarios are possible.
\begin{enumerate}
    \item \emph{Pure $\phi$ direction}:~In this case, the potential depends only on the $\phi$ field. It is bounded from below provided $\lambda_\pp > 0$.
    \item \emph{Pure $S$ direction}:~In this case, the potential depends only on the $S$ field. It is bounded from below provided $\lambda_S > 0$.
    \item \emph{General $\phi$ and $S$ directions}:~At large $\phi$ and $S$ field values, the quartic terms in Eq.~\eqref{eqn:bound-pot} dominate. In this case, the potential can be approximated by
    \begin{align*}
        V_{\textnormal{tree}} &\approx \frac{1}{4} \lambda_\pp \phi^4 + \frac{1}{4} \lambda_S S^4 + \frac{1}{4} \lambda_{\pp S} \phi^2 S^2 \\
        &= \frac{1}{4} \lambda_\pp \phi^4 + \frac{1}{4} \lambda_S \left(S^4 + \frac{\lambda_{\pp S}}{\lambda_S} \phi^2 S^2 \right) \\
        &= \frac{1}{4} \lambda_\pp \phi^4 + \frac{1}{4} \lambda_S \left(S^4 + \frac{\lambda_{\pp S}}{\lambda_S} \phi^2 S^2 + \frac{1}{4} \frac{\lambda_{\pp S}^2}{\lambda_S^2} \phi^4 \right) - \frac{1}{16} \frac{\lambda_{\pp S}^2}{\lambda_S} \phi^4 \\
        &= \frac{1}{4} \left(\lambda_\pp - \frac{1}{4} \frac{\lambda_{\pp S}^2}{\lambda_S} \right) \phi^4 + \frac{1}{4} \lambda_S \left(S^2 + \frac{1}{2} \frac{\lambda_{\pp S}}{\lambda_S} \phi^2 \right)^2.
    \end{align*}
    Thus, the potential is bounded from below provided $\lambda_S > 0$ \emph{and} $\lambda_{\pp S} > - 2 \sqrt{\lambda_{\pp} \lambda_S}$.
\end{enumerate}

\section{Mass eigenstate basis}\label{app:mass-eg}
The squared mass matrix $\calM^2$ is real and symmetric. It can be diagonalised by an orthogonal matrix $\mathcal{O}$. Thus, we define the mass eigenstates $(h, H)$ as
\begin{equation}\label{eqn:mixmat}
    \begin{pmatrix}
        h \\
        H
    \end{pmatrix} 
    = 
    \begin{pmatrix}
        \cos\alpha & -\sin\alpha \\
        \sin \alpha & \cos\alpha
    \end{pmatrix}
    \begin{pmatrix}
        \varphi \\
        s
    \end{pmatrix}.
\end{equation}
The interaction eigenstates $(\varphi, s)$ are given by
\begin{align*}
    \begin{pmatrix}
        \varphi \\
        s
    \end{pmatrix}
     = \mathcal{O}
     \begin{pmatrix}
         h \\
         H
     \end{pmatrix}, \quad \mathcal{O} = \begin{pmatrix}
         \cos\alpha & \sin\alpha \\
         -\sin\alpha & \cos\alpha
     \end{pmatrix}.
\end{align*}
Now, we consider the following matrix product
\begin{equation}\label{eqn:OtMO}
    \begin{pmatrix}
        \varphi & s 
    \end{pmatrix}
    \calM^2
    \begin{pmatrix}
        \varphi \\
        s
    \end{pmatrix} 
    =  
    \begin{pmatrix}
        h & H
    \end{pmatrix} \mathcal{O}^T \calM^2 \mathcal{O} 
    \begin{pmatrix}
        h \\
        H
    \end{pmatrix} \equiv
    \begin{pmatrix}
        h & H
    \end{pmatrix} \mathcal{D}
    \begin{pmatrix}
        h \\
        H
    \end{pmatrix}, 
\end{equation}
where $\mathcal{D} = \textnormal{diag}(m_{h}^2, m_H^2)$ is a diagonal squared mass matrix for the physical mass eigenstates.~Thus, the last equality in Eq.~\eqref{eqn:OtMO} requires
\begin{equation}\label{eqn:OTMO}
    \mathcal{O}^T \calM^2 \mathcal{O} = \mathcal{D}.
\end{equation}
The left-hand side of the above expression expands to
\begin{align*}
    \mathcal{O}^T \calM^2 \mathcal{O} &= 
    \begin{pmatrix}
        \cos\alpha & -\sin\alpha \\
        \sin\alpha & \cos\alpha
    \end{pmatrix}
    \begin{pmatrix}
        \calM_{\varphi \varphi}^2 & \calM_{\varphi s}^2 \\[2mm]
        \calM_{s \varphi}^2 & \calM_{ss}^2
    \end{pmatrix}
    \begin{pmatrix}
        \cos\alpha & \sin\alpha \\
        -\sin\alpha & \cos\alpha
    \end{pmatrix} \\
    &= 
    \begin{pmatrix}
        \cos\alpha & -\sin\alpha \\[1.5mm]
        \sin\alpha & \cos\alpha
    \end{pmatrix}
    \begin{pmatrix}
        \calM_{\varphi \varphi}^2 \cos\alpha - \calM_{\varphi s}^2 \sin\alpha &\quad \calM_{\varphi \varphi}^2 \sin\alpha + \calM_{\varphi s}^2 \cos\alpha \\[2mm]
        \calM_{s \varphi}^2 \cos\alpha - \calM_{ss}^2 \sin\alpha &\quad \calM_{s \varphi}^2 \sin\alpha + \calM_{ss}^2 \cos\alpha
    \end{pmatrix}.
\end{align*}
As $\calM_{\varphi s}^2 = \calM_{s \varphi}^2$, the elements of the $\mathcal{O}^T \calM^2 \mathcal{O}$ matrix are
\begin{align*}
    \left[\mathcal{O}^T \calM^2 \mathcal{O}\right]_{11} &= \calM_{\varphi \varphi}^2 \cos^2\alpha +  \calM_{ss}^2 \sin^2\alpha - \calM_{\varphi s}^2 \sin 2\alpha, \\[1.5mm]
    \left[\mathcal{O}^T \calM^2 \mathcal{O}\right]_{22} &= \calM_{\varphi \varphi}^2 \sin^2\alpha + \calM_{ss}^2 \cos^2\alpha + \calM_{\varphi s}^2 \sin 2\alpha, \\[1.5mm]
    \left[\mathcal{O}^T \calM^2 \mathcal{O}\right]_{12} = \left[\mathcal{O}^T \calM^2 \mathcal{O}\right]_{21} &= -\frac{1}{2} (\calM_{ss}^2 - \calM_{\varphi \varphi}^2) \sin 2\alpha + \calM_{\varphi s}^2 \cos 2\alpha.
\end{align*}
By equating these expressions to the elements of the diagonal matrix $\mathcal{D}$, we get
\begin{align}
    m_{h}^2 &= \calM_{\varphi \varphi}^2 \cos^2\alpha + \calM_{ss}^2 \sin^2\alpha - \calM_{\varphi s}^2 \sin 2\alpha, \label{eqn:phys-mass-1} \\[1.5mm]
    m_H^2 &= \calM_{\varphi \varphi}^2 \sin^2\alpha + \calM_{ss}^2 \cos^2\alpha + \calM_{\varphi s}^2 \sin 2\alpha, \\[1.5mm]
    0 &= -\frac{1}{2} (\calM_{ss}^2 - \calM_{\varphi \varphi}^2) \sin 2\alpha + \calM_{\varphi s}^2 \cos 2\alpha. \label{eqn:phys-mass-3}
\end{align}
The last equality can be conveniently expressed as
\begin{equation}
    \tan 2\alpha = \frac{2 \calM_{\varphi s}^2}{\calM_{ss}^2 - \calM_{\varphi \varphi}^2}.
\end{equation}

We can rewrite Eqs.~\eqref{eqn:phys-mass-1}--\eqref{eqn:phys-mass-3} as the following matrix product
\begin{equation*}
    \begin{pmatrix}
        m_h^2 \\[1.5mm]
        m_H^2 \\[1.5mm]
        0 
    \end{pmatrix}
    = 
    \begin{pmatrix}
        \cos^2 \alpha & \sin^2 \alpha & -2 \sin\alpha \cos\alpha \\[1.5mm]
        \sin^2 \alpha & \cos^2 \alpha & 2 \sin\alpha \cos\alpha \\[1.5mm]
        \sin\alpha \cos\alpha &\quad -\sin\alpha \cos\alpha &\quad \cos^2 \alpha - \sin^2 \alpha
    \end{pmatrix}
    \begin{pmatrix}
        \calM_{\varphi \varphi}^2 \\[1.5mm]
        \calM_{s s}^2 \\[1.5mm]
        \calM_{\varphi s}^2 
    \end{pmatrix}.    
\end{equation*}
By computing the inverse of the above $3 \times 3$ matrix (i.e., by taking $\alpha \rightarrow -\alpha$), we get
\begin{equation*}
    \begin{pmatrix}
        \calM_{\varphi \varphi}^2 \\[1.5mm]
        \calM_{s s}^2 \\[1.5mm]
        \calM_{\varphi s}^2 
    \end{pmatrix} 
    =
    \begin{pmatrix}
        \cos^2 \alpha & \sin^2\alpha & 2 \sin\alpha \cos\alpha \\[1.5mm]
        \sin^2\alpha & \cos^2 \alpha & - 2 \sin\alpha \cos\alpha \\[1.5mm]
        -\sin\alpha \cos\alpha &\quad \sin\alpha \cos\alpha &\quad \cos^2\alpha - \sin^2\alpha
    \end{pmatrix}
    \begin{pmatrix}
        m_h^2 \\[1.5mm]
        m_H^2 \\[1.5mm]
        0 
    \end{pmatrix}.
\end{equation*}
From the above matrix product and Eq.~\eqref{eqn:mass-matrix}, we get
\begin{align*}
     \lambda_\pp &= \frac{\calM_{\varphi \varphi}^2}{2 v_0^2} = \frac{1}{2 v_0^2} \left( m_h^2 \cos^2\alpha + m_H^2 \sin^2 \alpha \right), \\
     \mu_{\pp S} &= -\frac{2 s_0}{v_0^2} \left(\calM_{ss}^2 + \mu_3 s_0 - 2 \lambda_S s_0^2 \right) = -\frac{2 s_0}{v_0^2} \left(m_h^2 \sin^2 \alpha + m_H^2 \cos^2\alpha + \mu_3 s_0 - 2 \lambda_S s_0^2 \right), \\ 
     \lambda_{\pp S} &= \frac{1}{v_0 s_0} \left(\calM_{\varphi s}^2 - \mu_{\Phi S} v_0 \right)  = \frac{1}{v_0 s_0} \Big[(m_H^2 - m_h^2) \sin\alpha \cos\alpha - \mu_{\pp S} v_0 \Big].
\end{align*}

\section{Dark matter-nucleon coupling}\label{app:DM-nucleon}
The interaction eigenstates $(\varphi,s)$ can be written in terms of the mass eigenstates $(h, H)$ as 
\begin{equation*}
    \begin{pmatrix}
        \varphi \\
        s
    \end{pmatrix} 
    = 
    \begin{pmatrix}
        \cos\alpha & \sin\alpha \\
        -\sin \alpha & \cos\alpha
    \end{pmatrix}
    \begin{pmatrix}
        h \\
        H
    \end{pmatrix}.
\end{equation*}
Thus, the scalar-fermion DM and quark Yukawa term in the SM Lagrangian expands to
\begin{align*}
    \lagr_{\textnormal{DM--quark}} &= - g_S \ovr{\psi} \psi s - \sum_q \frac{m_q}{v_0}\,\varphi \ovr{q} q \\
    &= - g_S \ovr{\psi}\psi (- \sa h + \ca H) - \sum_q \frac{m_q}{v_0} (\ca h + \sa H) \ovr{q} q \\
    &= g_S \sa \ovr{\psi} \psi h - \frac{\ca}{v_0} \sum_q m_q \, h \ovr{q} q - g_S \ca \ovr{\psi} \psi H - \frac{\sa}{v_0} \sum_q m_q\, H \ovr{q} q.
\end{align*}

In a typical direct detection experiment, the momentum transfer $q$ is roughly on the order of a few MeV. Assuming that the mediator masses $m_{h/H}$ are well above this value,\footnote{This is clearly the case for a SM-like Higgs boson with mass $m_h = 125.13$\,GeV.}~i.e., $m_{h/H}^2 \gg q^2$, we can safely approach direct detection in the context of an effective field theory (EFT) and \emph{integrate out} the scalar mediators \cite{Berlin:2014tja}.~Thus, we can write down an effective DM-quark interaction Lagrangian as
\begin{equation}\label{eqn:DM-quark-lagr}
    \lagr_{\textnormal{DM--quark}}^{\textnormal{eff}} = - \sum_q G_q \,\ovr{\psi} \psi \, \ovr{q} q,
\end{equation}
where 
\begin{equation}\label{eqn:DM-quark-coup-s}
    G_q = \frac{g_S \sa \ca}{v_0} \left(\frac{1}{m_h^2} - \frac{1}{m_H^2} \right) m_q
\end{equation}
is the effective DM-quark coupling. 

In order to promote a DM-quark interaction to a DM-nucleon one, the quark contents of a nucleon must be taken into account.~For a scalar mediator (as in our model), the quark Yukawa couplings generally scales with the mass of an interacting fermion.~Thus, the dominant contribution comes from the strange quark content of a nucleon and from gluons via heavy quark loops. These contributions are parametrised by the hadronic matrix elements as
\begin{equation}
    f_{Tq}^{(\calN)} \equiv \frac{m_q}{m_\calN} \langle \calN | \ovr{q}q | \calN \rangle, 
\end{equation}
where $\calN \in (p, n)$.~For a pure scalar interaction, these matrix elements parametrise the contribution of a quark mass $m_q$ to the total mass of a nucleon $m_\calN$. For more details on these parameters and recent estimates, see Ref.~\cite{Workgroup:2017lvb} and references therein.

Using the heavy quark expansion \cite{Shifman:1978}, the contribution from gluons via heavy quark loops can be expressed in terms of the lighter quarks as
\begin{equation}
    f_{Tc}^{(\calN)} = f_{Tb}^{(\calN)} = f_{Tt}^{(\calN)} = \frac{2}{27} f_{TG}^{(\calN)} = \frac{2}{27} \left(1 - \sum_{q = u,\,d,\,s} f_{Tq}^{(\calN)} \right). 
\end{equation}
Thus, we can write
\begin{equation}
    \frac{G_\calN}{m_\calN} \equiv \sum_q \frac{G_q}{m_q} f_{Tq}^{(\calN)} = \sum_{q = u,\,d,\,s} \frac{G_q}{m_q} f_{Tq}^{(\calN)} + \frac{2}{27} \left(1 - \sum_{q = u,\,d,\,s} f_{Tq}^{(\calN)} \right) \sum_{q = c,\,b,\,t} \frac{G_q}{m_q}.
\end{equation}
Using Eq.~\eqref{eqn:DM-quark-coup-s}, the above expression expands to
\begin{align*}
    \frac{G_\calN}{m_\calN} &= \frac{g_S \sa \ca}{v_0} \left(\frac{1}{m_h^2} - \frac{1}{m_H^2} \right) \left[\sum_{q = u,\,d,\,s} f_{Tq}^{(\calN)} + \frac{2}{9} \left(1 - \sum_{q = u,\,d,\,s} f_{Tq}^{(\calN)} \right) \right] \\
    &= \frac{g_S \sa \ca}{v_0} \left(\frac{1}{m_h^2} - \frac{1}{m_H^2} \right) f_\calN, 
\end{align*}
where
\begin{equation*}
    f_\calN = \frac{2}{9} + \frac{7}{9} \sum_{q = u,\,d,\,s} f_{Tq}^{(\calN)}
\end{equation*}
is the Higgs-nucleon coupling \cite{Workgroup:2017lvb}. Thus, the effective DM-nucleon interaction Lagrangian can be written as
\begin{equation}
    \lagr_{\textnormal{DM--}\calN}^{\textnormal{eff}} = - \sum_{\calN = p,\,n} G_\calN \,\ovr{\psi} \psi \,\ovr{\calN} \calN,
\end{equation}
where 
\begin{equation}
    G_\calN = \frac{g_S \sa \ca}{v_0} \left(\frac{1}{m_h^2} - \frac{1}{m_H^2} \right) m_\calN f_\calN
\end{equation}
is the effective DM-nucleon coupling \cite{Agrawal:2010fh}.

For a SI DM-nucleon interaction, the DM-nucleus interaction is a coherent sum over the total number of protons $Z$ and neutrons $(A-Z)$ in the target nucleus $N$. Thus, the SI DM-nucleus cross-section is given by
\begin{equation}
    \sigma_{\textnormal{SI}}^{\psi N} = \frac{\mu_{\psi N}^2}{\pi} \left[Z G_p + (A - Z) G_n \right]^2,
\end{equation}
where $\mu_{\psi N} = m_\psi m_N/(m_\psi + m_N)$ is the DM-nucleus reduced mass.

\section{Effective potential}\label{app:effpot}
We include the following 1-loop corrections to the zero temperature potential in the cutoff regularisation and on-shell scheme \cite{Curtin:2014jma,Delaunay:2007wb}
\begin{equation}\label{eqn:V1}
 V_{\textnormal{1-loop}} (\phi,S)=\sum_{i=\phi,\,S,\,\chi}^{W,\,Z,\,t,\,\psi}\frac{n_i}{64\pi^2}\left[m_{i}^4 \left( \log\frac{m^2_{i}}{m^2_{0 i}}-\frac{3}{2}\right)+2 m^2_{i} m^2_{0i}\right] ,
\end{equation}
where $n_{\{\phi,\,S,\,\chi,\,W,\,Z,\,t,\,\psi\}}=\{1,\,1,\,3,\,6,\,3,\,-12,\,-4\}$.~The subscript ``0'' implies that the particle masses are calculated at the $T=0$ minimum, i.e., $(\phi,S)=(v_0,s_0)$. The $\phi$ and $S$ field dependent masses are given in Appendix~\ref{app:mass-eg}, whereas the rest are given by
\begin{align}
    m_W^2 &=\frac{g^2}{4}\phi^2 ,\quad  m_Z^2=\frac{g^2+g'^2}{4}\phi^2,\quad m_t^2=\frac{y_t^2}{2}\phi^2, \nonumber \\
    m_\chi^2 = -\mu_\Phi^2 &+\lambda_\Phi \phi^2+ \mu_{\Phi S} S + \frac{1}{2}\lambda_{\Phi S}S^2, \ \ \ \  m_\psi=\mu_\psi + g_S S.
\end{align}

The finite temperature corrections to the effective potential are given by
\begin{equation}\label{eqn:VT}
    V_T(\phi,S,T)= \frac{T^4}{2\pi^2} \left[\sum_{i= \phi,\,S,\,\chi}^{W,\,Z} n_i J_b\left(\frac{m^2_i}{T^2}\right)+\sum_{i= t,\,\psi} n_i J_f\left(\frac{m^2_i}{T^2}\right)\right],
\end{equation}
where 
\begin{equation}
J_{b/f }\left(\frac{m^2_i}{T^2}\right)=\int_0^\infty dk \, 
k^2\log\left[1\mp \exp\left(-\sqrt{\frac{k^2+m_i^2}{T^2}} \right) \right].
\end{equation}
The final important correction comes from resumming the multi-loop contributions to the boson longitudinal polarizations which are infrared divergent \cite{Arnold:1992rz,Carrington:1991hz}.~These are incorporate by supplementing the scalars and longitudinal polarizations of the gauge bosons with thermal mass corrections, in particular, by expanding Eq.~\eqref{eqn:VT} to the leading order in $m^2/T^2$ \cite{Arnold:1992rz}. For our model, they are given by
\begin{align}
    \Pi_\phi (T) & =\Pi_\chi(T)=T^2 \left( 
    \frac{g'^2}{16} + \frac{3 g}{16}+\frac{\lambda_\Phi}{2}+\frac{y_t^2}{4} + \frac{\lambda_{\Phi S}}{24}
    \right),  \nonumber \\
    \Pi_S(T)& =T^2 \left(\frac{\lambda_{\Phi S}}{3}+ \frac{\lambda_S}{4}+\frac{g_S^2}{6}  \right) , \quad
    \Pi_W(T) =\frac{11}{6} g^2 T^2.  
\end{align}

For the $\phi$ and $S$ fields, the corrected masses are the eigenvalues of the following squared mass matrix
\begin{equation}\label{eq:scalarmassmatrix}
    \calM^2 + \begin{pmatrix}
    \Pi_\phi (T) & 0 \\
    0  & \Pi_S (T)
\end{pmatrix},
\end{equation}
where $\calM^2$ is defined in Eq.~\eqref{eqn:mass-mat}.~For the $Z$ and $\gamma$ fields, namely $m^2_{Z/\gamma}+\Pi_{Z/\gamma}(T)$, the mass corrections are the eigenvalues of the following squared mass matrix
\begin{equation}
    \begin{pmatrix}
        \frac{1}{4} g^2 \phi^2+ \frac{11}{6} g^2T^2 & -\frac{1}{4}g' g \phi^2 \\[1.1mm]
        -\frac{1}{4}g' g \phi^2 & \frac{1}{4} g'^2 \phi^2+\frac{11}{6} g'^2 T^2
    \end{pmatrix}.
\end{equation}
In other cases, we simply use the following substitution
\begin{equation}
    m_i^2 \rightarrow m_i^2 +\Pi_i.
\end{equation}
Finally, the effective potential $V_{\textnormal{eff}}(\phi, S, T)$ is given by
\begin{equation}
    V_{\textnormal{eff}}(\phi,S,T) = V_{\textnormal{tree}}(\phi,S) + V_{\textnormal{1-loop}}(\phi,S) + V_{T}(\phi,S,T),
\end{equation}
where $V_{\textnormal{tree}}(\phi,S)$ is the tree-level scalar potential in Appendix~\ref{app:tree-potential}.

\bibliographystyle{JHEP}
\bibliography{UV_fermion}

\providecommand{\href}[2]{#2}\begingroup\raggedright\begin{thebibliography}{100}

\bibitem{Aad:2012tfa}
{\scshape ATLAS} Collaboration, G.~Aad et~al., \emph{{Observation of a new
  particle in the search for the Standard Model Higgs boson with the ATLAS
  detector at the LHC}},
  \href{https://doi.org/10.1016/j.physletb.2012.08.020}{\emph{Phys. Lett. B}
  {\bfseries 716} (2012) 1--29},
  [\href{https://arxiv.org/abs/1207.7214}{{\ttfamily 1207.7214}}].

\bibitem{Chatrchyan:2012xdj}
{\scshape CMS} Collaboration, S.~Chatrchyan et~al., \emph{{Observation of a new
  boson at a mass of 125 GeV with the CMS experiment at the LHC}},
  \href{https://doi.org/10.1016/j.physletb.2012.08.021}{\emph{Phys. Lett. B}
  {\bfseries 716} (2012) 30--61},
  [\href{https://arxiv.org/abs/1207.7235}{{\ttfamily 1207.7235}}].

\bibitem{Abbott:2016blz}
{\scshape Virgo, LIGO Scientific} Collaboration, B.~P. Abbott et~al.,
  \emph{{Observation of Gravitational Waves from a Binary Black Hole Merger}},
  \href{https://doi.org/10.1103/PhysRevLett.116.061102}{\emph{Phys. Rev. Lett.}
  {\bfseries 116} (2016) 061102},
  [\href{https://arxiv.org/abs/1602.03837}{{\ttfamily 1602.03837}}].

\bibitem{Abbott:2016nmj}
{\scshape Virgo, LIGO Scientific} Collaboration, B.~P. Abbott et~al.,
  \emph{{GW151226: Observation of Gravitational Waves from a 22-Solar-Mass
  Binary Black Hole Coalescence}},
  \href{https://doi.org/10.1103/PhysRevLett.116.241103}{\emph{Phys. Rev. Lett.}
  {\bfseries 116} (2016) 241103},
  [\href{https://arxiv.org/abs/1606.04855}{{\ttfamily 1606.04855}}].

\bibitem{Abbott:2017ylp}
{\scshape Virgo, LIGO Scientific} Collaboration, B.~P. Abbott et~al.,
  \emph{{First search for gravitational waves from known pulsars with Advanced
  LIGO}}, \href{https://doi.org/10.3847/1538-4357/aa9aee,
  10.3847/1538-4357/aa677f}{\emph{Astrophys. J.} {\bfseries 839} (2017) 12},
  [\href{https://arxiv.org/abs/1701.07709}{{\ttfamily 1701.07709}}].

\bibitem{Abbott:2017vtc}
{\scshape VIRGO, LIGO Scientific} Collaboration, B.~P. Abbott et~al.,
  \emph{{GW170104: Observation of a 50-Solar-Mass Binary Black Hole Coalescence
  at Redshift 0.2}},
  \href{https://doi.org/10.1103/PhysRevLett.118.221101}{\emph{Phys. Rev. Lett.}
  {\bfseries 118} (2017) 221101},
  [\href{https://arxiv.org/abs/1706.01812}{{\ttfamily 1706.01812}}].

\bibitem{Abbott:2017oio}
{\scshape Virgo, LIGO Scientific} Collaboration, B.~P. Abbott et~al.,
  \emph{{GW170814: A Three-Detector Observation of Gravitational Waves from a
  Binary Black Hole Coalescence}},
  \href{https://doi.org/10.1103/PhysRevLett.119.141101}{\emph{Phys. Rev. Lett.}
  {\bfseries 119} (2017) 141101},
  [\href{https://arxiv.org/abs/1709.09660}{{\ttfamily 1709.09660}}].

\bibitem{TheLIGOScientific:2017qsa}
{\scshape Virgo, LIGO Scientific} Collaboration, B.~P. Abbott et~al.,
  \emph{{GW170817: Observation of Gravitational Waves from a Binary Neutron
  Star Inspiral}},
  \href{https://doi.org/10.1103/PhysRevLett.119.161101}{\emph{Phys. Rev. Lett.}
  {\bfseries 119} (2017) 161101},
  [\href{https://arxiv.org/abs/1710.05832}{{\ttfamily 1710.05832}}].

\bibitem{Monitor:2017mdv}
{\scshape Virgo, Fermi-GBM, INTEGRAL, LIGO Scientific} Collaboration, B.~P.
  Abbott et~al., \emph{{Gravitational Waves and Gamma-rays from a Binary
  Neutron Star Merger: GW170817 and GRB 170817A}},
  \href{https://doi.org/10.3847/2041-8213/aa920c}{\emph{Astrophys. J.}
  {\bfseries 848} (2017) L13},
  [\href{https://arxiv.org/abs/1710.05834}{{\ttfamily 1710.05834}}].

\bibitem{Abbott:2017gyy}
{\scshape Virgo, LIGO Scientific} Collaboration, B.~P. Abbott et~al.,
  \emph{{GW170608: Observation of a 19-solar-mass Binary Black Hole
  Coalescence}},
  \href{https://doi.org/10.3847/2041-8213/aa9f0c}{\emph{Astrophys. J.}
  {\bfseries 851} (2017) L35},
  [\href{https://arxiv.org/abs/1711.05578}{{\ttfamily 1711.05578}}].

\bibitem{Caprini:2015zlo}
C.~Caprini et~al., \emph{{Science with the space-based interferometer eLISA.
  II: Gravitational waves from cosmological phase transitions}},
  \href{https://doi.org/10.1088/1475-7516/2016/04/001}{\emph{JCAP} {\bfseries
  1604} (2016) 001}, [\href{https://arxiv.org/abs/1512.06239}{{\ttfamily
  1512.06239}}].

\bibitem{Weir:2017wfa}
D.~J. Weir, \emph{{Gravitational waves from a first order electroweak phase
  transition: a brief review}},
  \href{https://doi.org/10.1098/rsta.2017.0126}{\emph{Phil. Trans. Roy. Soc.
  Lond. A} {\bfseries 376} (2018) 20170126},
  [\href{https://arxiv.org/abs/1705.01783}{{\ttfamily 1705.01783}}].

\bibitem{Caprini:2018mtu}
C.~Caprini and D.~G. Figueroa, \emph{{Cosmological Backgrounds of Gravitational
  Waves}}, \href{https://doi.org/10.1088/1361-6382/aac608}{\emph{Class. Quant.
  Grav.} {\bfseries 35} (2018) 163001},
  [\href{https://arxiv.org/abs/1801.04268}{{\ttfamily 1801.04268}}].

\bibitem{Kuzmin:1985mm}
V.~A. Kuzmin, V.~A. Rubakov and M.~E. Shaposhnikov, \emph{{On the Anomalous
  Electroweak Baryon Number Nonconservation in the Early Universe}},
  \href{https://doi.org/10.1016/0370-2693(85)91028-7}{\emph{Phys. Lett. B}
  {\bfseries 155} (1985) 36}.

\bibitem{Cohen:1993nk}
A.~G. Cohen, D.~B. Kaplan and A.~E. Nelson, \emph{{Progress in electroweak
  baryogenesis}},
  \href{https://doi.org/10.1146/annurev.ns.43.120193.000331}{\emph{Ann. Rev.
  Nucl. Part. Sci.} {\bfseries 43} (1993) 27--70},
  [\href{https://arxiv.org/abs/hep-ph/9302210}{{\ttfamily hep-ph/9302210}}].

\bibitem{Riotto:1999yt}
A.~Riotto and M.~Trodden, \emph{{Recent progress in baryogenesis}},
  \href{https://doi.org/10.1146/annurev.nucl.49.1.35}{\emph{Ann. Rev. Nucl.
  Part. Sci.} {\bfseries 49} (1999) 35--75},
  [\href{https://arxiv.org/abs/hep-ph/9901362}{{\ttfamily hep-ph/9901362}}].

\bibitem{Morrissey:2012db}
D.~E. Morrissey and M.~J. Ramsey-Musolf, \emph{{Electroweak baryogenesis}},
  \href{https://doi.org/10.1088/1367-2630/14/12/125003}{\emph{New J. Phys.}
  {\bfseries 14} (2012) 125003},
  [\href{https://arxiv.org/abs/1206.2942}{{\ttfamily 1206.2942}}].

\bibitem{Arnold:1992rz}
P.~B. Arnold and O.~Espinosa, \emph{{The Effective potential and first order
  phase transitions: Beyond leading-order}},
  \href{https://doi.org/10.1103/PhysRevD.50.6662,
  10.1103/PhysRevD.47.3546}{\emph{Phys. Rev. D} {\bfseries 47} (1993) 3546},
  [\href{https://arxiv.org/abs/hep-ph/9212235}{{\ttfamily hep-ph/9212235}}].

\bibitem{Kajantie:1996qd}
K.~Kajantie, M.~Laine, K.~Rummukainen and M.~E. Shaposhnikov, \emph{{A
  Nonperturbative analysis of the finite T phase transition in $SU(2) \times
  U(1)$ electroweak theory}},
  \href{https://doi.org/10.1016/S0550-3213(97)00164-8}{\emph{Nucl. Phys. B}
  {\bfseries 493} (1997) 413--438},
  [\href{https://arxiv.org/abs/hep-lat/9612006}{{\ttfamily hep-lat/9612006}}].

\bibitem{Curtin:2014jma}
D.~Curtin, P.~Meade and C.-T. Yu, \emph{{Testing Electroweak Baryogenesis with
  Future Colliders}},
  \href{https://doi.org/10.1007/JHEP11(2014)127}{\emph{JHEP} {\bfseries 11}
  (2014) 127}, [\href{https://arxiv.org/abs/1409.0005}{{\ttfamily 1409.0005}}].

\bibitem{Kotwal:2016tex}
A.~V. Kotwal, M.~J. Ramsey-Musolf, J.~M. No and P.~Winslow,
  \emph{{Singlet-catalyzed electroweak phase transitions in the 100 TeV
  frontier}}, \href{https://doi.org/10.1103/PhysRevD.94.035022}{\emph{Phys.
  Rev. D} {\bfseries 94} (2016) 035022},
  [\href{https://arxiv.org/abs/1605.06123}{{\ttfamily 1605.06123}}].

\bibitem{Choi:1993cv}
J.~Choi and R.~R. Volkas, \emph{{Real Higgs singlet and the electroweak phase
  transition in the Standard Model}},
  \href{https://doi.org/10.1016/0370-2693(93)91013-D}{\emph{Phys. Lett. B}
  {\bfseries 317} (1993) 385--391},
  [\href{https://arxiv.org/abs/hep-ph/9308234}{{\ttfamily hep-ph/9308234}}].

\bibitem{Ashoorioon:2009nf}
A.~Ashoorioon and T.~Konstandin, \emph{{Strong electroweak phase transitions
  without collider traces}},
  \href{https://doi.org/10.1088/1126-6708/2009/07/086}{\emph{JHEP} {\bfseries
  07} (2009) 086}, [\href{https://arxiv.org/abs/0904.0353}{{\ttfamily
  0904.0353}}].

\bibitem{Enqvist:2014zqa}
K.~Enqvist, S.~Nurmi, T.~Tenkanen and K.~Tuominen, \emph{{Standard Model with a
  real singlet scalar and inflation}},
  \href{https://doi.org/10.1088/1475-7516/2014/08/035}{\emph{JCAP} {\bfseries
  1408} (2014) 035}, [\href{https://arxiv.org/abs/1407.0659}{{\ttfamily
  1407.0659}}].

\bibitem{Kakizaki:2015wua}
M.~Kakizaki, S.~Kanemura and T.~Matsui, \emph{{Gravitational waves as a probe
  of extended scalar sectors with the first order electroweak phase
  transition}}, \href{https://doi.org/10.1103/PhysRevD.92.115007}{\emph{Phys.
  Rev. D} {\bfseries 92} (2015) 115007},
  [\href{https://arxiv.org/abs/1509.08394}{{\ttfamily 1509.08394}}].

\bibitem{Huang:2016odd}
F.~P. Huang, Y.~Wan, D.-G. Wang, Y.-F. Cai and X.~Zhang, \emph{{Hearing the
  echoes of electroweak baryogenesis with gravitational wave detectors}},
  \href{https://doi.org/10.1103/PhysRevD.94.041702}{\emph{Phys. Rev. D}
  {\bfseries 94} (2016) 041702},
  [\href{https://arxiv.org/abs/1601.01640}{{\ttfamily 1601.01640}}].

\bibitem{Hashino:2016rvx}
K.~Hashino, M.~Kakizaki, S.~Kanemura and T.~Matsui, \emph{{Synergy between
  measurements of gravitational waves and the triple-Higgs coupling in probing
  the first-order electroweak phase transition}},
  \href{https://doi.org/10.1103/PhysRevD.94.015005}{\emph{Phys. Rev. D}
  {\bfseries 94} (2016) 015005},
  [\href{https://arxiv.org/abs/1604.02069}{{\ttfamily 1604.02069}}].

\bibitem{Chala:2016ykx}
M.~Chala, G.~Nardini and I.~Sobolev, \emph{{Unified explanation for dark matter
  and electroweak baryogenesis with direct detection and gravitational wave
  signatures}}, \href{https://doi.org/10.1103/PhysRevD.94.055006}{\emph{Phys.
  Rev. D} {\bfseries 94} (2016) 055006},
  [\href{https://arxiv.org/abs/1605.08663}{{\ttfamily 1605.08663}}].

\bibitem{Tenkanen:2016idg}
T.~Tenkanen, K.~Tuominen and V.~Vaskonen, \emph{{A Strong Electroweak Phase
  Transition from the Inflaton Field}},
  \href{https://doi.org/10.1088/1475-7516/2016/09/037}{\emph{JCAP} {\bfseries
  1609} (2016) 037}, [\href{https://arxiv.org/abs/1606.06063}{{\ttfamily
  1606.06063}}].

\bibitem{Kobakhidze:2016mch}
A.~Kobakhidze, A.~Manning and J.~Yue, \emph{{Gravitational waves from the phase
  transition of a nonlinearly realized electroweak gauge symmetry}},
  \href{https://doi.org/10.1142/S0218271817501140}{\emph{Int. J. Mod. Phys. D}
  {\bfseries 26} (2017) 1750114},
  [\href{https://arxiv.org/abs/1607.00883}{{\ttfamily 1607.00883}}].

\bibitem{Huang:2016cjm}
P.~Huang, A.~J. Long and L.-T. Wang, \emph{{Probing the Electroweak Phase
  Transition with Higgs Factories and Gravitational Waves}},
  \href{https://doi.org/10.1103/PhysRevD.94.075008}{\emph{Phys. Rev. D}
  {\bfseries 94} (2016) 075008},
  [\href{https://arxiv.org/abs/1608.06619}{{\ttfamily 1608.06619}}].

\bibitem{Artymowski:2016tme}
M.~Artymowski, M.~Lewicki and J.~D. Wells, \emph{{Gravitational wave and
  collider implications of electroweak baryogenesis aided by non-standard
  cosmology}}, \href{https://doi.org/10.1007/JHEP03(2017)066}{\emph{JHEP}
  {\bfseries 03} (2017) 066},
  [\href{https://arxiv.org/abs/1609.07143}{{\ttfamily 1609.07143}}].

\bibitem{Hashino:2016xoj}
K.~Hashino, M.~Kakizaki, S.~Kanemura, P.~Ko and T.~Matsui, \emph{{Gravitational
  waves and Higgs boson couplings for exploring first order phase transition in
  the model with a singlet scalar field}},
  \href{https://doi.org/10.1016/j.physletb.2016.12.052}{\emph{Phys. Lett. B}
  {\bfseries 766} (2017) 49--54},
  [\href{https://arxiv.org/abs/1609.00297}{{\ttfamily 1609.00297}}].

\bibitem{Vaskonen:2016yiu}
V.~Vaskonen, \emph{{Electroweak baryogenesis and gravitational waves from a
  real scalar singlet}},
  \href{https://doi.org/10.1103/PhysRevD.95.123515}{\emph{Phys. Rev. D}
  {\bfseries 95} (2017) 123515},
  [\href{https://arxiv.org/abs/1611.02073}{{\ttfamily 1611.02073}}].

\bibitem{Baldes:2017rcu}
I.~Baldes, \emph{{Gravitational waves from the asymmetric-dark-matter
  generating phase transition}},
  \href{https://doi.org/10.1088/1475-7516/2017/05/028}{\emph{JCAP} {\bfseries
  1705} (2017) 028}, [\href{https://arxiv.org/abs/1702.02117}{{\ttfamily
  1702.02117}}].

\bibitem{Beniwal:2017eik}
A.~Beniwal, M.~Lewicki, J.~D. Wells, M.~White and A.~G. Williams,
  \emph{{Gravitational wave, collider and dark matter signals from a scalar
  singlet electroweak baryogenesis}},
  \href{https://doi.org/10.1007/JHEP08(2017)108}{\emph{JHEP} {\bfseries 08}
  (2017) 108}, [\href{https://arxiv.org/abs/1702.06124}{{\ttfamily
  1702.06124}}].

\bibitem{Kobakhidze:2017mru}
A.~Kobakhidze, C.~Lagger, A.~Manning and J.~Yue, \emph{{Gravitational waves
  from a supercooled electroweak phase transition and their detection with
  pulsar timing arrays}},
  \href{https://doi.org/10.1140/epjc/s10052-017-5132-y}{\emph{Eur. Phys. J. C}
  {\bfseries 77} (2017) 570},
  [\href{https://arxiv.org/abs/1703.06552}{{\ttfamily 1703.06552}}].

\bibitem{Cai:2017tmh}
R.-G. Cai, M.~Sasaki and S.-J. Wang, \emph{{The gravitational waves from the
  first-order phase transition with a dimension-six operator}},
  \href{https://doi.org/10.1088/1475-7516/2017/08/004}{\emph{JCAP} {\bfseries
  1708} (2017) 004}, [\href{https://arxiv.org/abs/1707.03001}{{\ttfamily
  1707.03001}}].

\bibitem{Croon:2018erz}
D.~Croon, V.~Sanz and G.~White, \emph{{Model Discrimination in Gravitational
  Wave spectra from Dark Phase Transitions}},
  \href{https://doi.org/10.1007/JHEP08(2018)203}{\emph{JHEP} {\bfseries 08}
  (2018) 203}, [\href{https://arxiv.org/abs/1806.02332}{{\ttfamily
  1806.02332}}].

\bibitem{Baldes:2018emh}
I.~Baldes and C.~Garcia-Cely, \emph{{Strong gravitational radiation from a
  simple dark matter model}},
  [\href{https://arxiv.org/abs/1809.01198}{{\ttfamily 1809.01198}}].

\bibitem{Hashino:2018wee}
K.~Hashino, R.~Jinno, M.~Kakizaki, S.~Kanemura, T.~Takahashi and M.~Takimoto,
  \emph{{Fingerprinting models of first-order phase transitions by the synergy
  between collider and gravitational-wave experiments}},
  [\href{https://arxiv.org/abs/1809.04994}{{\ttfamily 1809.04994}}].

\bibitem{Ahriche:2018rao}
A.~Ahriche, K.~Hashino, S.~Kanemura and S.~Nasri, \emph{{Gravitational Waves
  from Phase Transitions in Models with Charged Singlets}},
  \href{https://doi.org/10.1016/j.physletb.2018.12.013}{\emph{Phys. Lett. B}
  {\bfseries 789} (2019) 119--126},
  [\href{https://arxiv.org/abs/1809.09883}{{\ttfamily 1809.09883}}].

\bibitem{Silveira1985136}
V.~Silveira and A.~Zee, \emph{Scalar phantoms},
  \href{https://doi.org/https://doi.org/10.1016/0370-2693(85)90624-0}{\emph{Phys.
  Lett. B} {\bfseries 161} (1985) 136--140}.

\bibitem{PhysRevD.50.3637}
J.~McDonald, \emph{Gauge singlet scalars as cold dark matter},
  \href{https://doi.org/10.1103/PhysRevD.50.3637}{\emph{Phys. Rev. D}
  {\bfseries 50} (1994) 3637}.

\bibitem{Burgess:2000yq}
C.~P. Burgess, M.~Pospelov and T.~ter Veldhuis, \emph{{The Minimal model of
  nonbaryonic dark matter: A singlet scalar}},
  \href{https://doi.org/10.1016/S0550-3213(01)00513-2}{\emph{Nucl. Phys. B}
  {\bfseries 619} (2001) 709--728},
  [\href{https://arxiv.org/abs/hep-ph/0011335}{{\ttfamily hep-ph/0011335}}].

\bibitem{Espinosa:2008kw}
J.~R. Espinosa, T.~Konstandin, J.~M. No and M.~Quiros, \emph{{Some Cosmological
  Implications of Hidden Sectors}},
  \href{https://doi.org/10.1103/PhysRevD.78.123528}{\emph{Phys. Rev. D}
  {\bfseries 78} (2008) 123528},
  [\href{https://arxiv.org/abs/0809.3215}{{\ttfamily 0809.3215}}].

\bibitem{Alanne:2014bra}
T.~Alanne, K.~Tuominen and V.~Vaskonen, \emph{{Strong phase transition, dark
  matter and vacuum stability from simple hidden sectors}},
  \href{https://doi.org/10.1016/j.nuclphysb.2014.11.001}{\emph{Nucl. Phys. B}
  {\bfseries 889} (2014) 692--711},
  [\href{https://arxiv.org/abs/1407.0688}{{\ttfamily 1407.0688}}].

\bibitem{Martin-Lozano:2015dja}
V.~Martin~Lozano, J.~M. Moreno and C.~B. Park, \emph{{Resonant Higgs boson pair
  production in the $ hh\to b\overline{b}\ WW\to b\overline{b}{\ell}^{+}\nu
  {\ell}^{-}\overline{\nu} $ decay channel}},
  \href{https://doi.org/10.1007/JHEP08(2015)004}{\emph{JHEP} {\bfseries 08}
  (2015) 004}, [\href{https://arxiv.org/abs/1501.03799}{{\ttfamily
  1501.03799}}].

\bibitem{Falkowski:2015iwa}
A.~Falkowski, C.~Gross and O.~Lebedev, \emph{{A second Higgs from the Higgs
  portal}}, \href{https://doi.org/10.1007/JHEP05(2015)057}{\emph{JHEP}
  {\bfseries 05} (2015) 057},
  [\href{https://arxiv.org/abs/1502.01361}{{\ttfamily 1502.01361}}].

\bibitem{Buttazzo:2015bka}
D.~Buttazzo, F.~Sala and A.~Tesi, \emph{{Singlet-like Higgs bosons at present
  and future colliders}},
  \href{https://doi.org/10.1007/JHEP11(2015)158}{\emph{JHEP} {\bfseries 11}
  (2015) 158}, [\href{https://arxiv.org/abs/1505.05488}{{\ttfamily
  1505.05488}}].

\bibitem{Heikinheimo:2016yds}
M.~Heikinheimo, T.~Tenkanen, K.~Tuominen and V.~Vaskonen, \emph{{Observational
  Constraints on Decoupled Hidden Sectors}},
  \href{https://doi.org/10.1103/PhysRevD.96.109902,
  10.1103/PhysRevD.94.063506}{\emph{Phys. Rev. D} {\bfseries 94} (2016)
  063506}, [\href{https://arxiv.org/abs/1604.02401}{{\ttfamily 1604.02401}}].

\bibitem{Balazs:2016tbi}
C.~Balazs, A.~Fowlie, A.~Mazumdar and G.~White, \emph{{Gravitational waves at
  aLIGO and vacuum stability with a scalar singlet extension of the Standard
  Model}}, \href{https://doi.org/10.1103/PhysRevD.95.043505}{\emph{Phys. Rev.
  D} {\bfseries 95} (2017) 043505},
  [\href{https://arxiv.org/abs/1611.01617}{{\ttfamily 1611.01617}}].

\bibitem{Lewis:2017dme}
I.~M. Lewis and M.~Sullivan, \emph{{Benchmarks for Double Higgs Production in
  the Singlet Extended Standard Model at the LHC}},
  \href{https://doi.org/10.1103/PhysRevD.96.035037}{\emph{Phys. Rev. D}
  {\bfseries 96} (2017) 035037},
  [\href{https://arxiv.org/abs/1701.08774}{{\ttfamily 1701.08774}}].

\bibitem{Ghorbani:2017jls}
P.~H. Ghorbani, \emph{{Electroweak Baryogenesis and Dark Matter via a
  Pseudoscalar vs. Scalar}},
  \href{https://doi.org/10.1007/JHEP08(2017)058}{\emph{JHEP} {\bfseries 08}
  (2017) 058}, [\href{https://arxiv.org/abs/1703.06506}{{\ttfamily
  1703.06506}}].

\bibitem{Chen:2017qcz}
C.-Y. Chen, J.~Kozaczuk and I.~M. Lewis, \emph{{Non-resonant Collider
  Signatures of a Singlet-Driven Electroweak Phase Transition}},
  \href{https://doi.org/10.1007/JHEP08(2017)096}{\emph{JHEP} {\bfseries 08}
  (2017) 096}, [\href{https://arxiv.org/abs/1704.05844}{{\ttfamily
  1704.05844}}].

\bibitem{Kamon:2017yfx}
T.~Kamon, P.~Ko and J.~Li, \emph{{Characterizing Higgs portal dark matter
  models at the ILC}},
  \href{https://doi.org/10.1140/epjc/s10052-017-5240-8}{\emph{Eur. Phys. J. C}
  {\bfseries 77} (2017) 652},
  [\href{https://arxiv.org/abs/1705.02149}{{\ttfamily 1705.02149}}].

\bibitem{Ettefaghi:2017vbh}
M.~Ettefaghi and R.~Moazzemi, \emph{{Analyzing of singlet fermionic dark matter
  via the updated direct detection data}},
  \href{https://doi.org/10.1140/epjc/s10052-017-4894-6}{\emph{Eur. Phys. J. C}
  {\bfseries 77} (2017) 343},
  [\href{https://arxiv.org/abs/1705.07571}{{\ttfamily 1705.07571}}].

\bibitem{Baker:2017zwx}
M.~J. Baker, M.~Breitbach, J.~Kopp and L.~Mittnacht, \emph{{Dynamic Freeze-In:
  Impact of Thermal Masses and Cosmological Phase Transitions on Dark Matter
  Production}}, \href{https://doi.org/10.1007/JHEP03(2018)114}{\emph{JHEP}
  {\bfseries 03} (2018) 114},
  [\href{https://arxiv.org/abs/1712.03962}{{\ttfamily 1712.03962}}].

\bibitem{Baum:2017enm}
S.~Baum, M.~Carena, N.~R. Shah and C.~E.~M. Wagner, \emph{{Higgs portals for
  thermal Dark Matter. EFT perspectives and the NMSSM}},
  \href{https://doi.org/10.1007/JHEP04(2018)069}{\emph{JHEP} {\bfseries 04}
  (2018) 069}, [\href{https://arxiv.org/abs/1712.09873}{{\ttfamily
  1712.09873}}].

\bibitem{Bernal:2018ins}
N.~Bernal, C.~Cosme and T.~Tenkanen, \emph{{Phenomenology of Self-Interacting
  Dark Matter in a Matter-Dominated Universe}},
  \href{https://doi.org/10.1140/epjc/s10052-019-6608-8}{\emph{Eur. Phys. J. C}
  {\bfseries 79} (2019) 99},
  [\href{https://arxiv.org/abs/1803.08064}{{\ttfamily 1803.08064}}].

\bibitem{Kim:2008pp}
Y.~G. Kim, K.~Y. Lee and S.~Shin, \emph{{Singlet fermionic dark matter}},
  \href{https://doi.org/10.1088/1126-6708/2008/05/100}{\emph{JHEP} {\bfseries
  05} (2008) 100}, [\href{https://arxiv.org/abs/0803.2932}{{\ttfamily
  0803.2932}}].

\bibitem{Baek:2011aa}
S.~Baek, P.~Ko and W.-I. Park, \emph{{Search for the Higgs portal to a singlet
  fermionic dark matter at the LHC}},
  \href{https://doi.org/10.1007/JHEP02(2012)047}{\emph{JHEP} {\bfseries 02}
  (2012) 047}, [\href{https://arxiv.org/abs/1112.1847}{{\ttfamily 1112.1847}}].

\bibitem{Baek:2012uj}
S.~Baek, P.~Ko, W.-I. Park and E.~Senaha, \emph{{Vacuum structure and stability
  of a singlet fermion dark matter model with a singlet scalar messenger}},
  \href{https://doi.org/10.1007/JHEP11(2012)116}{\emph{JHEP} {\bfseries 11}
  (2012) 116}, [\href{https://arxiv.org/abs/1209.4163}{{\ttfamily 1209.4163}}].

\bibitem{Espinosa:2011ax}
J.~R. Espinosa, T.~Konstandin and F.~Riva, \emph{{Strong Electroweak Phase
  Transitions in the Standard Model with a Singlet}},
  \href{https://doi.org/10.1016/j.nuclphysb.2011.09.010}{\emph{Nucl. Phys. B}
  {\bfseries 854} (2012) 592--630},
  [\href{https://arxiv.org/abs/1107.5441}{{\ttfamily 1107.5441}}].

\bibitem{Fairbairn:2013uta}
M.~Fairbairn and R.~Hogan, \emph{{Singlet Fermionic Dark Matter and the
  Electroweak Phase Transition}},
  \href{https://doi.org/10.1007/JHEP09(2013)022}{\emph{JHEP} {\bfseries 09}
  (2013) 022}, [\href{https://arxiv.org/abs/1305.3452}{{\ttfamily 1305.3452}}].

\bibitem{Li:2014wia}
T.~Li and Y.-F. Zhou, \emph{{Strongly first order phase transition in the
  singlet fermionic dark matter model after LUX}},
  \href{https://doi.org/10.1007/JHEP07(2014)006}{\emph{JHEP} {\bfseries 07}
  (2014) 006}, [\href{https://arxiv.org/abs/1402.3087}{{\ttfamily 1402.3087}}].

\bibitem{Ade:2015xua}
{\scshape Planck} Collaboration, P.~A.~R. Ade et~al., \emph{{Planck 2015
  results. XIII. Cosmological parameters}},
  \href{https://doi.org/10.1051/0004-6361/201525830}{\emph{Astron. Astrophys.}
  {\bfseries 594} (2016) A13},
  [\href{https://arxiv.org/abs/1502.01589}{{\ttfamily 1502.01589}}].

\bibitem{Aprile:2018dbl}
{\scshape XENON} Collaboration, E.~Aprile et~al., \emph{{Dark Matter Search
  Results from a One Tonne$\times$Year Exposure of XENON1T}},
  \href{https://doi.org/10.1103/PhysRevLett.121.111302}{\emph{Phys. Rev. Lett.}
  {\bfseries 121} (2018) 111302},
  [\href{https://arxiv.org/abs/1805.12562}{{\ttfamily 1805.12562}}].

\bibitem{Haller:2018nnx}
J.~Haller, A.~Hoecker, R.~Kogler, K.~M?nig, T.~Peiffer and J.~Stelzer,
  \emph{{Update of the global electroweak fit and constraints on
  two-Higgs-doublet models}},
  \href{https://doi.org/10.1140/epjc/s10052-018-6131-3}{\emph{Eur. Phys. J. C}
  {\bfseries 78} (2018) 675},
  [\href{https://arxiv.org/abs/1803.01853}{{\ttfamily 1803.01853}}].

\bibitem{Bechtle:2013wla}
P.~Bechtle, O.~Brein, S.~Heinemeyer, O.~Stal, T.~Stefaniak, G.~Weiglein et~al.,
  \emph{{HiggsBounds-4: Improved Tests of Extended Higgs Sectors against
  Exclusion Bounds from LEP, the Tevatron and the LHC}},
  \href{https://doi.org/10.1140/epjc/s10052-013-2693-2}{\emph{Eur. Phys. J. C}
  {\bfseries 74} (2014) 2693},
  [\href{https://arxiv.org/abs/1311.0055}{{\ttfamily 1311.0055}}].

\bibitem{Bechtle:2013xfa}
P.~Bechtle, S.~Heinemeyer, O.~Stal, T.~Stefaniak and G.~Weiglein,
  \emph{{HiggsSignals: Confronting arbitrary Higgs sectors with measurements at
  the Tevatron and the LHC}},
  \href{https://doi.org/10.1140/epjc/s10052-013-2711-4}{\emph{Eur. Phys. J. C}
  {\bfseries 74} (2014) 2711},
  [\href{https://arxiv.org/abs/1305.1933}{{\ttfamily 1305.1933}}].

\bibitem{Cline:2013gha}
J.~M. Cline, K.~Kainulainen, P.~Scott and C.~Weniger, \emph{{Update on scalar
  singlet dark matter}}, \href{https://doi.org/10.1103/PhysRevD.92.039906,
  10.1103/PhysRevD.88.055025}{\emph{Phys. Rev. D} {\bfseries 88} (2013)
  055025}, [\href{https://arxiv.org/abs/1306.4710}{{\ttfamily 1306.4710}}].

\bibitem{Beniwal:2015sdl}
A.~Beniwal, F.~Rajec, C.~Savage, P.~Scott, C.~Weniger, M.~White et~al.,
  \emph{{Combined analysis of effective Higgs portal dark matter models}},
  \href{https://doi.org/10.1103/PhysRevD.93.115016}{\emph{Phys. Rev. D}
  {\bfseries 93} (2016) 115016},
  [\href{https://arxiv.org/abs/1512.06458}{{\ttfamily 1512.06458}}].

\bibitem{He:2016mls}
X.-G. He and J.~Tandean, \emph{{New LUX and PandaX-II Results Illuminating the
  Simplest Higgs-Portal Dark Matter Models}},
  \href{https://doi.org/10.1007/JHEP12(2016)074}{\emph{JHEP} {\bfseries 12}
  (2016) 074}, [\href{https://arxiv.org/abs/1609.03551}{{\ttfamily
  1609.03551}}].

\bibitem{Escudero:2016gzx}
M.~Escudero, A.~Berlin, D.~Hooper and M.-X. Lin, \emph{{Toward (Finally!)
  Ruling Out Z and Higgs Mediated Dark Matter Models}},
  \href{https://doi.org/10.1088/1475-7516/2016/12/029}{\emph{JCAP} {\bfseries
  1612} (2016) 029}, [\href{https://arxiv.org/abs/1609.09079}{{\ttfamily
  1609.09079}}].

\bibitem{Wu:2016mbe}
H.~Wu and S.~Zheng, \emph{{Scalar Dark Matter: Real vs Complex}},
  \href{https://doi.org/10.1007/JHEP03(2017)142}{\emph{JHEP} {\bfseries 03}
  (2017) 142}, [\href{https://arxiv.org/abs/1610.06292}{{\ttfamily
  1610.06292}}].

\bibitem{Banerjee:2016vrp}
S.~Banerjee and N.~Chakrabarty, \emph{{A revisit to scalar dark matter with
  radiative corrections}},  [\href{https://arxiv.org/abs/1612.01973}{{\ttfamily
  1612.01973}}].

\bibitem{Casas:2017jjg}
J.~A. Casas, D.~G. Cerdeño, J.~M. Moreno and J.~Quilis, \emph{{Reopening the
  Higgs portal for single scalar dark matter}},
  \href{https://doi.org/10.1007/JHEP05(2017)036}{\emph{JHEP} {\bfseries 05}
  (2017) 036}, [\href{https://arxiv.org/abs/1701.08134}{{\ttfamily
  1701.08134}}].

\bibitem{Athron:2017kgt}
{\scshape GAMBIT} Collaboration, P.~Athron et~al., \emph{{Status of the scalar
  singlet dark matter model}},
  \href{https://doi.org/10.1140/epjc/s10052-017-5113-1}{\emph{Eur. Phys. J. C}
  {\bfseries 77} (2017) 568},
  [\href{https://arxiv.org/abs/1705.07931}{{\ttfamily 1705.07931}}].

\bibitem{Hoferichter:2017olk}
M.~Hoferichter, P.~Klos, J.~Menéndez and A.~Schwenk, \emph{{Improved limits
  for Higgs-portal dark matter from LHC searches}},
  \href{https://doi.org/10.1103/PhysRevLett.119.181803}{\emph{Phys. Rev. Lett.}
  {\bfseries 119} (2017) 181803},
  [\href{https://arxiv.org/abs/1708.02245}{{\ttfamily 1708.02245}}].

\bibitem{Athron:2018ipf}
P.~Athron, J.~M. Cornell, F.~Kahlhoefer, J.~Mckay, P.~Scott and S.~Wild,
  \emph{{Impact of vacuum stability, perturbativity and XENON1T on global fits
  of $\mathbb {Z}_2$ and $\mathbb {Z}_3$ scalar singlet dark matter}},
  \href{https://doi.org/10.1140/epjc/s10052-018-6314-y}{\emph{Eur. Phys. J. C}
  {\bfseries 78} (2018) 830},
  [\href{https://arxiv.org/abs/1806.11281}{{\ttfamily 1806.11281}}].

\bibitem{Semenov:2014rea}
A.~Semenov, \emph{{LanHEP: A package for automatic generation of Feynman rules
  from the Lagrangian. Version 3.2}},
  \href{https://doi.org/10.1016/j.cpc.2016.01.003}{\emph{Comput. Phys. Commun.}
  {\bfseries 201} (2016) 167--170},
  [\href{https://arxiv.org/abs/1412.5016}{{\ttfamily 1412.5016}}].

\bibitem{Belanger:2014vza}
G.~Belanger, F.~Boudjema, A.~Pukhov and A.~Semenov, \emph{{micrOMEGAs4.1: two
  dark matter candidates}},
  \href{https://doi.org/10.1016/j.cpc.2015.03.003}{\emph{Comput. Phys. Commun.}
  {\bfseries 192} (2015) 322--329},
  [\href{https://arxiv.org/abs/1407.6129}{{\ttfamily 1407.6129}}].

\bibitem{Belyaev:2012qa}
A.~Belyaev, N.~D. Christensen and A.~Pukhov, \emph{{CalcHEP 3.4 for collider
  physics within and beyond the Standard Model}},
  \href{https://doi.org/10.1016/j.cpc.2013.01.014}{\emph{Comput. Phys. Commun.}
  {\bfseries 184} (2013) 1729--1769},
  [\href{https://arxiv.org/abs/1207.6082}{{\ttfamily 1207.6082}}].

\bibitem{Workgroup:2017htr}
{\scshape GAMBIT} Collaboration, G.~D. Martinez, J.~McKay, B.~Farmer, P.~Scott,
  E.~Roebber, A.~Putze et~al., \emph{{Comparison of statistical sampling
  methods with ScannerBit, the GAMBIT scanning module}},
  \href{https://doi.org/10.1140/epjc/s10052-017-5274-y}{\emph{Eur. Phys. J. C}
  {\bfseries 77} (2017) 761},
  [\href{https://arxiv.org/abs/1705.07959}{{\ttfamily 1705.07959}}].

\bibitem{Cline:2012hg}
J.~M. Cline and K.~Kainulainen, \emph{{Electroweak baryogenesis and dark matter
  from a singlet Higgs}},
  \href{https://doi.org/10.1088/1475-7516/2013/01/012}{\emph{JCAP} {\bfseries
  1301} (2013) 012}, [\href{https://arxiv.org/abs/1210.4196}{{\ttfamily
  1210.4196}}].

\bibitem{Agrawal:2010fh}
P.~Agrawal, Z.~Chacko, C.~Kilic and R.~K. Mishra, \emph{{A Classification of
  Dark Matter Candidates with Primarily Spin-Dependent Interactions with
  Matter}},  [\href{https://arxiv.org/abs/1003.1912}{{\ttfamily 1003.1912}}].

\bibitem{Esch:2013rta}
S.~Esch, M.~Klasen and C.~E. Yaguna, \emph{{Detection prospects of singlet
  fermionic dark matter}},
  \href{https://doi.org/10.1103/PhysRevD.88.075017}{\emph{Phys. Rev. D}
  {\bfseries 88} (2013) 075017},
  [\href{https://arxiv.org/abs/1308.0951}{{\ttfamily 1308.0951}}].

\bibitem{Bagherian:2014iia}
Z.~Bagherian, M.~M. Ettefaghi, Z.~Haghgouyan and R.~Moazzemi, \emph{{A new
  parameter space study of the fermionic cold dark matter model}},
  \href{https://doi.org/10.1088/1475-7516/2014/10/033}{\emph{JCAP} {\bfseries
  1410} (2014) 033}, [\href{https://arxiv.org/abs/1406.2927}{{\ttfamily
  1406.2927}}].

\bibitem{Franarin:2014yua}
T.~H. Franarin, C.~A.~Z. Vasconcellos and D.~Hadjimichef, \emph{{On the
  possibility of a 130 GeV gamma-ray line from annihilating singlet fermionic
  dark matter}}, \href{https://doi.org/10.1002/asna.201412087}{\emph{Astron.
  Nachr.} {\bfseries 335} (2014) 647--652},
  [\href{https://arxiv.org/abs/1404.0406}{{\ttfamily 1404.0406}}].

\bibitem{Kim:2016csm}
Y.~G. Kim, K.~Y. Lee, C.~B. Park and S.~Shin, \emph{{Secluded singlet fermionic
  dark matter driven by the Fermi gamma-ray excess}},
  \href{https://doi.org/10.1103/PhysRevD.93.075023}{\emph{Phys. Rev. D}
  {\bfseries 93} (2016) 075023},
  [\href{https://arxiv.org/abs/1601.05089}{{\ttfamily 1601.05089}}].

\bibitem{Ghorbani:2014qpa}
K.~Ghorbani, \emph{{Fermionic dark matter with pseudo-scalar Yukawa
  interaction}},
  \href{https://doi.org/10.1088/1475-7516/2015/01/015}{\emph{JCAP} {\bfseries
  1501} (2015) 015}, [\href{https://arxiv.org/abs/1408.4929}{{\ttfamily
  1408.4929}}].

\bibitem{Balazs:2015boa}
C.~Balazs, T.~Li, C.~Savage and M.~White, \emph{{Interpreting the Fermi-LAT
  gamma ray excess in the simplified framework}},
  \href{https://doi.org/10.1103/PhysRevD.92.123520}{\emph{Phys. Rev. D}
  {\bfseries 92} (2015) 123520},
  [\href{https://arxiv.org/abs/1505.06758}{{\ttfamily 1505.06758}}].

\bibitem{Athron:2018hpc}
{\scshape GAMBIT} Collaboration, P.~Athron et~al., \emph{{Global analyses of
  Higgs portal singlet dark matter models using GAMBIT}},
  \href{https://doi.org/10.1140/epjc/s10052-018-6513-6}{\emph{Eur. Phys. J. C}
  {\bfseries 79} (2019) 38},
  [\href{https://arxiv.org/abs/1808.10465}{{\ttfamily 1808.10465}}].

\bibitem{Kim:2018uov}
Y.~G. Kim, C.~B. Park and S.~Shin, \emph{{Collider probes of singlet fermionic
  dark matter scenarios for the Fermi gamma-ray excess}},
  \href{https://doi.org/10.1007/JHEP12(2018)036}{\emph{JHEP} {\bfseries 12}
  (2018) 036}, [\href{https://arxiv.org/abs/1809.01143}{{\ttfamily
  1809.01143}}].

\bibitem{Workgroup:2017lvb}
{\scshape The GAMBIT Dark Matter Workgroup} Collaboration, T.~Bringmann et~al.,
  \emph{{DarkBit: A GAMBIT module for computing dark matter observables and
  likelihoods}},
  \href{https://doi.org/10.1140/epjc/s10052-017-5155-4}{\emph{Eur. Phys. J. C}
  {\bfseries 77} (2017) 831},
  [\href{https://arxiv.org/abs/1705.07920}{{\ttfamily 1705.07920}}].

\bibitem{Grojean:2006bp}
C.~Grojean and G.~Servant, \emph{{Gravitational Waves from Phase Transitions at
  the Electroweak Scale and Beyond}},
  \href{https://doi.org/10.1103/PhysRevD.75.043507}{\emph{Phys. Rev. D}
  {\bfseries 75} (2007) 043507},
  [\href{https://arxiv.org/abs/hep-ph/0607107}{{\ttfamily hep-ph/0607107}}].

\bibitem{Quiros:1999jp}
M.~Quiros, \emph{{Finite temperature field theory and phase transitions}},  in
  \emph{{Proceedings, Summer School in High-energy physics and cosmology:
  Trieste, Italy, June 29-July 17, 1998}}, pp.~187--259, 1999,
  [\href{https://arxiv.org/abs/hep-ph/9901312}{{\ttfamily hep-ph/9901312}}].

\bibitem{Funakubo:2009eg}
K.~Funakubo and E.~Senaha, \emph{{Electroweak phase transition, critical
  bubbles and sphaleron decoupling condition in the MSSM}},
  \href{https://doi.org/10.1103/PhysRevD.79.115024}{\emph{Phys. Rev. D}
  {\bfseries 79} (2009) 115024},
  [\href{https://arxiv.org/abs/0905.2022}{{\ttfamily 0905.2022}}].

\bibitem{Katz:2014bha}
A.~Katz and M.~Perelstein, \emph{{Higgs Couplings and Electroweak Phase
  Transition}}, \href{https://doi.org/10.1007/JHEP07(2014)108}{\emph{JHEP}
  {\bfseries 07} (2014) 108},
  [\href{https://arxiv.org/abs/1401.1827}{{\ttfamily 1401.1827}}].

\bibitem{Fuyuto:2014yia}
K.~Fuyuto and E.~Senaha, \emph{{Improved sphaleron decoupling condition and the
  Higgs coupling constants in the real singlet-extended standard model}},
  \href{https://doi.org/10.1103/PhysRevD.90.015015}{\emph{Phys. Rev. D}
  {\bfseries 90} (2014) 015015},
  [\href{https://arxiv.org/abs/1406.0433}{{\ttfamily 1406.0433}}].

\bibitem{Ellis:2018mja}
J.~Ellis, M.~Lewicki and J.~M. No, \emph{{On the Maximal Strength of a
  First-Order Electroweak Phase Transition and its Gravitational Wave Signal}},
   [\href{https://arxiv.org/abs/1809.08242}{{\ttfamily 1809.08242}}].

\bibitem{Bodeker:2009qy}
D.~Bodeker and G.~D. Moore, \emph{{Can electroweak bubble walls run away?}},
  \href{https://doi.org/10.1088/1475-7516/2009/05/009}{\emph{JCAP} {\bfseries
  0905} (2009) 009}, [\href{https://arxiv.org/abs/0903.4099}{{\ttfamily
  0903.4099}}].

\bibitem{Kozaczuk:2015owa}
J.~Kozaczuk, \emph{{Bubble Expansion and the Viability of Singlet-Driven
  Electroweak Baryogenesis}},
  \href{https://doi.org/10.1007/JHEP10(2015)135}{\emph{JHEP} {\bfseries 10}
  (2015) 135}, [\href{https://arxiv.org/abs/1506.04741}{{\ttfamily
  1506.04741}}].

\bibitem{Kurup:2017dzf}
G.~Kurup and M.~Perelstein, \emph{{Dynamics of Electroweak Phase Transition In
  Singlet-Scalar Extension of the Standard Model}},
  \href{https://doi.org/10.1103/PhysRevD.96.015036}{\emph{Phys. Rev. D}
  {\bfseries 96} (2017) 015036},
  [\href{https://arxiv.org/abs/1704.03381}{{\ttfamily 1704.03381}}].

\bibitem{No:2011fi}
J.~M. No, \emph{{Large Gravitational Wave Background Signals in Electroweak
  Baryogenesis Scenarios}},
  \href{https://doi.org/10.1103/PhysRevD.84.124025}{\emph{Phys. Rev. D}
  {\bfseries 84} (2011) 124025},
  [\href{https://arxiv.org/abs/1103.2159}{{\ttfamily 1103.2159}}].

\bibitem{Caprini:2011uz}
C.~Caprini and J.~M. No, \emph{{Supersonic Electroweak Baryogenesis: Achieving
  Baryogenesis for Fast Bubble Walls}},
  \href{https://doi.org/10.1088/1475-7516/2012/01/031}{\emph{JCAP} {\bfseries
  1201} (2012) 031}, [\href{https://arxiv.org/abs/1111.1726}{{\ttfamily
  1111.1726}}].

\bibitem{Katz:2016adq}
A.~Katz and A.~Riotto, \emph{{Baryogenesis and Gravitational Waves from Runaway
  Bubble Collisions}},
  \href{https://doi.org/10.1088/1475-7516/2016/11/011}{\emph{JCAP} {\bfseries
  1611} (2016) 011}, [\href{https://arxiv.org/abs/1608.00583}{{\ttfamily
  1608.00583}}].

\bibitem{Peskin:1991sw}
M.~E. Peskin and T.~Takeuchi, \emph{{Estimation of oblique electroweak
  corrections}}, \href{https://doi.org/10.1103/PhysRevD.46.381}{\emph{Phys.
  Rev. D} {\bfseries 46} (1992) 381--409}.

\bibitem{Grimus:2008nb}
W.~Grimus, L.~Lavoura, O.~M. Ogreid and P.~Osland, \emph{{The Oblique
  parameters in multi-Higgs-doublet models}},
  \href{https://doi.org/10.1016/j.nuclphysb.2008.04.019}{\emph{Nucl. Phys. B}
  {\bfseries 801} (2008) 81--96},
  [\href{https://arxiv.org/abs/0802.4353}{{\ttfamily 0802.4353}}].

\bibitem{Profumo:2014opa}
S.~Profumo, M.~J. Ramsey-Musolf, C.~L. Wainwright and P.~Winslow,
  \emph{{Singlet-catalyzed electroweak phase transitions and precision Higgs
  boson studies}},
  \href{https://doi.org/10.1103/PhysRevD.91.035018}{\emph{Phys. Rev. D}
  {\bfseries 91} (2015) 035018},
  [\href{https://arxiv.org/abs/1407.5342}{{\ttfamily 1407.5342}}].

\bibitem{Djouadi:2005gi}
A.~Djouadi, \emph{{The Anatomy of electro-weak symmetry breaking. I: The Higgs
  boson in the standard model}},
  \href{https://doi.org/10.1016/j.physrep.2007.10.004}{\emph{Phys. Rept.}
  {\bfseries 457} (2008) 1--216},
  [\href{https://arxiv.org/abs/hep-ph/0503172}{{\ttfamily hep-ph/0503172}}].

\bibitem{Aad:2015gba}
{\scshape ATLAS} Collaboration, G.~Aad et~al., \emph{{Measurements of the Higgs
  boson production and decay rates and coupling strengths using pp collision
  data at $\sqrt{s}=7$ and 8 TeV in the ATLAS experiment}},
  \href{https://doi.org/10.1140/epjc/s10052-015-3769-y}{\emph{Eur. Phys. J. C}
  {\bfseries 76} (2016) 6}, [\href{https://arxiv.org/abs/1507.04548}{{\ttfamily
  1507.04548}}].

\bibitem{Chatrchyan:2013iaa}
{\scshape CMS} Collaboration, S.~Chatrchyan et~al., \emph{{Measurement of Higgs
  boson production and properties in the WW decay channel with leptonic final
  states}}, \href{https://doi.org/10.1007/JHEP01(2014)096}{\emph{JHEP}
  {\bfseries 01} (2014) 096},
  [\href{https://arxiv.org/abs/1312.1129}{{\ttfamily 1312.1129}}].

\bibitem{Chatrchyan:2013mxa}
{\scshape CMS} Collaboration, S.~Chatrchyan et~al., \emph{{Measurement of the
  properties of a Higgs boson in the four-lepton final state}},
  \href{https://doi.org/10.1103/PhysRevD.89.092007}{\emph{Phys. Rev. D}
  {\bfseries 89} (2014) 092007},
  [\href{https://arxiv.org/abs/1312.5353}{{\ttfamily 1312.5353}}].

\bibitem{Khachatryan:2014ira}
{\scshape CMS} Collaboration, V.~Khachatryan et~al., \emph{{Observation of the
  diphoton decay of the Higgs boson and measurement of its properties}},
  \href{https://doi.org/10.1140/epjc/s10052-014-3076-z}{\emph{Eur. Phys. J. C}
  {\bfseries 74} (2014) 3076},
  [\href{https://arxiv.org/abs/1407.0558}{{\ttfamily 1407.0558}}].

\bibitem{Chatrchyan:2014vua}
{\scshape CMS} Collaboration, S.~Chatrchyan et~al., \emph{{Evidence for the
  direct decay of the 125 GeV Higgs boson to fermions}},
  \href{https://doi.org/10.1038/nphys3005}{\emph{Nature Phys.} {\bfseries 10}
  (2014) 557--560}, [\href{https://arxiv.org/abs/1401.6527}{{\ttfamily
  1401.6527}}].

\bibitem{Stal:2013hwa}
O.~Stal and T.~Stefaniak, \emph{{Constraining extended Higgs sectors with
  HiggsSignals}}, {\emph{PoS} {\bfseries EPS-HEP2013} (2013) 314},
  [\href{https://arxiv.org/abs/1310.4039}{{\ttfamily 1310.4039}}].

\bibitem{Cowan:2010js}
G.~Cowan, K.~Cranmer, E.~Gross and O.~Vitells, \emph{{Asymptotic formulae for
  likelihood-based tests of new physics}},
  \href{https://doi.org/10.1140/epjc/s10052-011-1554-0,
  10.1140/epjc/s10052-013-2501-z}{\emph{Eur. Phys. J. C} {\bfseries 71} (2011)
  1554}, [\href{https://arxiv.org/abs/1007.1727}{{\ttfamily 1007.1727}}].

\bibitem{Wilks:1938dza}
S.~S. Wilks, \emph{{The Large-Sample Distribution of the Likelihood Ratio for
  Testing Composite Hypotheses}},
  \href{https://doi.org/10.1214/aoms/1177732360}{\emph{Annals Math. Statist.}
  {\bfseries 9} (1938) 60--62}.

\bibitem{Scott:2012qh}
P.~Scott, \emph{{Pippi - painless parsing, post-processing and plotting of
  posterior and likelihood samples}},
  \href{https://doi.org/10.1140/epjp/i2012-12138-3}{\emph{Eur. Phys. J. Plus}
  {\bfseries 127} (2012) 138},
  [\href{https://arxiv.org/abs/1206.2245}{{\ttfamily 1206.2245}}].

\bibitem{Akerib:2018lyp}
{\scshape LUX-ZEPLIN} Collaboration, D.~S. Akerib et~al., \emph{{Projected WIMP
  Sensitivity of the LUX-ZEPLIN (LZ) Dark Matter Experiment}},
  [\href{https://arxiv.org/abs/1802.06039}{{\ttfamily 1802.06039}}].

\bibitem{Turner:1992tz}
M.~S. Turner, E.~J. Weinberg and L.~M. Widrow, \emph{{Bubble nucleation in
  first order inflation and other cosmological phase transitions}},
  \href{https://doi.org/10.1103/PhysRevD.46.2384}{\emph{Phys. Rev. D}
  {\bfseries 46} (1992) 2384--2403}.

\bibitem{Guth:1982pn}
A.~H. Guth and E.~J. Weinberg, \emph{{Could the Universe Have Recovered from a
  Slow First Order Phase Transition?}},
  \href{https://doi.org/10.1016/0550-3213(83)90307-3}{\emph{Nucl. Phys. B}
  {\bfseries 212} (1983) 321--364}.

\bibitem{Enqvist:1991xw}
K.~Enqvist, J.~Ignatius, K.~Kajantie and K.~Rummukainen, \emph{{Nucleation and
  bubble growth in a first order cosmological electroweak phase transition}},
  \href{https://doi.org/10.1103/PhysRevD.45.3415}{\emph{Phys. Rev. D}
  {\bfseries 45} (1992) 3415--3428}.

\bibitem{Kamionkowski:1993fg}
M.~Kamionkowski, A.~Kosowsky and M.~S. Turner, \emph{{Gravitational radiation
  from first order phase transitions}},
  \href{https://doi.org/10.1103/PhysRevD.49.2837}{\emph{Phys. Rev. D}
  {\bfseries 49} (1994) 2837--2851},
  [\href{https://arxiv.org/abs/astro-ph/9310044}{{\ttfamily
  astro-ph/9310044}}].

\bibitem{Huber:2008hg}
S.~J. Huber and T.~Konstandin, \emph{{Gravitational Wave Production by
  Collisions: More Bubbles}},
  \href{https://doi.org/10.1088/1475-7516/2008/09/022}{\emph{JCAP} {\bfseries
  0809} (2008) 022}, [\href{https://arxiv.org/abs/0806.1828}{{\ttfamily
  0806.1828}}].

\bibitem{Jinno:2016vai}
R.~Jinno and M.~Takimoto, \emph{{Gravitational waves from bubble collisions: An
  analytic derivation}},
  \href{https://doi.org/10.1103/PhysRevD.95.024009}{\emph{Phys. Rev. D}
  {\bfseries 95} (2017) 024009},
  [\href{https://arxiv.org/abs/1605.01403}{{\ttfamily 1605.01403}}].

\bibitem{Jinno:2017fby}
R.~Jinno and M.~Takimoto, \emph{{Gravitational waves from bubble dynamics:
  Beyond the Envelope}},
  \href{https://doi.org/10.1088/1475-7516/2019/01/060}{\emph{JCAP} {\bfseries
  1901} (2019) 060}, [\href{https://arxiv.org/abs/1707.03111}{{\ttfamily
  1707.03111}}].

\bibitem{Bodeker:2017cim}
D.~Bodeker and G.~D. Moore, \emph{{Electroweak Bubble Wall Speed Limit}},
  \href{https://doi.org/10.1088/1475-7516/2017/05/025}{\emph{JCAP} {\bfseries
  1705} (2017) 025}, [\href{https://arxiv.org/abs/1703.08215}{{\ttfamily
  1703.08215}}].

\bibitem{Hindmarsh:2013xza}
M.~Hindmarsh, S.~J. Huber, K.~Rummukainen and D.~J. Weir, \emph{{Gravitational
  waves from the sound of a first order phase transition}},
  \href{https://doi.org/10.1103/PhysRevLett.112.041301}{\emph{Phys. Rev. Lett.}
  {\bfseries 112} (2014) 041301},
  [\href{https://arxiv.org/abs/1304.2433}{{\ttfamily 1304.2433}}].

\bibitem{Hindmarsh:2015qta}
M.~Hindmarsh, S.~J. Huber, K.~Rummukainen and D.~J. Weir, \emph{{Numerical
  simulations of acoustically generated gravitational waves at a first order
  phase transition}},
  \href{https://doi.org/10.1103/PhysRevD.92.123009}{\emph{Phys. Rev. D}
  {\bfseries 92} (2015) 123009},
  [\href{https://arxiv.org/abs/1504.03291}{{\ttfamily 1504.03291}}].

\bibitem{Hindmarsh:2016lnk}
M.~Hindmarsh, \emph{{Sound shell model for acoustic gravitational wave
  production at a first-order phase transition in the early Universe}},
  \href{https://doi.org/10.1103/PhysRevLett.120.071301}{\emph{Phys. Rev. Lett.}
  {\bfseries 120} (2018) 071301},
  [\href{https://arxiv.org/abs/1608.04735}{{\ttfamily 1608.04735}}].

\bibitem{Hindmarsh:2017gnf}
M.~Hindmarsh, S.~J. Huber, K.~Rummukainen and D.~J. Weir, \emph{{Shape of the
  acoustic gravitational wave power spectrum from a first order phase
  transition}}, \href{https://doi.org/10.1103/PhysRevD.96.103520}{\emph{Phys.
  Rev. D} {\bfseries 96} (2017) 103520},
  [\href{https://arxiv.org/abs/1704.05871}{{\ttfamily 1704.05871}}].

\bibitem{Caprini:2009yp}
C.~Caprini, R.~Durrer and G.~Servant, \emph{{The stochastic gravitational wave
  background from turbulence and magnetic fields generated by a first-order
  phase transition}},
  \href{https://doi.org/10.1088/1475-7516/2009/12/024}{\emph{JCAP} {\bfseries
  0912} (2009) 024}, [\href{https://arxiv.org/abs/0909.0622}{{\ttfamily
  0909.0622}}].

\bibitem{Kosowsky:2001xp}
A.~Kosowsky, A.~Mack and T.~Kahniashvili, \emph{{Gravitational radiation from
  cosmological turbulence}},
  \href{https://doi.org/10.1103/PhysRevD.66.024030}{\emph{Phys. Rev. D}
  {\bfseries 66} (2002) 024030},
  [\href{https://arxiv.org/abs/astro-ph/0111483}{{\ttfamily
  astro-ph/0111483}}].

\bibitem{Gogoberidze:2007an}
G.~Gogoberidze, T.~Kahniashvili and A.~Kosowsky, \emph{{The Spectrum of
  Gravitational Radiation from Primordial Turbulence}},
  \href{https://doi.org/10.1103/PhysRevD.76.083002}{\emph{Phys. Rev. D}
  {\bfseries 76} (2007) 083002},
  [\href{https://arxiv.org/abs/0705.1733}{{\ttfamily 0705.1733}}].

\bibitem{Niksa:2018ofa}
P.~Niksa, M.~Schlederer and G.~Sigl, \emph{{Gravitational Waves produced by
  Compressible MHD Turbulence from Cosmological Phase Transitions}},
  \href{https://doi.org/10.1088/1361-6382/aac89c}{\emph{Class. Quant. Grav.}
  {\bfseries 35} (2018) 144001},
  [\href{https://arxiv.org/abs/1803.02271}{{\ttfamily 1803.02271}}].

\bibitem{Bartolo:2016ami}
N.~Bartolo et~al., \emph{{Science with the space-based interferometer LISA. IV:
  Probing inflation with gravitational waves}},
  \href{https://doi.org/10.1088/1475-7516/2016/12/026}{\emph{JCAP} {\bfseries
  1612} (2016) 026}, [\href{https://arxiv.org/abs/1610.06481}{{\ttfamily
  1610.06481}}].

\bibitem{Yagi:2011wg}
K.~Yagi and N.~Seto, \emph{{Detector configuration of DECIGO/BBO and
  identification of cosmological neutron-star binaries}},
  \href{https://doi.org/10.1103/PhysRevD.95.109901,
  10.1103/PhysRevD.83.044011}{\emph{Phys. Rev. D} {\bfseries 83} (2011)
  044011}, [\href{https://arxiv.org/abs/1101.3940}{{\ttfamily 1101.3940}}].

\bibitem{TheLIGOScientific:2014jea}
{\scshape LIGO Scientific} Collaboration, J.~Aasi et~al., \emph{{Advanced
  LIGO}}, \href{https://doi.org/10.1088/0264-9381/32/7/074001}{\emph{Class.
  Quant. Grav.} {\bfseries 32} (2015) 074001},
  [\href{https://arxiv.org/abs/1411.4547}{{\ttfamily 1411.4547}}].

\bibitem{TheLIGOScientific:2016wyq}
{\scshape Virgo, LIGO Scientific} Collaboration, B.~P. Abbott et~al.,
  \emph{{GW150914: Implications for the stochastic gravitational wave
  background from binary black holes}},
  \href{https://doi.org/10.1103/PhysRevLett.116.131102}{\emph{Phys. Rev. Lett.}
  {\bfseries 116} (2016) 131102},
  [\href{https://arxiv.org/abs/1602.03847}{{\ttfamily 1602.03847}}].

\bibitem{Thrane:2013oya}
E.~Thrane and J.~D. Romano, \emph{{Sensitivity curves for searches for
  gravitational-wave backgrounds}},
  \href{https://doi.org/10.1103/PhysRevD.88.124032}{\emph{Phys. Rev. D}
  {\bfseries 88} (2013) 124032},
  [\href{https://arxiv.org/abs/1310.5300}{{\ttfamily 1310.5300}}].

\bibitem{vanHaasteren:2011ni}
R.~van Haasteren et~al., \emph{{Placing limits on the stochastic
  gravitational-wave background using European Pulsar Timing Array data}},
  \href{https://doi.org/10.1111/j.1365-2966.2011.18613.x,
  10.1111/j.1365-2966.2012.20916.x}{\emph{Mon. Not. Roy. Astron. Soc.}
  {\bfseries 414} (2011) 3117--3128},
  [\href{https://arxiv.org/abs/1103.0576}{{\ttfamily 1103.0576}}].

\bibitem{Janssen:2014dka}
G.~Janssen et~al., \emph{{Gravitational wave astronomy with the SKA}},
  {\emph{PoS} {\bfseries AASKA14} (2015) 037},
  [\href{https://arxiv.org/abs/1501.00127}{{\ttfamily 1501.00127}}].

\bibitem{Evans:2016mbw}
{\scshape LIGO Scientific} Collaboration, B.~P. Abbott et~al., \emph{{Exploring
  the Sensitivity of Next Generation Gravitational Wave Detectors}},
  \href{https://doi.org/10.1088/1361-6382/aa51f4}{\emph{Class. Quant. Grav.}
  {\bfseries 34} (2017) 044001},
  [\href{https://arxiv.org/abs/1607.08697}{{\ttfamily 1607.08697}}].

\bibitem{Punturo:2010zz}
M.~Punturo et~al., \emph{{The Einstein Telescope: A third-generation
  gravitational wave observatory}},
  \href{https://doi.org/10.1088/0264-9381/27/19/194002}{\emph{Class. Quant.
  Grav.} {\bfseries 27} (2010) 194002}.

\bibitem{Hild:2010id}
S.~Hild et~al., \emph{{Sensitivity Studies for Third-Generation Gravitational
  Wave Observatories}},
  \href{https://doi.org/10.1088/0264-9381/28/9/094013}{\emph{Class. Quant.
  Grav.} {\bfseries 28} (2011) 094013},
  [\href{https://arxiv.org/abs/1012.0908}{{\ttfamily 1012.0908}}].

\bibitem{Ellis:2016jkw}
J.~Ellis, \emph{{TikZ-Feynman: Feynman diagrams with TikZ}},
  \href{https://doi.org/10.1016/j.cpc.2016.08.019}{\emph{Comput. Phys. Commun.}
  {\bfseries 210} (2017) 103--123},
  [\href{https://arxiv.org/abs/1601.05437}{{\ttfamily 1601.05437}}].

\bibitem{Berlin:2014tja}
A.~Berlin, D.~Hooper and S.~D. McDermott, \emph{{Simplified Dark Matter Models
  for the Galactic Center Gamma-Ray Excess}},
  \href{https://doi.org/10.1103/PhysRevD.89.115022}{\emph{Phys. Rev. D}
  {\bfseries 89} (2014) 115022},
  [\href{https://arxiv.org/abs/1404.0022}{{\ttfamily 1404.0022}}].

\bibitem{Shifman:1978}
M.~Shifman, A.~Vainshtein and V.~Zakharov, \emph{Remarks on higgs-boson
  interactions with nucleons},
  \href{https://doi.org/http://dx.doi.org/10.1016/0370-2693(78)90481-1}{\emph{Phys.
  Lett. B} {\bfseries 78} (1978) 443 -- 446}.

\bibitem{Delaunay:2007wb}
C.~Delaunay, C.~Grojean and J.~D. Wells, \emph{{Dynamics of Non-renormalizable
  Electroweak Symmetry Breaking}},
  \href{https://doi.org/10.1088/1126-6708/2008/04/029}{\emph{JHEP} {\bfseries
  04} (2008) 029}, [\href{https://arxiv.org/abs/0711.2511}{{\ttfamily
  0711.2511}}].

\bibitem{Carrington:1991hz}
M.~E. Carrington, \emph{{The Effective potential at finite temperature in the
  Standard Model}}, \href{https://doi.org/10.1103/PhysRevD.45.2933}{\emph{Phys.
  Rev. D} {\bfseries 45} (1992) 2933--2944}.

\end{thebibliography}\endgroup

\end{document}